\documentclass[11pt,lettersize]{article}
\usepackage[margin=1in]{geometry}
\usepackage{amsmath,amsfonts}
\usepackage{array}
\usepackage[caption=false,font=normalsize,labelfont=sf,textfont=sf]{subfig}
\usepackage{textcomp}
\usepackage{stfloats}
\usepackage{url}
\usepackage{verbatim}
\usepackage{graphicx}
\usepackage{afterpage}
\usepackage{cite}
\renewcommand{\index}[1]{}

\usepackage{subfig}
\usepackage{multirow}
\usepackage{algorithm}
\usepackage[noend]{algpseudocode}

\begin{document}

\title{Long-Term and Short-Term Transistor Aging in Deep Neural Networks: Impact and Mitigation}

\author{Alireza Sarmadi$^*$, Virinchi Roy Surabhi$^*$, Prashanth Krishnamurthy,\\ Hussam Amrouch, Ramesh Karri, Farshad Khorrami
\thanks{The first, second, third, fifth, and sixth authors are with the Dept. of Electrical and Computer Engineering, New York University (NYU) Tandon School of Engineering, New York, USA. The fourth author is with the School of Computation, Information and Technology, Technical University of Munich (TUM), Munich, Germany. Email addresses: as11986@nyu.edu, vs1930@nyu.edu, prashanth.krishnamurthy@nyu.edu, amrouch@tum.de, rkarri@nyu.edu, khorrami@nyu.edu}%
\thanks{This work was supported in part by the U.S. Office of Naval Research under Grants N00014-18-1-2672,  N00014-21-1-2390, and N00014-22-1-2314.}
\thanks{$^*$Equal contribution.}
}
\date{}

\maketitle


\begin{abstract}
   Deep neural networks (DNNs)\index{deep neural networks (DNNs)} are used in a variety of real-world applications including, for example, image classification and speech recognition. %
   The inference accuracy of DNN implemented on hardware in integrated circuits (ICs) %
   degrades under phenomena such as transistor aging\index{transistor aging}. %
   Aging slows down the switching speed of transistors, resulting in system-level timing violations due to unsustainable clocks. To maintain reliability for the entire projected lifetime, %
   designers add guardbands to prevent timing violations; however, adding large timing guardbands causes losses in performance (speed or throughput). This article provides a detailed discussion of the effects of long-term and short-term transistor aging on DNN inference accuracy.  Furthermore, to mitigate aging effects on DNN's accuracy and keep them at bay, a methodology for aging-aware retraining\index{aging-aware resilience} %
   is presented in order to generate a resilient DNN even when aggressive (i.e., smaller than required)  guardbands are used. This improves the inference accuracy of the DNNs even in the presence of  aging-induced degradation. These effects are discussed in this article along with mitigation strategies on a hardware implementation of a DNN for image classification on an off-the-shelf image dataset.
  The application of short-term aging as an excitation mechanism for the detection of hardware Trojans in integrated circuits is also briefly discussed.
\end{abstract}

\section{Introduction}
Deep neural networks (DNNs) have aided advances in real-time image classification,  visual recognition, and other tasks. Training and inference of large DNNs require high computational capacity \cite{sze2017efficient}.
To reduce the need for power-hungry and expensive programmable CPUs and GPUs and enable a more efficient hardware-based realization, researchers have proposed specialized hardware implementations such as systolic arrays and multiply accumulate (MAC) units~\cite{chen2017eyeriss,moons2016tops,han2016eie,whatmough2017soc}. These   hardware implementations leverage design margins and slacks to achieve area and energy efficiency without compromising the throughput. On the other hand, data quantization and compression techniques \cite{han2015deep,gysel2016hardware} lower energy consumption by simplifying parameters, utilizing the extra timing margins in critical paths during the design phase \cite{whatmough2017soc,moons2017energy}. %
Other  hardware implementation approaches use  aggressive precision scaling, voltage scaling, data compression, and tolerance of timing error rates to boost performance and energy efficiency~\cite{chen2017eyeriss,moons2016tops,han2016eie,whatmough2017soc}. These hardware implementations use advanced CMOS technologies including FinFET. Long-term aging\index{aging!long-term}\index{aging} effects on hardware implementations have been studied in some recent works \cite{salamin2021reliability,liu2019analysis}.  While they study the effects of aging on individual MAC units that can be used for implementing DNNs,
 they do not investigate the effects on the entire DNN when it is completely realized in hardware. For applications requiring very high throughput or low latency, entire DNNs have to be implemented in hardware. This article explores the effects of aging on DNNs implemented fully in hardware, including multiple layers and non-linear activation functions on ICs. While many existing solutions focus on specific components like MAC units and systolic arrays due to their efficiency and feasibility, the discussion in this article broadens the analysis to the entire DNN to provide deeper insights into aging-related challenges and mitigation strategies for complete hardware-based implementations.%

Threshold voltage shift\index{threshold voltage} in FinFET\index{FinFET} transistors affects the switching speed of logic gates, limiting the maximum operating frequency. Hot carrier injection (HCI)\index{hot carrier injection (HCI)} and negative bias temperature instability (NBTI)\index{negative bias temperature instability (NBTI)} are two key aging mechanisms that have been identified to be the principal causes of performance degradation in n-type and p-type devices~\cite{wang2007compact}.
These aging mechanisms  increase timing degradation in combinational circuits.
Moreover,  high temperatures exacerbates the problem because the bulk of processes behind transistor aging are exponentially dependent on operating temperature \cite{mahapatra2013comparative,sootkaneung2017thermal}. Aging effects have become a key short-term issue as has been demonstrated in recent studies \cite{vansanten2016agingaware}. Short-term aging\index{aging!short-term} causes degradation within a short period of time.
When transistor aging is paired with voltage scaling, the impact of aging-induced deterioration (i.e., $\Delta V_{th}$ increase) at high voltage is even higher once the voltage is scaled down to a lower level. This is because transistor delay is proportional to ($V_{dd}$ - $V_{th}$), and therefore the same amount of $V_{th}$ has a considerably higher influence on current and hence on transistor delay at lower $V_{dd}$. Short-term aging occurs in advanced technology nodes ($\leq$ 45nm). It is fair to speculate that the accuracy of DNN calculations will decrease with time (long-term) or for a short time (short-term) in energy-efficient hardware implementations intended to run on the edge of the timing margin to maximize performance. This article discusses the deleterious impact that short-term aging effects have on the accuracy of DNNs.

Timing errors\index{timing errors} occur in circuits when the operating frequency becomes unsustainable (i.e., results in setup timing violations) due to aging-induced degradations. To  mitigate short-term and long-term aging effects, a timing guardband\index{timing guardband} must be applied on top of the critical path delay during the design phase. However, this leads to efficiency loss of the chip over its lifetime, even when these degradations are minimal or nonexistent. The cost of guardband is incurred from the outset, even if it is not necessary at the beginning of chip operation (i.e., at time zero). To employ a narrower guardband and hence recover DNN inference accuracy, aging-induced  degradations must be mitigated. Neglecting  aging-induced faults on inference outputs  of hardware implemented DNNs could be disastrous since they will result in incorrect outputs of the DNN and misclassification, potentially causing significant real-world impact depending on the application.

This article first discusses the susceptibility of outputs of DNN layers to aging and quantify the decline in inference accuracy as a function of aging (both classical long-term aging as well as short-term aging). Thereafter, techniques are presented to counteract inference accuracy degradation induced by aging. %
It is seen that a significant accuracy loss can be caused by aging; for example, in FinFET technology, at 0.6V, an aging level that corresponds to a 20\% duty cycle leads to a 50\% classification accuracy drop. %

This article provides the following items:
\begin{enumerate}
\item The impact of aging\index{aging} on DNN inference accuracy is investigated by performing a detailed analysis of a gate-level implementation of a complete DNN. The analysis shows that transistor aging can significantly degrade DNN accuracy. While earlier studies \cite{liu2019analysis,salamin2021reliability} have focused primarily on aging effects in individual components like MAC units, the entire DNN implemented in hardware is considered in this article, including multiple layers and non-linear activation functions. This approach allows us to capture the cumulative aging effects across all layers and components, providing a more comprehensive understanding of the overall system's behavior. Although MAC units and systolic arrays are widely used due to their efficiency, studying the entire DNN helps to reveal potential vulnerabilities in fully hardware-accelerated deep learning systems, which may not be apparent when only isolated components are examined.

\item The inference accuracy\index{inference accuracy} deterioration of DNN due to short-term aging is examined. In advanced nodes ($\leq$ 45nm), fast voltage switching from higher to lower levels leads to the accumulation of aging degradations at higher voltages onto lower voltages. This causes transient errors in the circuit. Therefore, it is essential to investigate short-term aging in conjunction with long-term aging in DNNs.
\item Mitigation strategies\index{aging-aware mitigation} are presented for inference accuracy degradation due to aging effects while preserving high clock frequency. The approach %
leverages a combination of techniques, such as addition of gradient noise and/or aging-aware feature noise during training. This improves DNN's inference accuracy without the need to reduce clock frequency or equivalently implement a larger guardband, which compromises DNN throughput. It is seen that the mitigations presented in this article reduce accuracy drops and therefore enable operation at higher clock frequencies under aging conditions without sacrificing accuracy.
\item While this article focuses primarily on aging effects in the context of their impact on DNNs, the application of short-term aging as an excitation mechanism for the detection of hardware Trojans in integrated circuits is also briefly discussed.
\end{enumerate}

The article is structured as follows. Section \ref{sec:related_work} summarizes related work. Section \ref{sec:aging} discusses transistor aging and the standard cell library creation process. Section \ref{sec:effects_age_DNN} analyzes the effects of aging on DNN's inference accuracy. Section %
\ref{sec:mitigation_gradnoise_imitator} presents mitigations to improve the DNN's inference accuracy.
Section~\ref{sec:trojan_detection} discusses the application of short-term aging for hardware Trojan detection.
Finally, Section \ref{sec:conclusion} draws conclusions.

\section{Related Work}
\label{sec:related_work}
\subsection{Resilience Studies on Hardware Implementations}
Timing errors\index{timing errors}  induced by dynamic voltage and temperature variations in adder and multiplier units are investigated in ~\cite{jiao2017assessment}. The study assumes that when a timing error occurs, the compute unit returns a random result. They evaluate the accuracy of a multi-layer perceptron (MLP) and a convolutional neural network (CNN) at various timing error rates. The susceptibility of MLPs and CNNs to bit-faults in their weights is investigated in \cite{arechiga2018effect}. Random bit flips are inserted into randomly picked weights in a trained network. %
The study shows that different numbers of inaccurate weights impact the resulting effects on overall accuracy. Another study in \cite{arechiga2018robustness} assesses the robustness of VGG16, ResNet50, and InceptionV3 by inserting faults into the weights and biases of all three DNNs.
ResNet50 and InceptionV3 are seen to be more robust to errors than VGG16. \cite{hong2019terminal} investigates the sensitivity of DNN parameters to single-bit flips. A parameter is susceptible if switching at least one of its bits causes a significant loss of DNN accuracy. They investigate criticality of bit locations for LeNet5 (on the MNIST dataset) and AlexNet (on CIFAR10). \cite{schorn2018accurate} provide a metric for assessing neurons' error robustness in DNNs. A change in the output of a neuron that contributes a minor amount to the DNN output has less influence on accuracy. %
According to \cite{neggaz2018reliability}, convolutional layers are more resistant to soft errors than fully connected layers. This is because the subsequent layers such as maxpooling mask the output of convolutional layers.

Reagen et al.~\cite{reagen2018ares} propose a DNN fault-injection framework that introduces defects either offline or during inference, reducing overhead by treating bit-flips as element-wise operations and adding noise linearly. Li et al. \cite{li2017understanding} examine the fault-tolerance capabilities of four CNNs (AlexNet, CaffeNet, NiN, and ConvNet), analyzing transient faults in buffers and datapaths, and exploring various floating and fixed point formats, as well as silent data corruption methods. Salami et al. \cite{salami2018resilience} study the fault-resilience of a 6-layer fully connected neural network using an RTL-level neural network and investigate both transient and permanent defects, noting that permanent faults cause more errors. Lastly, Sabbagh et al. \cite{sabbagh2019evaluating} analyze the effect of pruning and quantization on the fault-resilience of two DNNs (LeNet5 on MNIST and VGG16 on CIFAR-10), comparing five DNN variants, including original, unstructured pruned, structured pruned, binary-quantized, and structured+pruned models.

\subsection{Improving Resilience of DNNs in Hardware Implementations\index{neural network resiliency}}

Different components within a DNN have varying impacts on output errors. Layers with fewer neurons have more crucial contributions from each neuron~\cite{schorn2018accurate}. Faults in the mantissa minimally affect DNN accuracy, while certain exponent bits significantly impact accuracy~\cite{hong2019terminal}. Mitigation strategies leverage this knowledge to reduce protection overhead and increase efficacy. At the DNN algorithm level, critical neuron influence can be decreased by adjusting weight values or layer locations~~\cite{schorn2018accurate}. At the architecture level, critical components are stored in areas with enhanced protection or lower error rates. Sensitive weight filters are mapped to the MAC array column with the lowest mean timing error rates~\cite{choi2019sensitivity}.

Several studies~\cite{zhao2017aep,jia2018calibrating,hacene2019training,kim2018matic} consider fault patterns of inference hardware implementations during training. To ensure training convergence, some approaches store both floating point and fixed point weights~\cite{kim2018matic}, or inexact and exact weights~\cite{zhao2017aep}. Others perform digital domain backward pass and analog domain forward pass~\cite{jia2018calibrating}. Retraining is often used to mitigate error impacts~ \cite{temam2012defect,jia2018calibrating,zhang2018analyzing,nguyen2019stdrc}. By selectively retraining fault-prone fully connected layers, the overhead of comprehensive retraining is avoided~\cite{jia2018calibrating}.

Strategies for handling timing errors include increasing voltage upon detecting a timing error~ \cite{pandey2019greentpu}, ignoring multiplication with timing errors and forwarding unmodified partial sums~\cite{zhang2018analyzing,choi2019sensitivity}, identifying input sequences causing timing errors and increasing voltage accordingly~\cite{zhang2018analyzing}, using failure-prone cells as in-situ canaries~\cite{kim2018matic}, and increasing SRAM cell voltage during read/write operations~\cite{chandramoorthy2019resilient}. These approaches do not consider transistor aging effects on DNN inference accuracy. Long-term aging effects on DNN inference accuracy are studied in \cite{salamin2021reliability,liu2019analysis}, where decreasing clock frequency is shown to improve aged DNN inference accuracy. Another proposed method improves DNN resiliency to aging through input compression using quantization methods, although this considers aging effects only in MAC units.

This article considers the aging effects on DNNs entirely implemented in hardware, instead of studying the effects on only a MAC unit. Further, the effects of short-term aging along with long-term aging on the inference accuracy of hardware-implemented DNNs are discussed. Finally, a combination of mitigations at the DNN algorithm level that applies aging-aware training to robustify the DNN to aging degradations is presented. The application of short-term aging for hardware Trojan detection is also briefly discussed.

\section{Aging in Integrated Circuits}
\label{sec:aging}
As semiconductor technology advances into the nanoscale realm, electric fields become stronger with each new generation, allowing transistors to switch faster. While a transistor is in operation, it experiences stress from electric fields, which can cause defects of various kinds, including interface traps and oxide traps. Bias Temperature Instability (BTI)\index{bias temperature instability (BTI)} is a key transistor aging factor that is responsible for the creation of defects over time. The accumulation of these defects can cause changes in the transistor's electrical properties including its threshold voltage ($V_{th}$) and carrier mobility ($\mu$)\index{carrier mobility}. %
The result is a slower transistor switching speed, which leads to increased delays in the circuit path and potential errors in the circuit's outputs caused by timing violations induced by a high clock frequency. To prevent the effects of aging\index{aging}, timing guard bands\index{timing guardband} are added to the worst-case path delay of the circuit to accommodate for delay increases caused by aging during the circuit's operation. The aging of transistors depends on several factors, including voltage, temperature, and duty cycle. When the voltage and/or temperature is higher, the mechanisms that cause defects to form are accelerated, resulting in faster and more severe degradation. Transistors that have a high-duty cycle, meaning those that are stressed for a greater percentage of the time, degrade faster.

\subsection{Long-term Aging\index{aging!long-term}}
$\Delta V_{th}$ for a new chip is equal to 0mV. However, after about 10 years, it approaches 50mV (end of life). This level of degradation is so severe that the transistor can no longer function properly. This effect is known as traditional long-term aging. It typically takes in the order of around 10 years to occur. In this study, the duty cycle is used to indicate the level of aging, where a 0\% duty cycle indicates no aging and a 100\% duty cycle indicates worst-case aging. The change in $\Delta V_{th}$ as the transistor ages, varies non-linearly with duty cycle and time \cite{mahapatra2016fundamentals}.

\subsection{Timing Errors Caused by Aging}
A transistor's delay is inversely proportional to its ON current ($I_{ON}$ ). $I_{ON}$ depends on $V_{th}$ and $\mu$ (Eq.~\ref{eq:I_d}) \cite{amrouch2016reliability}. When $V_{th}$ increases and $\mu$ decreases because of aging, $I_{ON}$ decreases, and subsequently transistor delay increases.

\begin{equation}
\label{eq:I_d}
\text{Delay} \propto \frac{1}{I_{ON}} \text{ ; }
I_{ON} \approx \frac{\mu}{2} \cdot C_{ox} \cdot \frac{W}{L}  \cdot  (V_{dd} - V_{th})
\end{equation}
where, $C_{ox}$ is oxide capacitance, $W$ is width, and $L$ is length of transistor. %

Even if the timing guardband is sufficient, aged transistors will still experience slower operation. However, if the guardband is insufficient, this slowdown can lead to an increase in errors at the circuit outputs, as suggested by Eq.~\ref{eq:timing_errors}. This happens because the switching frequency becomes unsustainable, causing timing violations in critical paths, which in turn propagate to the circuit outputs, resulting in errors. Denoting the propagation delay of each element in the critical path (CP) by $t_{d_i}$, the total propagation delay of the critical path is
\begin{equation}
\label{eq:timing_errors}
t_{CP} =  \sum\limits_{d_i \in CP}^{} t_{d_i}.
\end{equation}
If $t_{CP}$ under the aging condition is bigger than the clock period $t_{clock}$, then \textit{timing errors } result.

\subsection{Short-term Aging\index{aging!short-term}}
Eq.~\ref{eq:I_d} shows that the impact of aging on $\Delta V_{th}$  depends largely on $V_{dd}$. When $V_{dd}$ is scaled down, the same increment in the value of $V_{th}$ because of aging (e.g., 25mV) will lead to a significantly larger transistor propagation delay. Under aging conditions, the timing errors exhibited by the circuit are sensitive to aging-induced deterioration ($V_{th}$, $\mu$) and $V_{dd}$. This pairing amplifies the effects of aging, transforming the aging phenomenon from a solely long-term issue (i.e., a deterioration that could take months to take place) to a short-term issue (i.e., a deterioration that can result in timing errors in circuits within hours).

As described above, if voltage scaling\index{voltage scaling} happens when the transistor is aged, the impact of aging-induced degradation  at high voltage is amplified when the voltage is reduced to a lower level \cite{vansanten2016agingaware}. This is referred to as short-term aging, and it differs from traditional long-term aging that occurs over extended periods of time (e.g., months and years), short-term aging occurs significantly faster and results in increased timing errors. However, intentionally excited short-term aging can be applied to detect hardware Trojans by observing the bit error patterns of an IC and comparing with machine learning models of the statistical properties of bit error patterns of a non-Trojaned simulated baseline, yielding  a Trojan detection accuracy of more than 95\% \cite{surabhi2022trojan}. Short-term aging has been applied to detect Trojans in ICs implemented with the 45 nm CMOS technology node \cite{surabhi2020hardware,surabhi2020exposing} as well as the
14 nm FinFET technology node \cite{surabhi2022trojan}. Short-term aging can also be used to estimate an IC's age and to detect recycled ICs \cite{surabhi2023golden}.

\subsection{Aging-Aware Standard Cell Libraries\index{aging-aware standard cell libraries}}

Aging is caused by various defect-generating processes at the nanoscale. To study and characterize how these defects spread to the system level, where they cause timing issues, it is necessary to carefully navigate through intermediate abstraction levels. Moreover, real digital circuits often have multiple paths with similar latencies, making it challenging to  quantify how aging-induced degradation\index{aging-induced degradation} will violate timing paths and how these violations translate into circuit output errors. This requires precise simulation of how normal cells behave when experiencing age-related degradation. Investigations in this area rely on commercial static timing analysis tool flows,
making them consistent with the current  design flow.

All circuit analyses discussed in this article use the 14nm FinFET\index{FinFET} technology. The FinFET transistor models used are calibrated to match the specifications of  commercially available technology. Additionally, they have been verified through comparison with semiconductor measurement data for the 14nm FinFET node. The industrial FinFET technology compact model (BSIM-CMG) \cite{chauhan2015finfet} has been adjusted to replicate the Intel 14nm FinFET measurement data \cite{mishra2018simulation} for both nFinFET and pFinFET transistors. The calibration of the transistor model has been verified by comparing it with test data under different voltage biases, including gate and drain voltages. The values of the calibrated transistor model match the test data accurately.
To estimate the effect of short-term aging on circuit path delays and output error rates, cross-layer modeling is used. The analysis begins at the transistor level, where the impact of BTI on the transistor's threshold voltage is estimated using an advanced physics-based aging model \cite{parihar2017bti}. As the degradation depends on the stress that the transistor undergoes (i.e., duty cycle), $V_{th}$ is estimated in 1\% steps, ranging from 0 (fresh, no aging) to 100\% (continuous stress of aging). The predicted degradation is incorporated into  BSIM-CMG \cite{chauhan2015finfet}. This way, precise circuit SPICE simulations\index{SPICE simulation} can be conducted to determine standard cell delays.

The open-source FinFET PDK \cite{finfet_pdk} is used to extract the post-layout SPICE netlist of standard cells from it. The SPICE netlist\index{SPICE netlist} is used to create an aging-aware standard cell library\index{standard cell library}. For each standard cell, the delay using HSPICE is measured under varying input signal slews as well as output load capacitances.  The library generation process is performed at both nominal voltage (0.8V) and lower voltage (0.6V) to capture the combined impact of voltage and aging on circuit path delay.

\section{Effects of Aging on Deep Neural Networks}
\label{sec:effects_age_DNN}
\begin{figure}[!b]
  \centering
  \includegraphics[width=0.95\linewidth]{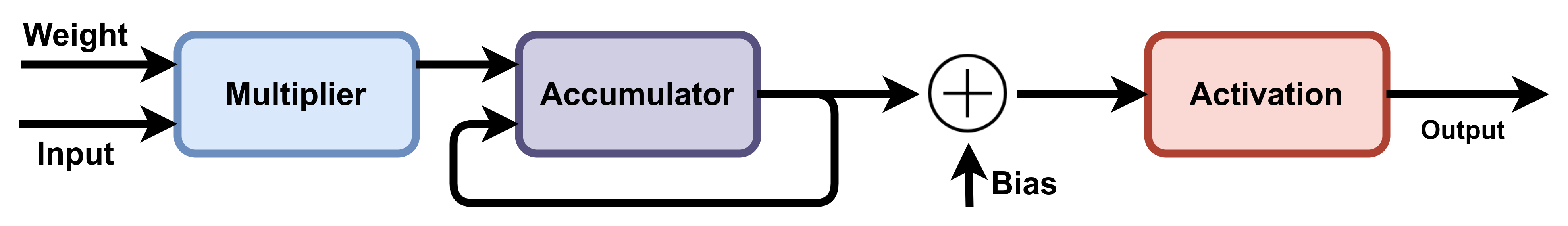}
  \caption{Neuron architecture.}
  \label{fig:neuron_architecture}
\end{figure}

A DNN\index{deep neural networks (DNNs)} can be viewed as a composition of a large number of interconnected non-linear functions. Each function is represented by a neuron. This neuron processes an input vector by performing a linear transformation utilizing a weight vector. Subsequently, a non-linear activation function, represented by $\sigma$, is applied to the resulting value. The mathematical representation of the computation occurring within a neuron is as follows:
\begin{equation}
    \label{eq:perceptron}
    y(x) = \sigma(\mathbf{w}.\mathbf{x} + b),
\end{equation}
where $y$ is the output, $\mathbf{x}$ is the input vector,  $\mathbf{w}$ is the weight vector, and $b$ is the bias. In the Eq. \ref{eq:perceptron}, the products between corresponding weights and inputs are accumulated. A MAC unit\index{multiply-accumulate (MAC) units} can carry out this operation and the architecture of the MAC unit is shown in Figure \ref{fig:neuron_architecture}. Thus, the number of MAC units required in a fully connected DNN is equal to the number of neurons in the DNN.

As a DNN implemented on hardware ages over time, the increase in the transistor's threshold voltage, $V_{th}$, reduces the driving current and hence increases the path delay. Although the computation error rate due to timing degradation from transistor aging in an arithmetic circuit is minimal, the substantial accumulations in DNNs (as illustrated in Eq. \ref{eq:perceptron}) increase the likelihood of errors in each output component. The timing violation more frequently impacts the higher-order bits of these intermediate products. An error in the most significant bit results in a large magnitude shift in the two's complement number representation. Consequently, during the accumulation process, the inaccurate outputs dominate the intermediate products.

In a DNN, each layer of neurons is connected to the neurons in the following layer. For example, consider a fully connected neural network\index{fully connected neural network} in which every neuron in layer $i$ is connected to every neuron in layer $i+1$. Suppose there are $n$ layers in a fully connected neural network. Let the weights of neurons in the layer $i$ be represented by the matrix $W_i$ and the biases by $\mathbf{b_i}$. Let the input to the neural network be $\mathbf{x'}$. Therefore, the output of the DNN can be represented as

\begin{equation}
    \label{eq:FCNN}
    Y = \sigma (W_n\sigma(.... \sigma(W_2\sigma(W_1\mathbf{x'} + \mathbf{b_1}) + \mathbf{b_2}) +....)+\mathbf{b_n}.
\end{equation}

It is evident from Eq. \ref{eq:FCNN} that the errors propagate through the layers. The errors accumulate in successive layers and can have increased impact as the number of layers increases. With increased aging resulting in larger numbers of errors in the neuron outputs, the final output of the DNN changes, resulting in incorrect classifications and therefore reduced accuracy. In Eq. \ref{eq:perceptron}, each operation, whether multiplication or accumulation, requires one clock cycle. By increasing the clock period to counteract the delay resulting from aging, sufficient time is provided for the calculations to be executed. Therefore, one approach to reduce errors caused by aging is to decrease the clock frequency, as discussed in Section \ref{sec:mitigation_clockfrequency}. This allows more time for the calculations performed by MAC units to be completed.
Another method to minimize errors is selecting weights $\mathbf{w}$ and biases $b$ in a manner that enhances DNN resilience to bit errors\index{aging-aware resilience} (Section \ref{sec:mitigation_gradnoise_imitator}). This article addresses the aforementioned mitigation strategies, demonstrating that DNNs can be trained to decrease their sensitivity to bit errors and increase overall robustness. In the remainder of the article, a representative DNN hardware implementation (Section \ref{sec:case_study}) is addressed to analyze the impact of aging on DNN inference accuracy\index{inference accuracy}.

\subsection{Case Study: Aging Effects on Hardware-Implemented DNN for Image Classification}
\label{sec:case_study}
\begin{figure*}[h]
	\centering
	\subfloat[]{\includegraphics[width=0.6\linewidth]{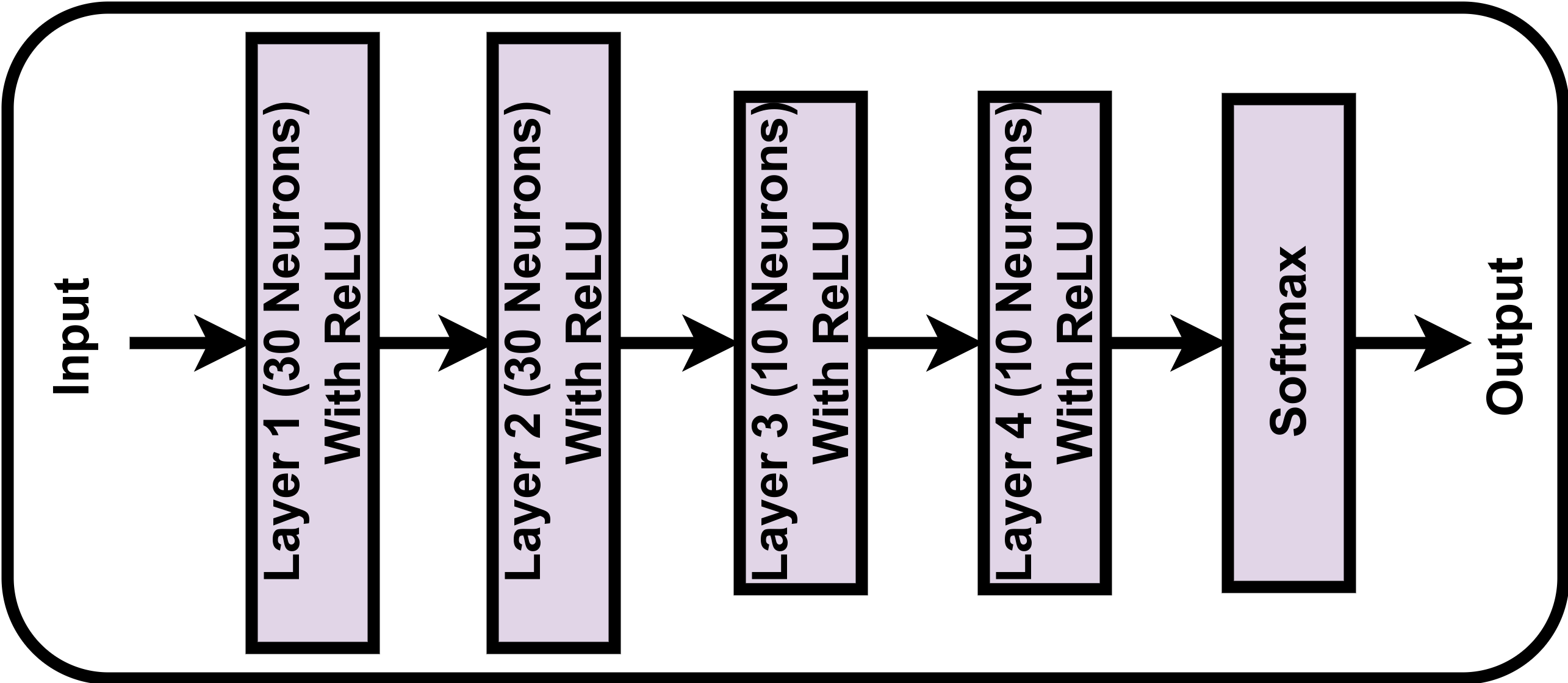}}
 \hfill
	\subfloat[]{\includegraphics[width=0.37\linewidth]{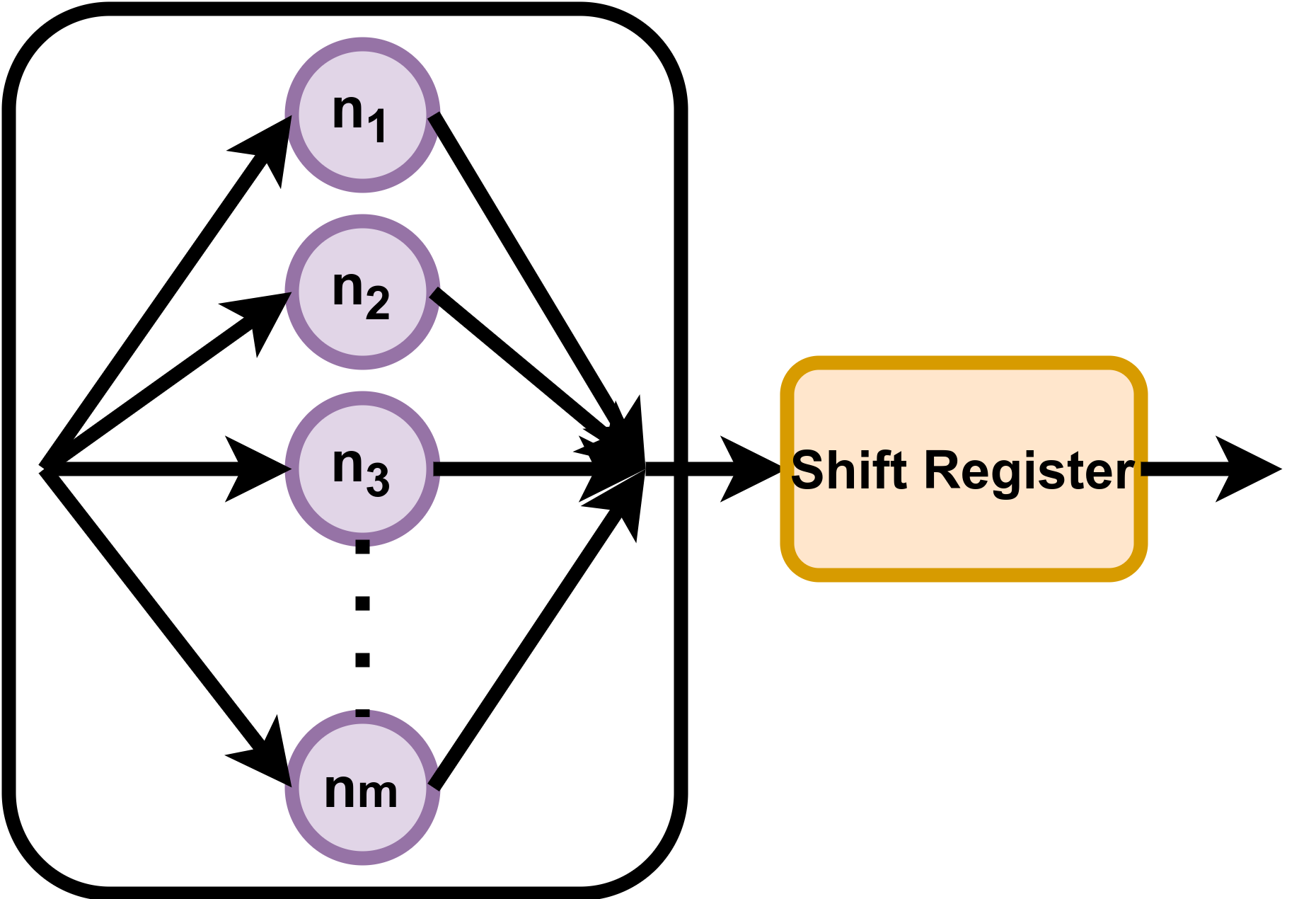}}
	\caption{(a) Neural network architecture. (b) Layer architecture with $m$ neurons.}
	\label{fig:neuralnetwork_layer}
\end{figure*}

\begin{figure}[!t]
  \centering
  \includegraphics[width=\linewidth]{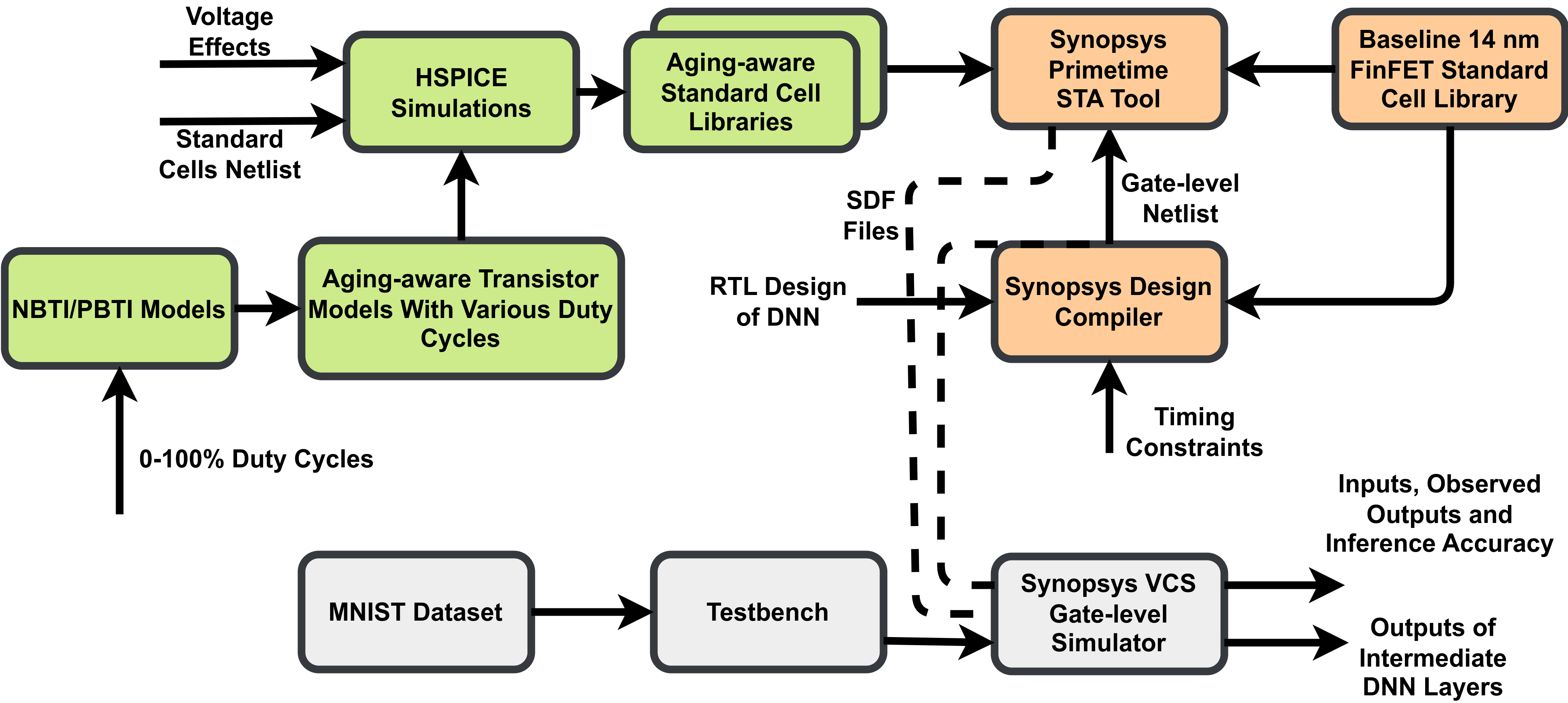}
  \caption{%
  Overview of the framework. The tool flow has three main steps. First, short-term aging standard cell libraries are created, taking into account both transistor aging and voltage effects. Next, these libraries are integrated into standard static timing analysis tool flows\index{static timing analysis (STA)} to obtain accurate delay information for the circuit netlist under study. Lastly, gate-level simulations are conducted to extract input-output pairs and outputs from each layer of the DNN.}
  \label{fig:synopsys_toolflow}
\end{figure}

To investigate the impact of aging on DNNs implemented in hardware, consider a representative example of a 4-layer fully connected DNN for an image classification task (identifying handwritten digits in images in the MNIST dataset\index{MNIST dataset}). The DNN features rectified linear unit (ReLU) activations\index{rectified linear unit (ReLU)} in each layer and the four layers of the DNN contain 30, 30, 10, and 10 neurons, respectively. The output layer employs a softmax activation function\index{softmax activation}. The architecture of the DNN is depicted in Figure \ref{fig:neuralnetwork_layer} (a). The DNN is initially implemented at the Register Transfer Level (RTL)\index{RTL}, followed by synthesis using a 14 nm FinFET standard cell library to produce a gate-level netlist\index{gate-level netlist}. This gate-level netlist, combined with aging-aware standard cell libraries, generates Standard Delay Format (SDF) files\index{standard delay format (SDF)}. Gate-level timing simulations\index{gate-level simulation} are then conducted using the SDF files. Figure \ref{fig:synopsys_toolflow} illustrates the overall tool flow. The subsequent sections provide further details on each step of this process.

A 4-layer fully connected DNN is considered in this article as it provides a representative structure that is complex enough to analyze the propagation of aging-induced errors across layers while also being relatively small to reduce computational requirements in the analysis. Also, a fully connected network ensures that every neuron in one layer is connected to all neurons in the next, which amplifies potential impact of timing errors and helps to illustrate cumulative degradation in accuracy. While CNNs are also commonly used for tasks like image classification due to their versatility in handling spatial features, a fully connected network is selected to focus on the core aging effects on a general network architecture. Nevertheless, the analysis and observations are not specific to a particular network structure and are generally relevant to a wide range of DNNs since the main phenomenon being studied is the cumulative effect of errors due to aging along with potential mitigation techniques to address these aging-related impacts.

This case study focuses on analyzing the aging effects across an entire neural network implemented directly in hardware, rather than limiting the investigation to a specific hardware accelerator like a systolic mac array. The goal is to examine how aging-induced degradation and delay increases in standard cells propagate through the different layers of the neural network within the underlying circuit's netlist. This is particularly important for edge-AI applications, where high throughput is critical, and implementing the entire neural network in hardware—rather than relying on a CPU connected to an accelerator—is essential. Nevertheless, the methodology presented in this article is not limited to this specific scenario and can be analogously applied to other contexts, such as hardware accelerators DNNs.

\subsubsection{RTL Design of the DNN\index{RTL}}
The DNN is implemented in Verilog\index{Verilog}, featuring shift registers after every layer to optimize data flow. The data is represented using a 16-bit fixed-point system\index{fixed-point representation}, comprising a 10-bit integer part and a 6-bit fractional part. Two's complement representation is employed for the data to accommodate negative numbers. For read/write operations, an AXI4-Lite interface\index{AXI4-Lite} is utilized.

Each neuron in the network has its own interface for configuring weights and biases. With separate data interfaces for every neuron, the network can scale and improve clock performance without increasing latency. A control circuit retrieves relevant weight values when inputs are provided to the network. Inputs and weight values are then multiplied and accumulated before the bias value is added. The output of the Multiply-Accumulate (MAC) unit~\index{multiply-accumulate (MAC) units} is subsequently passed to the activation unit (ReLU). The MAC unit's architecture is depicted in Figure \ref{fig:neuron_architecture}.

As each neuron has only a single data port and a fully connected layer necessitates connectivity to every neuron from the previous layer, data from each layer is initially stored in a shift register \cite{vipin2019zynet}. The data is then transferred to the subsequent layer. The architecture of a layer is depicted in Figure \ref{fig:neuralnetwork_layer}. The well-known MNIST dataset\index{MNIST dataset} for handwritten digit recognition is utilized. The network's weights and biases are generated using a software implementation in PyTorch.

\subsubsection{Logic Synthesis}
The RTL\index{RTL} of the entire network is synthesized using the 14nm FinFET standard cell library without aging at 0.8V, employing Synopsys Design Compiler for logic synthesis\index{logic synthesis} to produce a gate-level netlist\index{gate-level netlist}. The resulting netlist contains a total of 1,323,232 cells. Static Timing Analysis (STA)\index{static timing analysis (STA)} is performed on the netlist using aging-aware standard cell libraries\index{aging-aware resilience} (Section \ref{sec:aging}) and Synopsys Primetime \cite{primetime}. STA is executed at various aging levels, ranging from 0\% to 100\% in 5\% increments, across two voltage levels (0.8V and 0.6V). Synopsys Primetime generates Standard Delay Format (SDF) files\index{standard delay format (SDF)} for different operating conditions, which are combinations of aging and voltage levels. These SDF files are utilized to simulate the netlist behavior under the respective operating conditions. Additionally, the STA tool provides the nominal clock period for each operating condition.

\subsubsection{Simulations and Analysis of Aging Effects on DNNs}

Gate-level netlist simulations\index{gate-level simulation} are conducted using Synopsys VCS by annotating the delay values supplied by the SDF files\index{standard delay format (SDF)}. In total, there are $21$ aging states $\times$ $2$ voltage levels, resulting in 42 SDF files that cover the operating conditions. The nominal clock periods for 0.8V and 0.6V at zero aging are 0.185 ns and 0.305 ns, respectively. The gate-level simulations are carried out at the reported nominal clock periods.

\begin{figure*}[h]
	\centering
	\subfloat[Normalized Accuracy]{\includegraphics[width=0.5\linewidth]{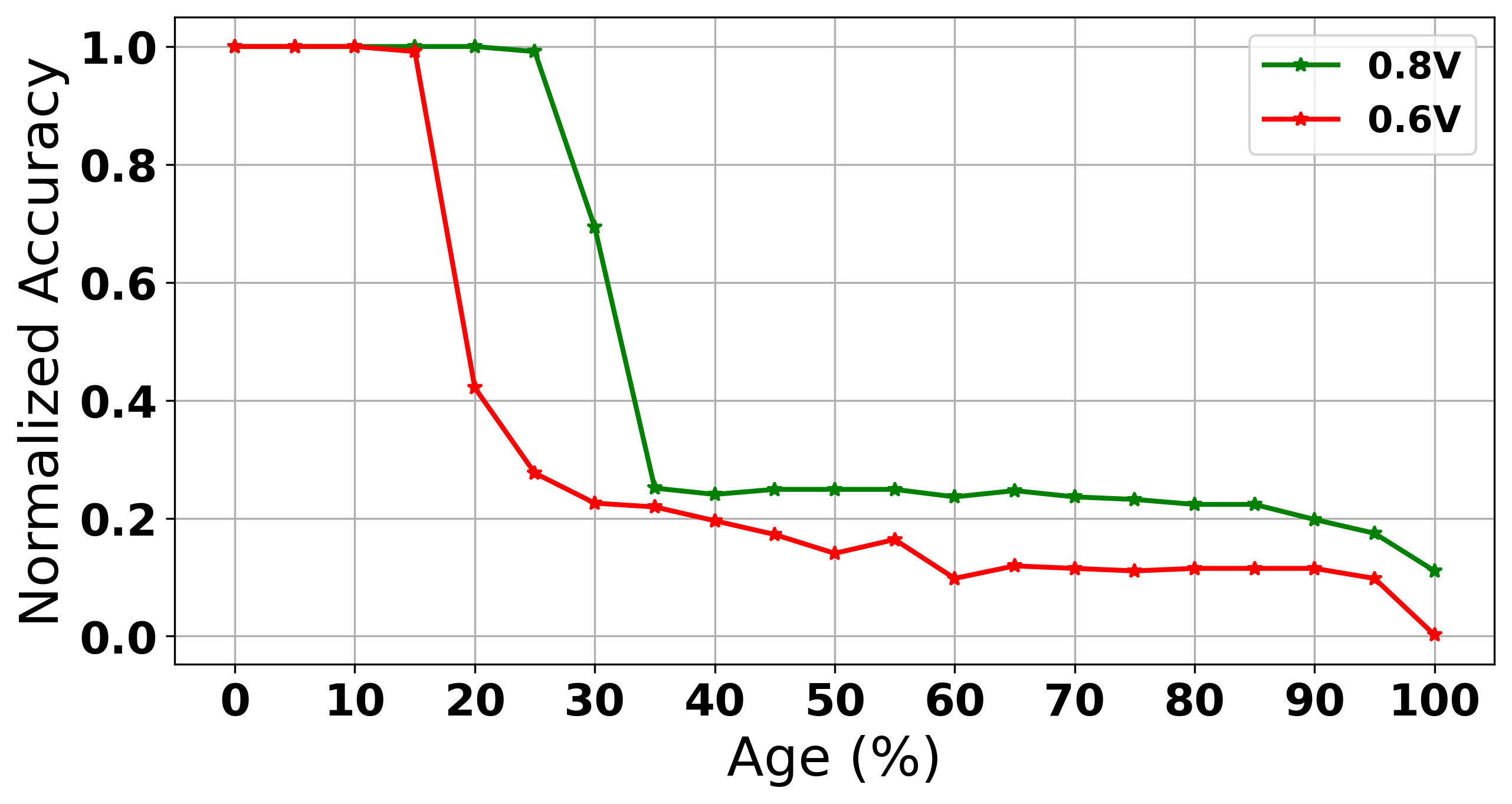}}
 \hfill
	\subfloat[Actual Accuracy]{\includegraphics[width=0.5\linewidth]{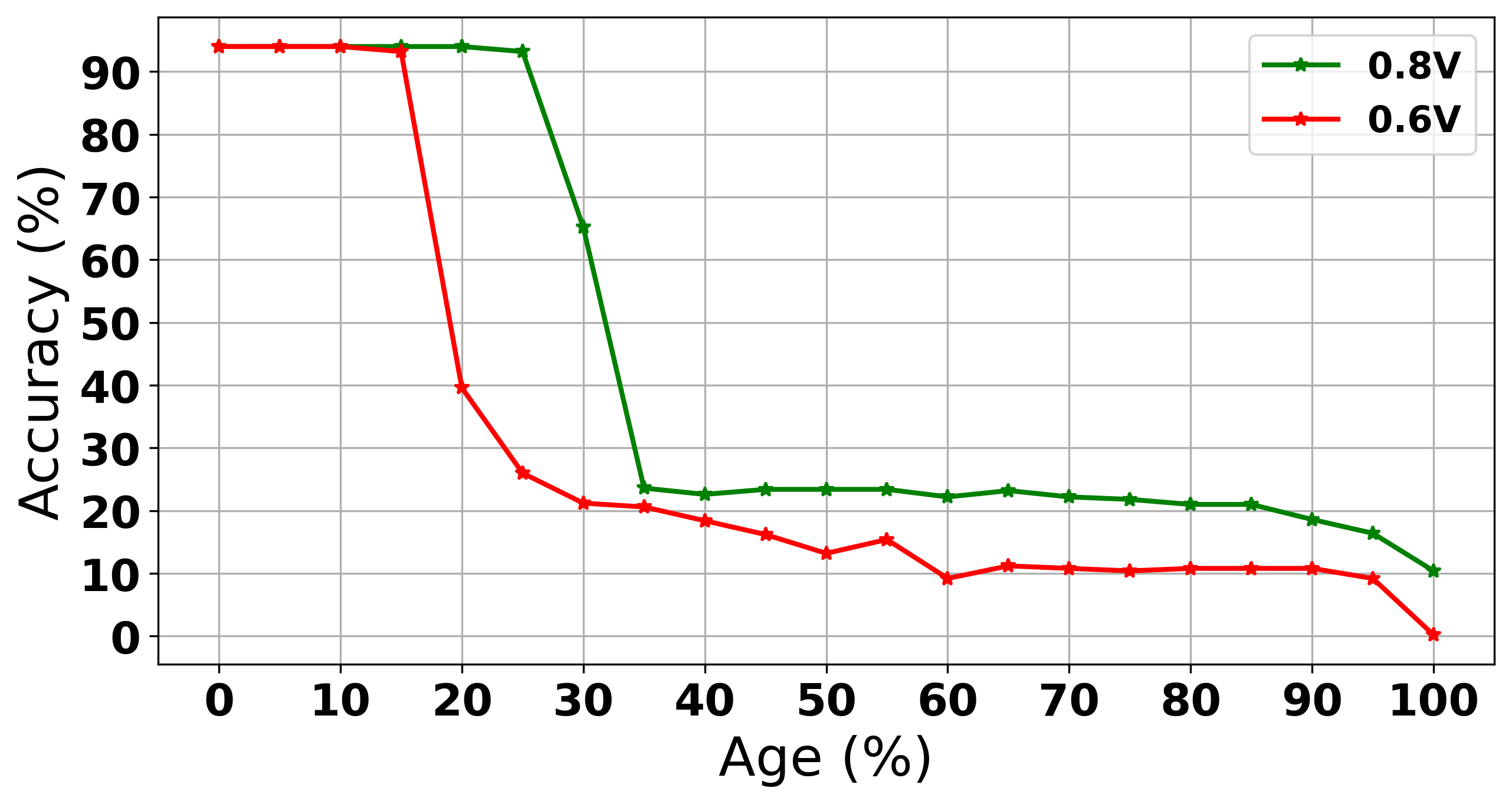}}
	\caption{Accuracy drop of the DNN due to transistor aging for 0.8V and 0.6V. The nominal clock periods (reference clocks used for the IC design in the absence of aging), 0.185 ns for 0.8V and 0.305 ns for 0.6V, are used.}
	\label{fig:acc_nom_clk_freq}
\end{figure*}

For this experiment, 500 images are randomly selected from the MNIST dataset, ensuring an equal number of images in each class. The images are passed through the gate-level netlist\index{gate-level netlist} annotating the delay from the SDF files\index{standard delay format (SDF)} from various aging levels and voltage levels, and the observed outputs are recorded. Figure \ref{fig:acc_nom_clk_freq} displays the normalized accuracies (left) and actual accuracies (right) as obtained from the gate-level timing simulations. It shows the decline in DNN accuracy on MNIST dataset classification as aging levels increase. The green curve corresponds to 0.8V, while the red curve represents 0.6V. At 0.8V, the accuracy drops substantially around 30\% aging. The accuracy decreases by 30\% at 30\% aging and up to 70\% when the hardware implemented DNN ages by 35\%. With 0.6V, a significant decline in accuracy is observed around just 20\% aging, with a 55\% drop in accuracy. The accuracy continues to decline significantly with increasing aging levels.

\begin{figure*}
	\centering
	\subfloat[]{\includegraphics[width=0.5\linewidth]{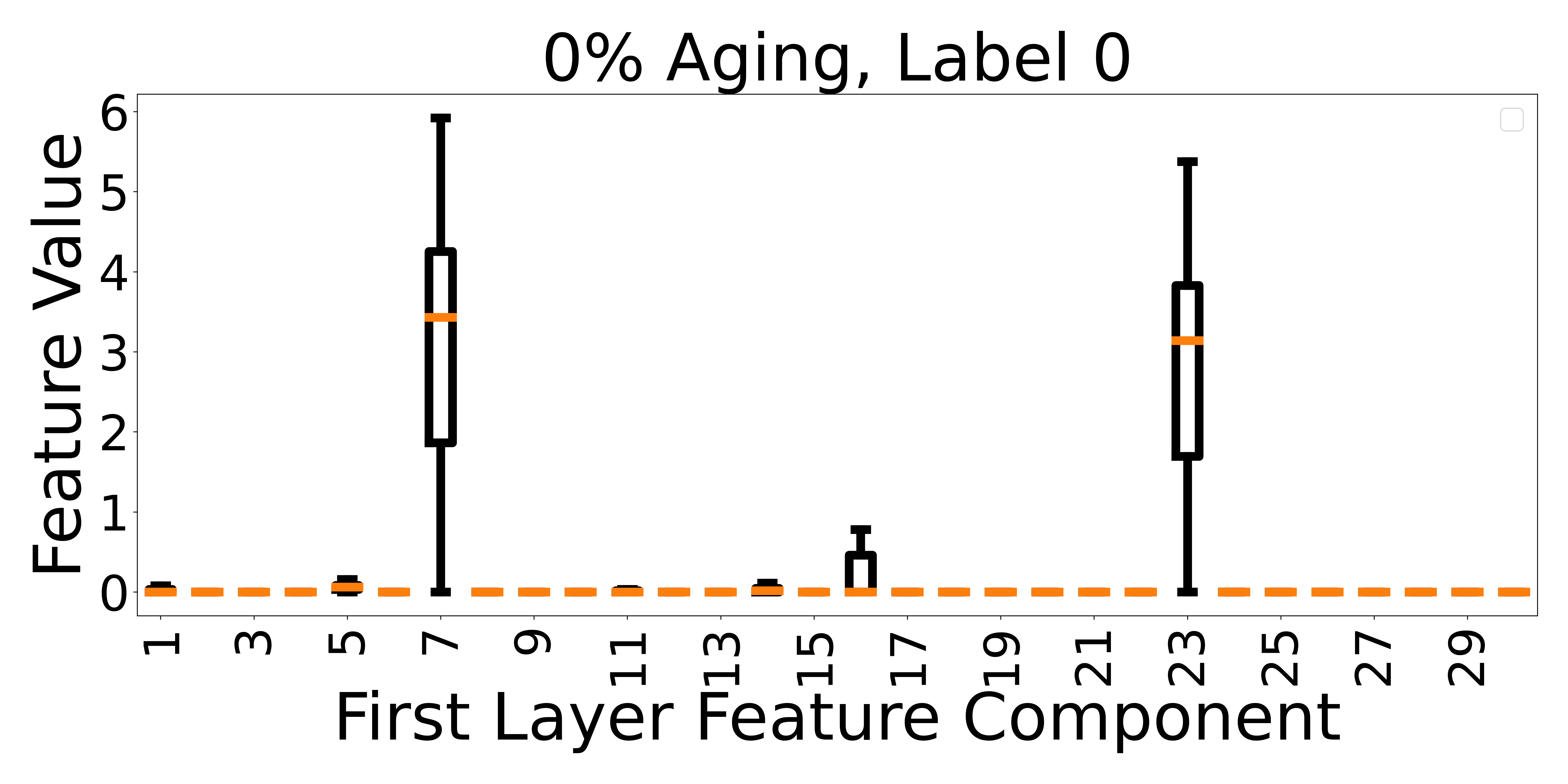}}
    \hfill
 	\subfloat[]{\includegraphics[width=0.5\linewidth]{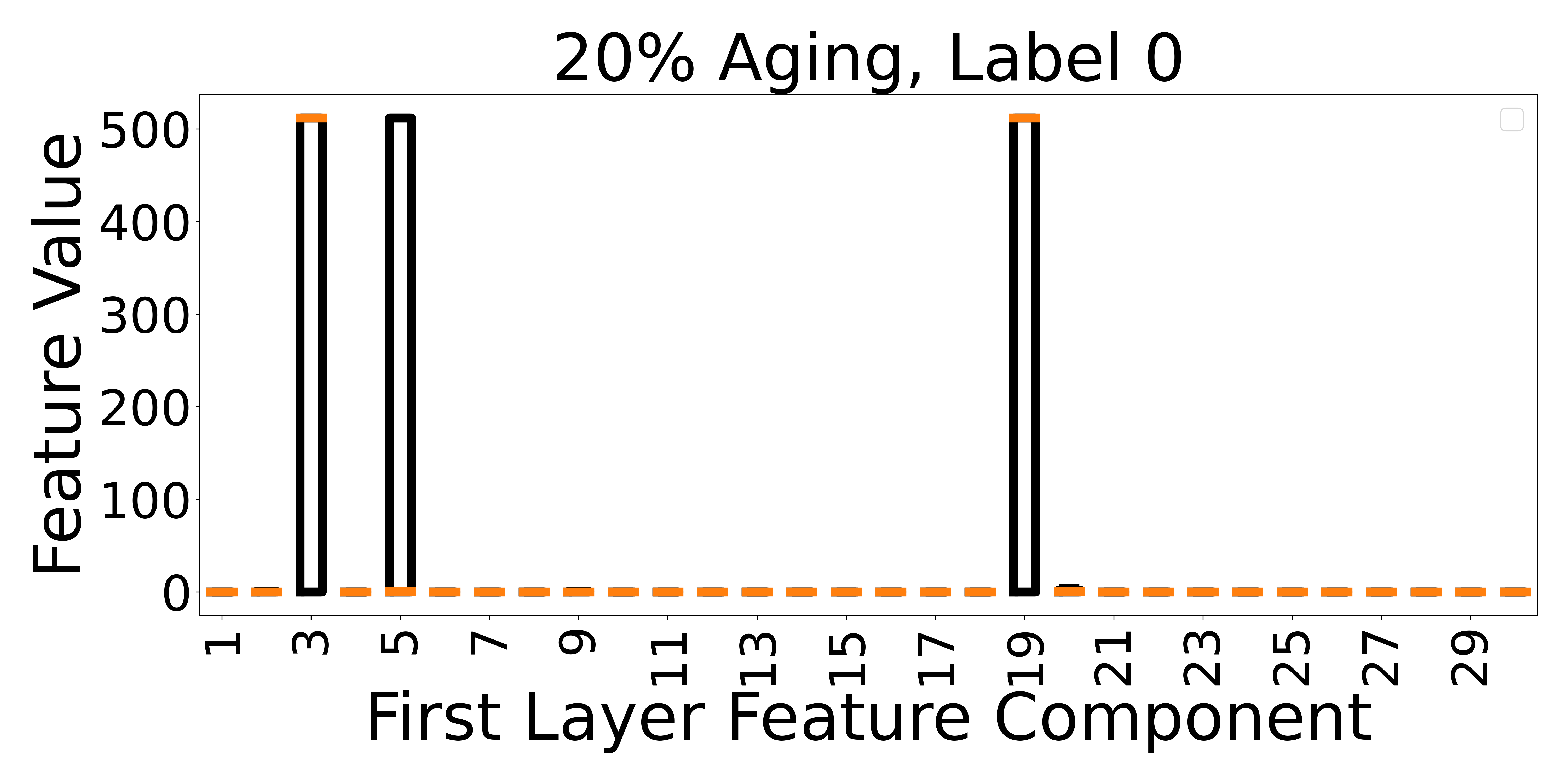}}
    \\
	\subfloat[]{\includegraphics[width=0.5\linewidth]{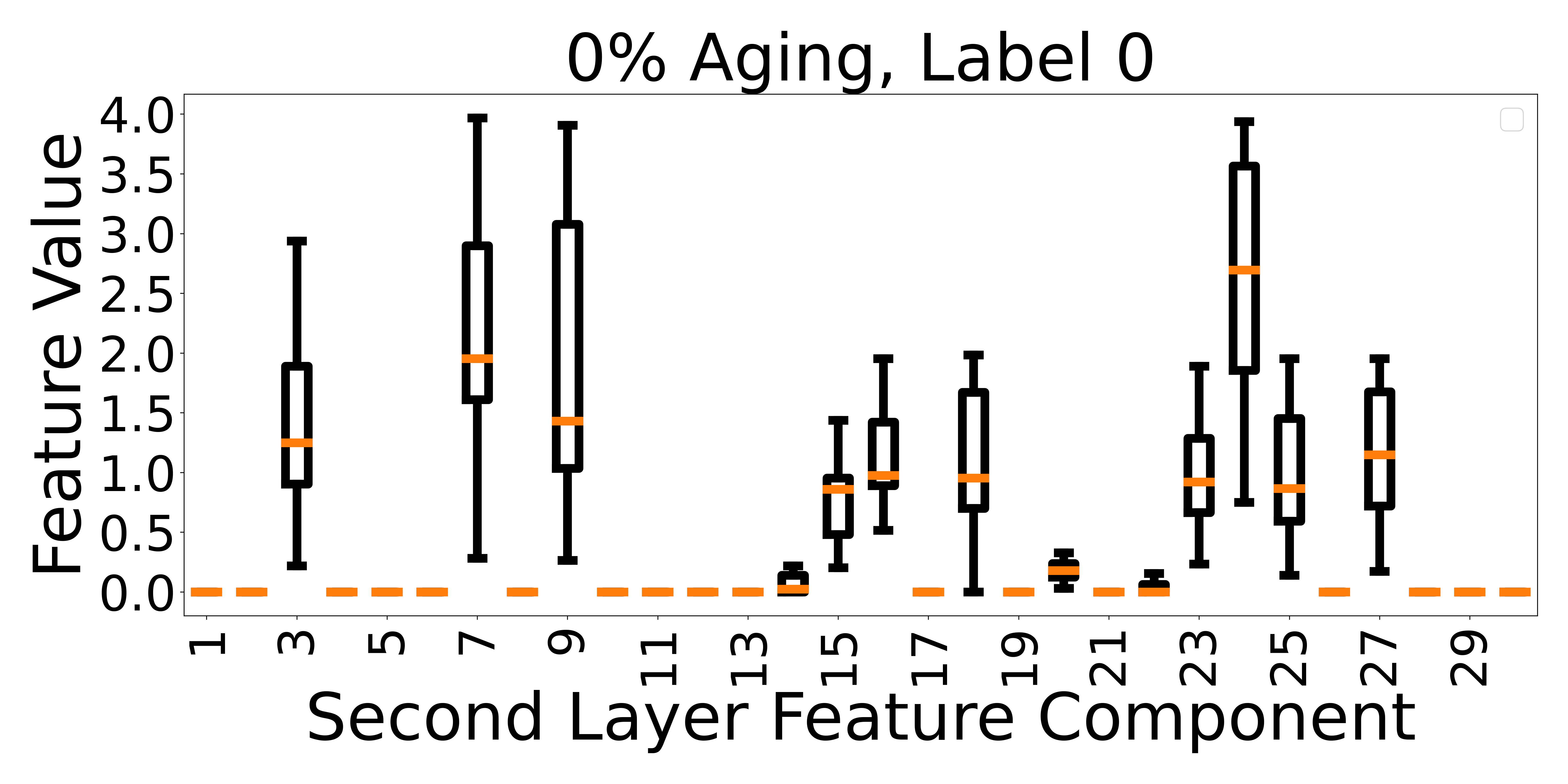}}
    \hfill
	\subfloat[]{\includegraphics[width=0.5\linewidth]{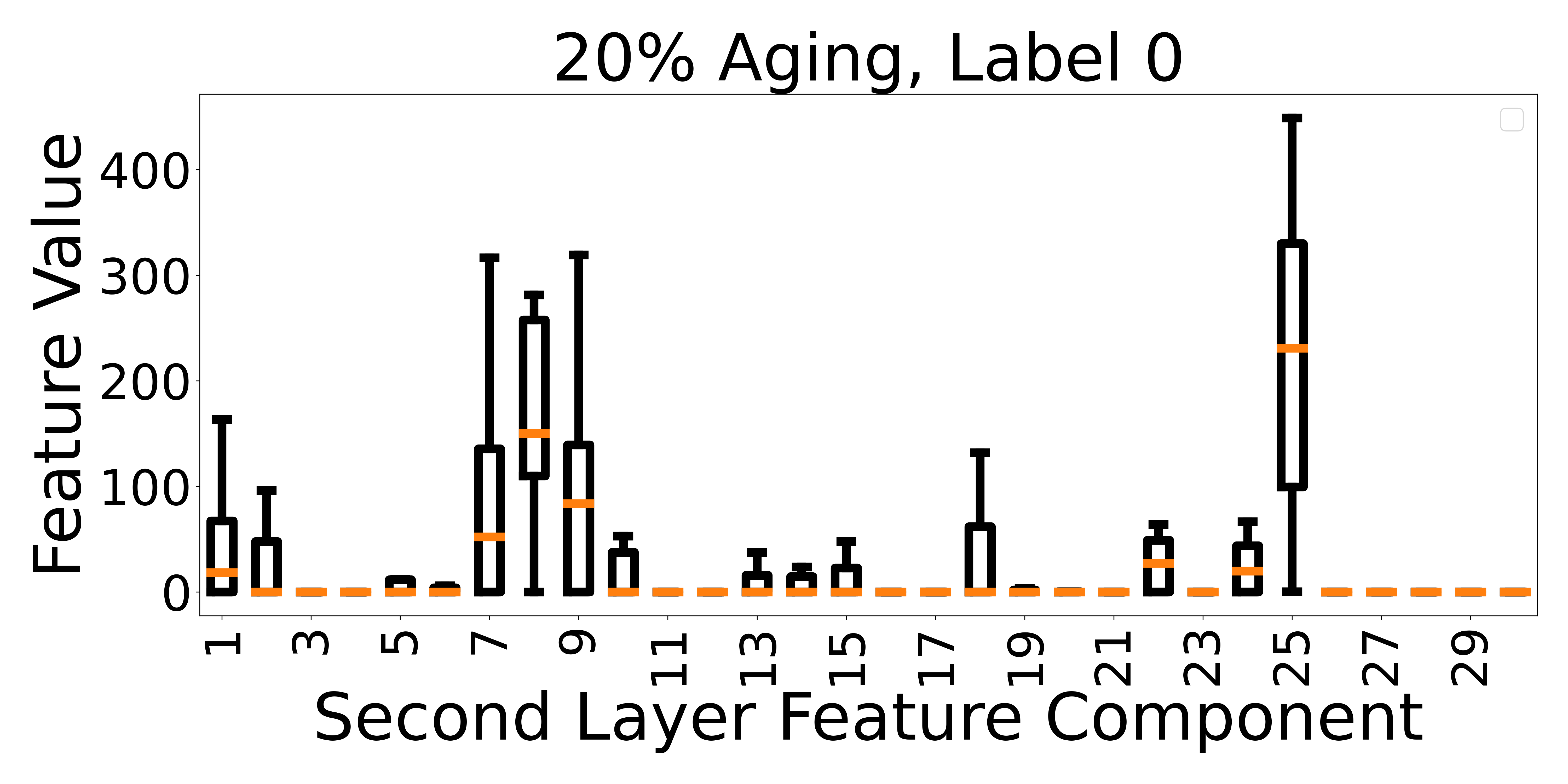}}
    \\
 	\subfloat[]{\includegraphics[width=0.5\linewidth]{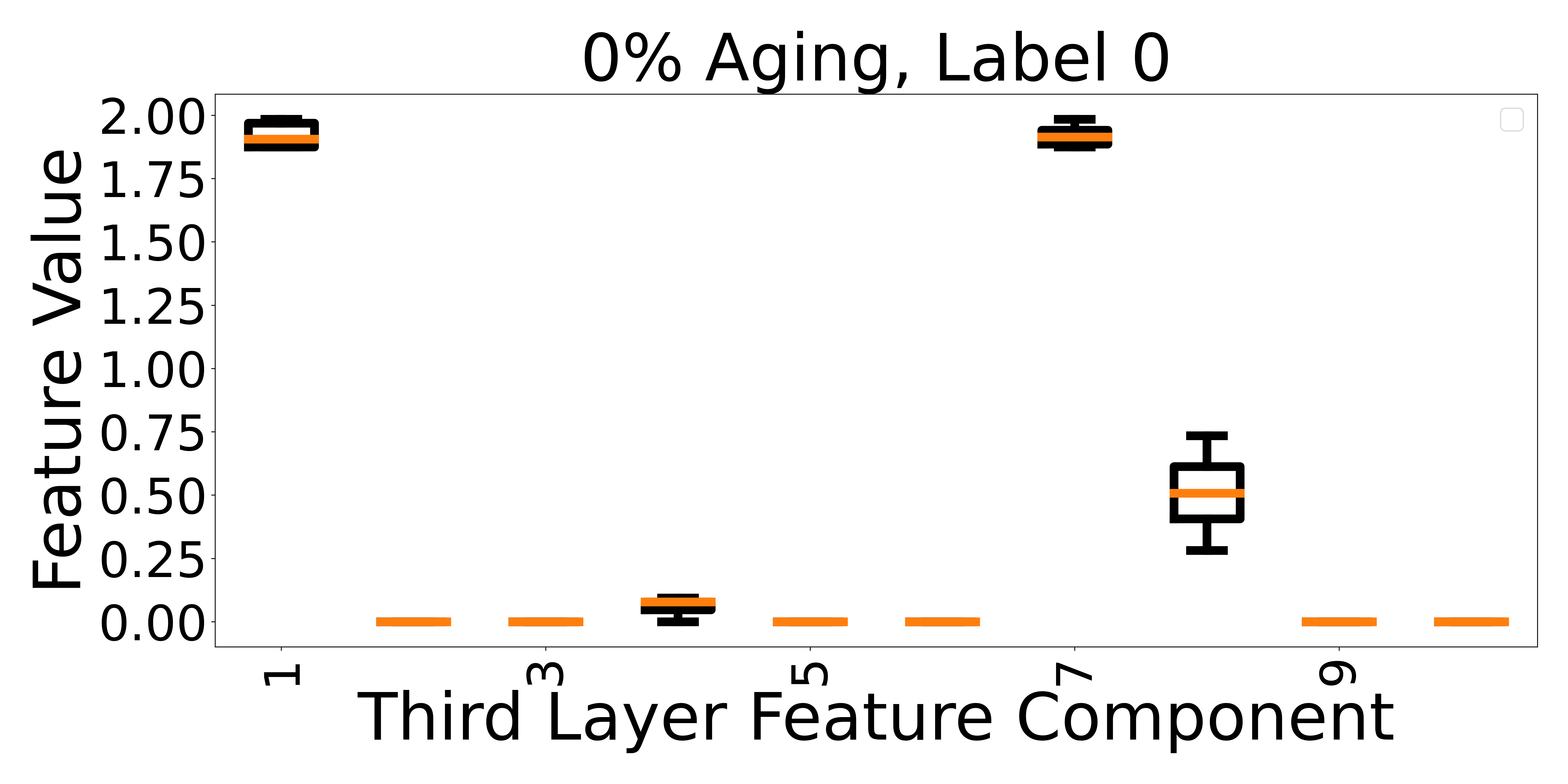}}
    \hfill
 	\subfloat[]{\includegraphics[width=0.5\linewidth]{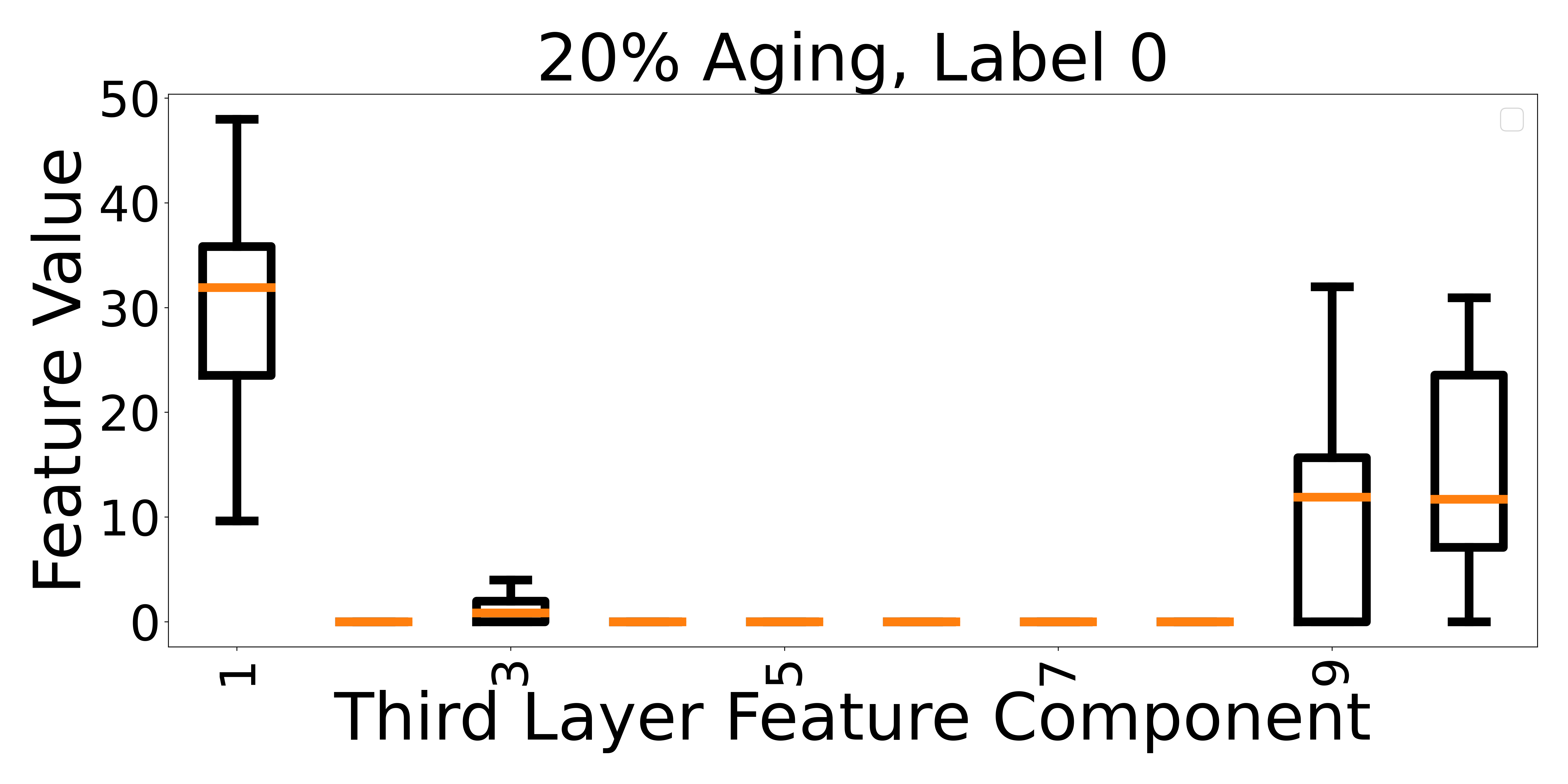}}
    \\
	\subfloat[]{\includegraphics[width=0.5\linewidth]{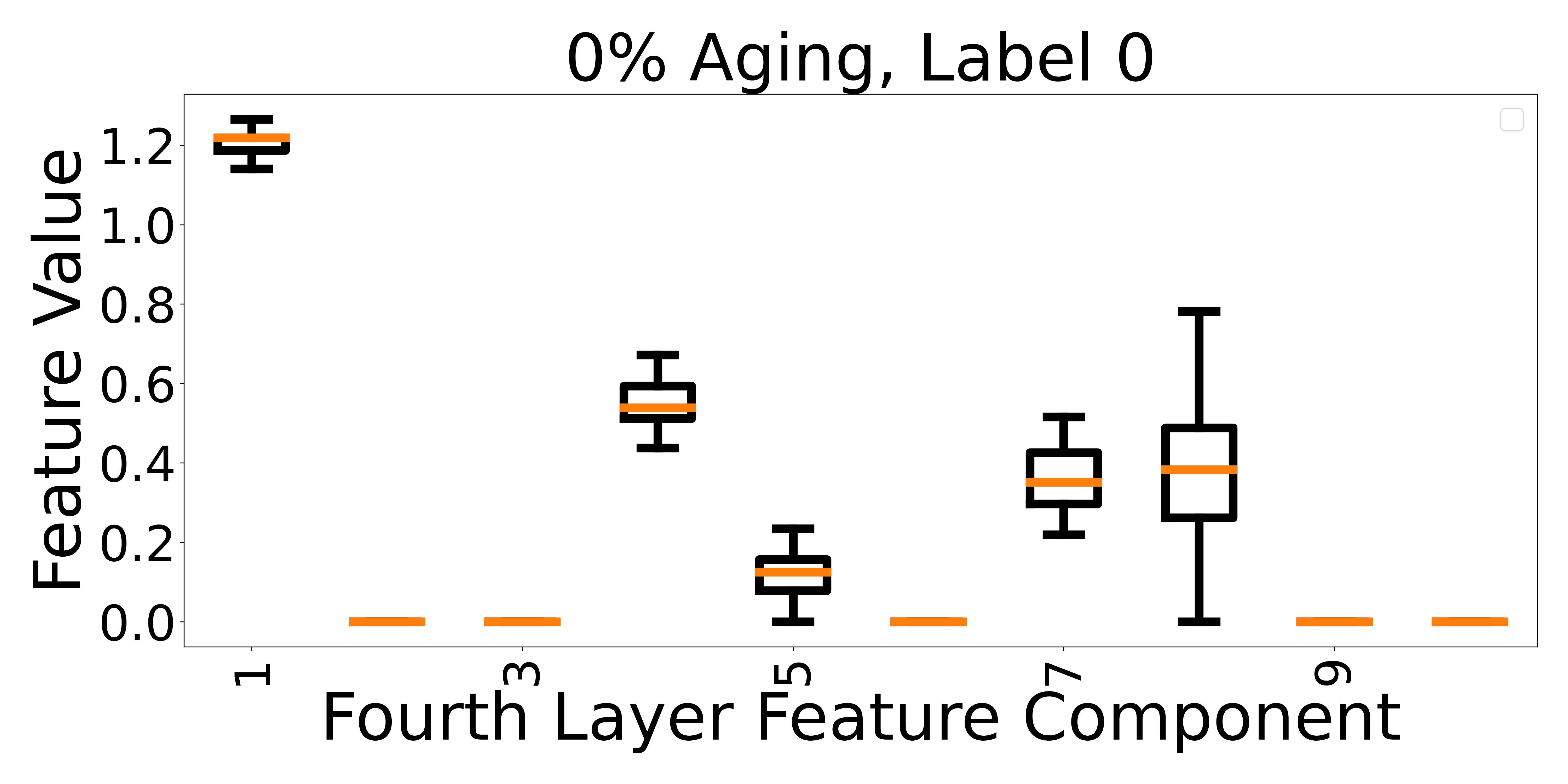}}
    \hfill
	\subfloat[]{\includegraphics[width=0.5\linewidth]{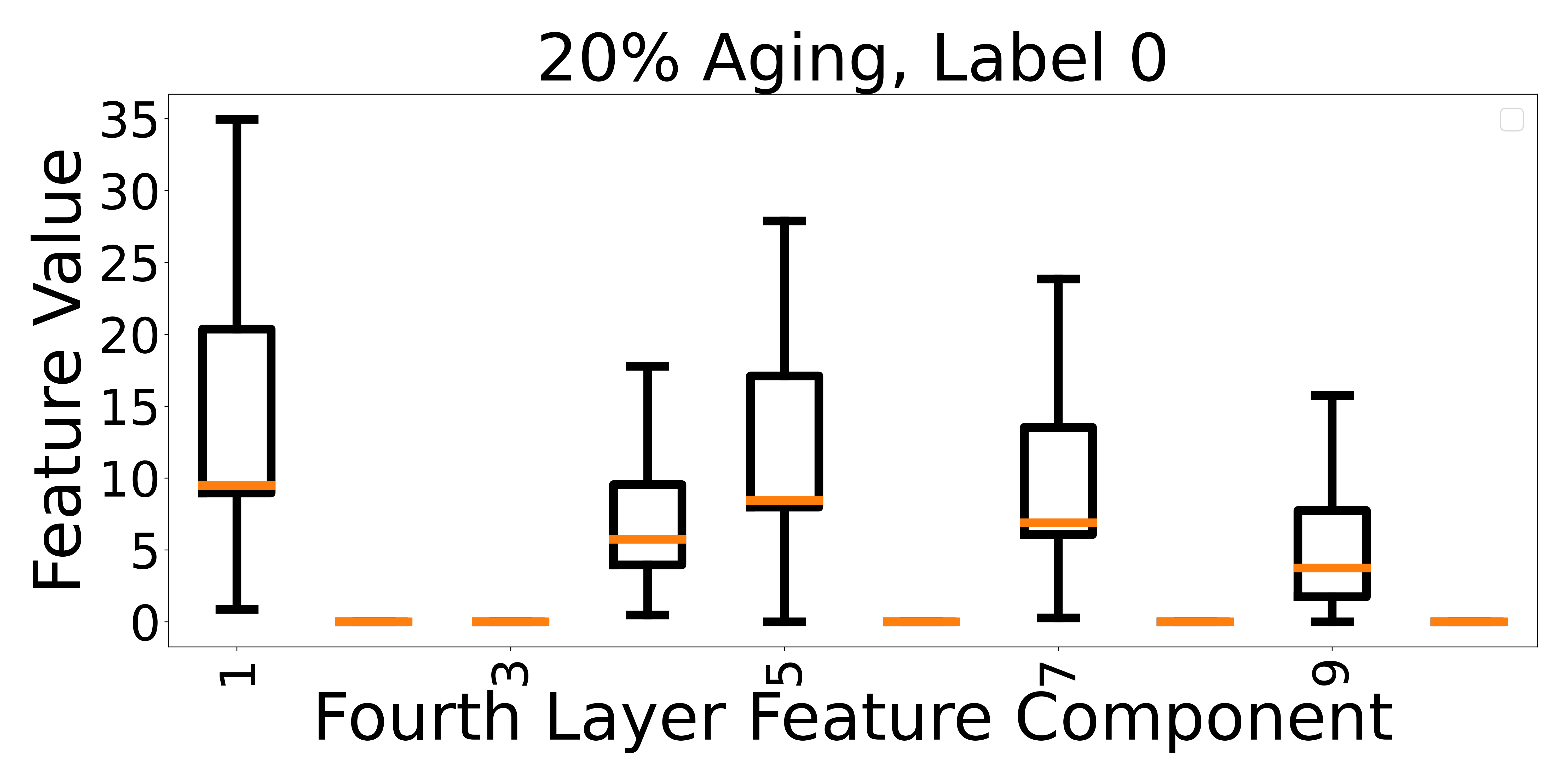}}
    \hfill
	\caption{
 Aging has a considerable effect on feature value statistics\index{feature statistics}. The statistics for each layer's feature vectors are depicted for samples with label 0. Each row corresponds to a specific layer, with the left column displaying 0\% aging and the right column showcasing 20\% aging. The features are plotted for 0.6V at the nominal clock period of 0.305 ns.
 }
	\label{fig:feature_stats}
\end{figure*}

The sudden drop in accuracy is due to the depth of the neural network. Timing errors\index{timing errors} resulting from aging occur at each layer and accumulate as the data reaches the output layer. As soon as a sufficient number of errors accumulate, the accuracy of the DNN drops sharply. This indicates that the classification accuracy worsens over time with more layers due to aging. To further examine this effect, feature vector statistics for samples with labels 0 at each layer of the network are presented in Figure \ref{fig:feature_stats} for 0.6V at zero aging and 20\% aging. The x-axis corresponds to the neurons of the layer. For each neuron, box plots for the features are plotted, with the label zero at zero aging on the top row and at 20\% aging on the bottom row. A significant shift between feature statistics at zero aging and 20\% aging is evident, demonstrating the considerable impact of aging on the DNN's behavior. It can be seen that on the last layer of the zero aging case, the features of the zeroth neuron are dominating, suggesting that the samples are indeed classified as label 0. However, when the hardware implementation of DNN is aged by 20\%, the distribution of features changes significantly, and the features of label 0 no longer dominate. This effect can be observed in the other layers too. Errors caused due to aging change the distribution of features at every layer, causing a significant drop in the overall classification accuracy.

\begin{figure}[h]
  \centering
  \subfloat[0.8V]{\includegraphics[width=0.5\linewidth]{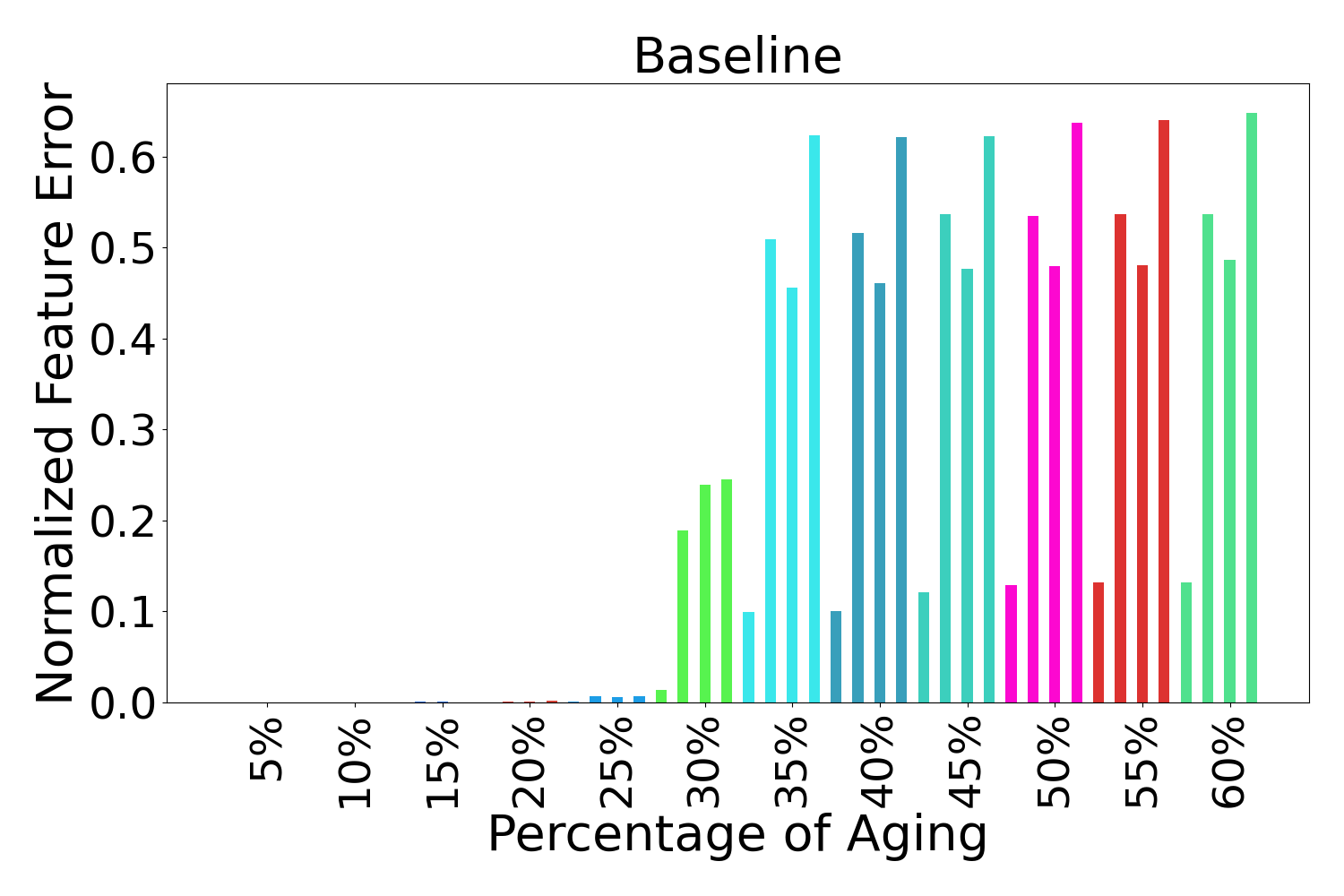}}
  \hfill
  \subfloat[0.6V]{\includegraphics[width=0.5\linewidth]{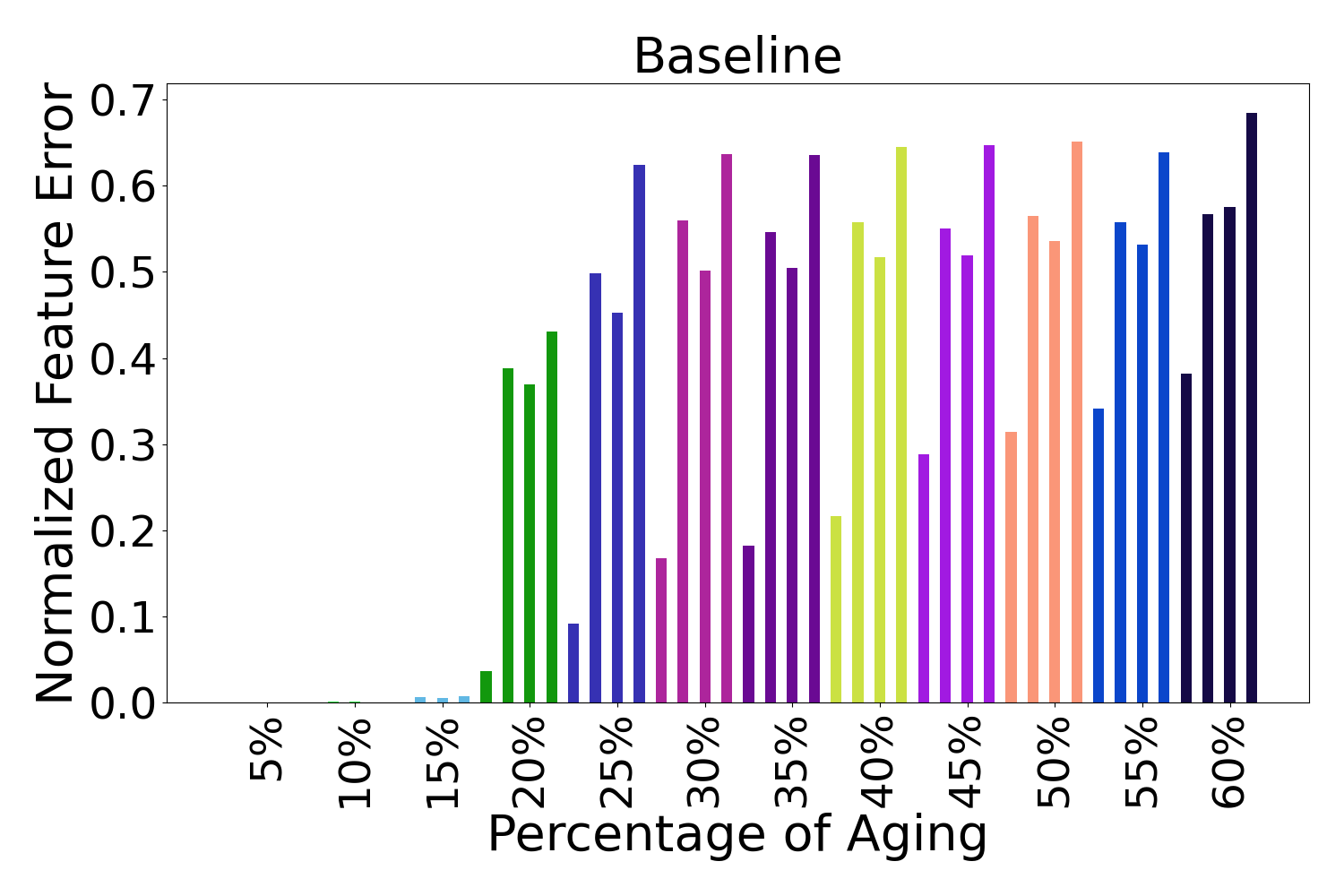}}
  \caption{Feature errors propagation through different layers for various aging levels. Each bar represents the average number of feature element mismatches when compared to 0\% aging. The X-axis shows the aging level from 0\% being no aging and 100\% indicating worst-case aging. Each color set corresponds to a particular aging level and for each color, there are four bars showing the errors for the four feature vectors.
  The tests are performed at the nominal clock periods (0.185 ns and 0.305 ns for 0.8V and 0.6V, respectively). The nominal clock period at 0.6V is higher than at 0.8V due to scaling down of frequency with reduction in voltage to account for higher path delays at lower voltages.}
  \label{fig:feature_errors}
\end{figure}

Figure \ref{fig:feature_errors} shows the errors in outputs (post-activation) at each layer across different aging states and two voltage levels, 0.8V (left) and 0.6V (right). The errors that occur at every layer propagate to subsequent layers. The number of mismatches at each layer increases with the depth of the network. There is a slight decrease when going from the second layer to the third layer. This is because of the effect of the activation function. As ReLU makes every negative number zero, this cancels out some of the errors that are accumulated from the previous layers. More layers in a DNN model result in lower classification accuracy as devices age, which negates the accuracy benefit of deeper network design after a certain time of operation.

\subsection{Mitigation Using Clock Frequency Reduction  (i.e., Increased Timing Guardbands)}
\label{sec:mitigation_clockfrequency}

\begin{figure*}
	\centering
	\subfloat[0.8V]{\includegraphics[width=0.5\linewidth]{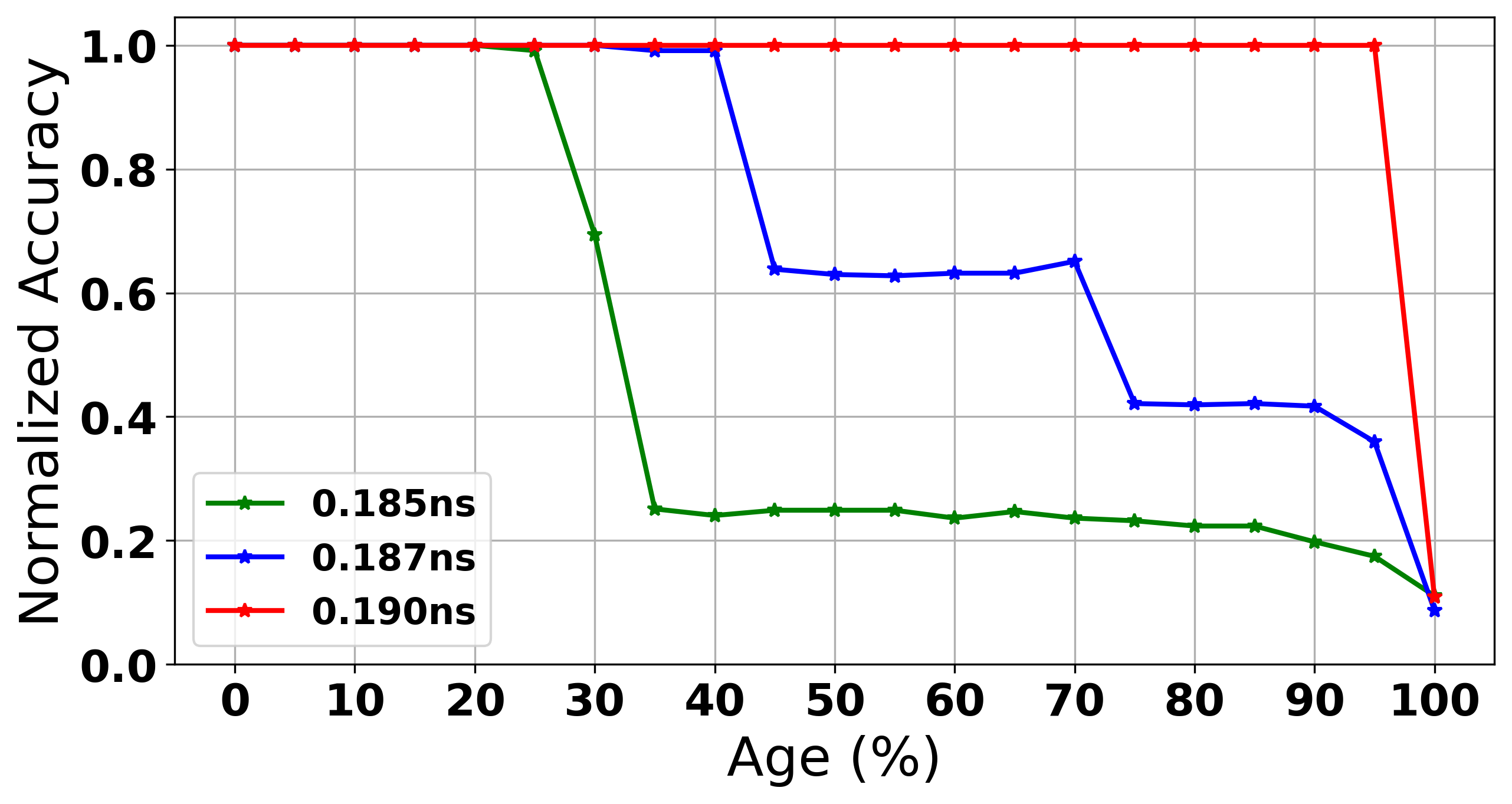}}
 \hfill
	\subfloat[0.6V]{\includegraphics[width=0.5\linewidth]{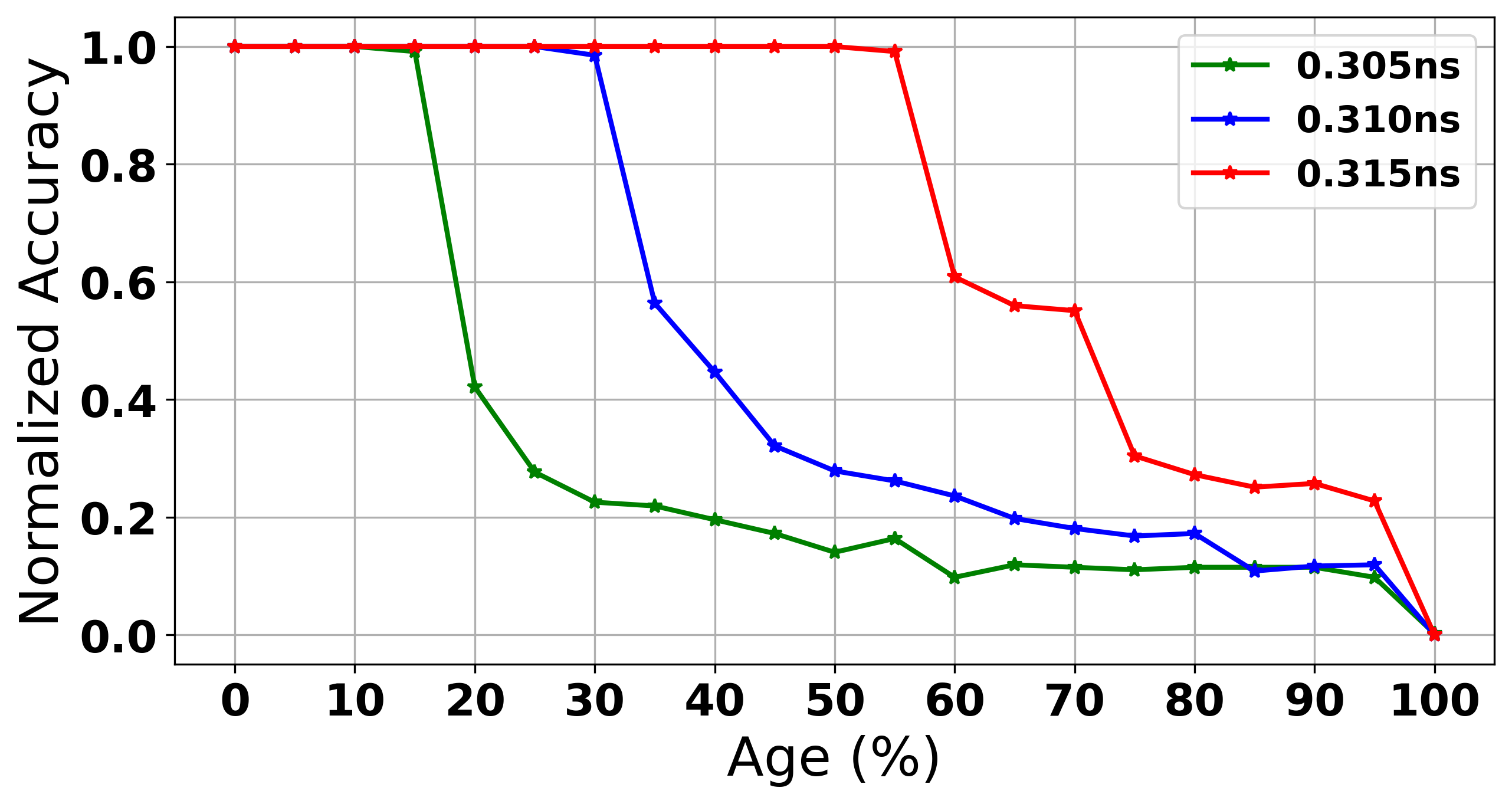}}
	\caption{Normalized accuracy drops with increasing aging level for different clock periods.
  The tests are performed at the nominal clock periods (0.185 ns and 0.305 ns for 0.8V and 0.6V, respectively).
  }
	\label{fig:acc_multi_freq}
\end{figure*}

Transistor aging effects become apparent as delays at the system level. One approach to mitigate aging effects is to decrease the input clock frequency. Reducing the input frequency allows more time for data to propagate, minimizing timing errors at the output. Figure \ref{fig:acc_multi_freq} demonstrates the DNN's inference accuracy at various aging levels for 0.8V and 0.6V. For 0.8V, the nominal clock period is 0.185 ns. When the clock period is increased to 0.190 ns, the accuracy remains stable until almost the end of life. In comparison to the nominal frequency, the aging effect on the DNN's accuracy is reduced. In the case of 0.6V, the nominal clock period is 0.305 ns. By increasing the clock period to 0.315 ns, the DNN maintains its inference accuracy for an extended period. Further increasing the clock period can delay the aging impacts even more through clock frequency scaling\index{clock frequency scaling}.

In summary,
incorporating a timing guardband\index{timing guardband} into a deep learning hardware implementation can mitigate the impact of aging. Increasing the clock period to 0.190 ns (for 0.8 V) and to 0.315 ns (for 0.6 V) can delay the decline in accuracy, but it eventually does lead to accuracy loss over time. By using a larger guardband, the DNN can maintain its inference accuracy for an extended period. However, implementing such a larger guardband slows down the operation of the DNN and therefore contradicts the initial objective of aggressively increasing or sustaining high throughput while reducing hardware redundancy for enhanced energy efficiency.

\section{Aging-Aware Resilient Deep Networks}
\label{sec:mitigation_gradnoise_imitator}
In this section, mitigations are discussed to enable robustifying a DNN to aging\index{aging-aware resilience}.
\subsection{Problem Formulation}
Consider a DNN $f(.;\theta)$ performing classification on a set of samples $D=\{(x_i,y_i) | f(x_i;\theta)=y_i; i=0,\cdots, N \}$ where $x_i$ is the input,  $y_i$ is the corresponding label, and $N$ is the number of samples. Due to aging, there is degradation in network inference accuracy. Therefore, for the same set of samples, the output observations are $D_d=\{(x_i, y_{i,d}) | f_d(x_i;\theta)=y_{i,d}\}$ where $f_d$ is the degraded version of the network (i.e., the behavior of the hardware implementation of the DNN under a specific aging level). The objective is to find a set of parameters $\theta$ for the network so that it performs similarly for normal and degraded versions:
\begin{equation}
    \begin{aligned}
        f(x_i;\theta)   = y_i \hspace{0.2in} \mbox{and} \hspace{0.2in}
        f_d(x_i;\theta) = y_i.
    \end{aligned}
\end{equation}
To achieve the aforementioned objective, the following optimization problem is introduced:
\begin{equation}
    \min_{\theta} \sum_{i=0}^N l(f(x_i;\theta),y_i)+l(f_d(x_i;\theta),y_i)
    \label{eq:main_prob}
\end{equation}
where $l(.,.)$ represents the loss function. Solving Eq.~\ref{eq:main_prob} with gradient-based methods is computationally infeasible due to the need to calculate gradients through the hardware-implemented DNNs. The findings in Sec.~\ref{sec:effects_age_DNN} indicate that accuracy degradation due to aging mainly results from bit flips occurring during bit-wise multiplications on the hardware. To build robustness to such inaccuracies in computation (i.e., a noisy computation environment), one approach would be to mimic the noisy computation environment during training. The noise could be added to features and/or gradient during training. Three strategies are presented to address this issue.
\subsection{Feature Noise During Training}
\label{subsec:feature_noise}
\begin{figure}[!h]
	\centering
	\subfloat[0.8V]{\includegraphics[width=0.5\linewidth]{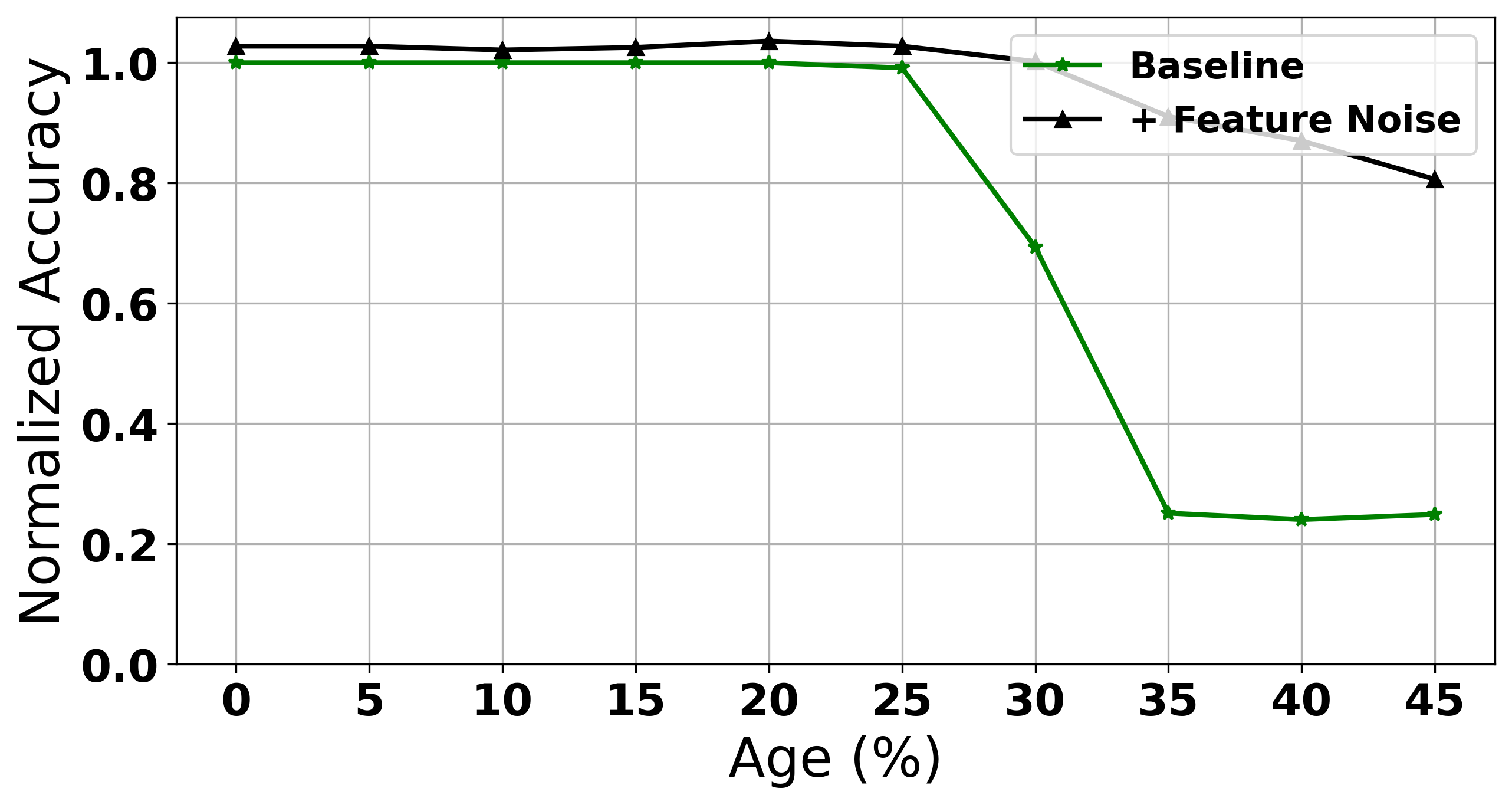}}
 \hfill
	\subfloat[0.6V]{\includegraphics[width=0.5\linewidth]{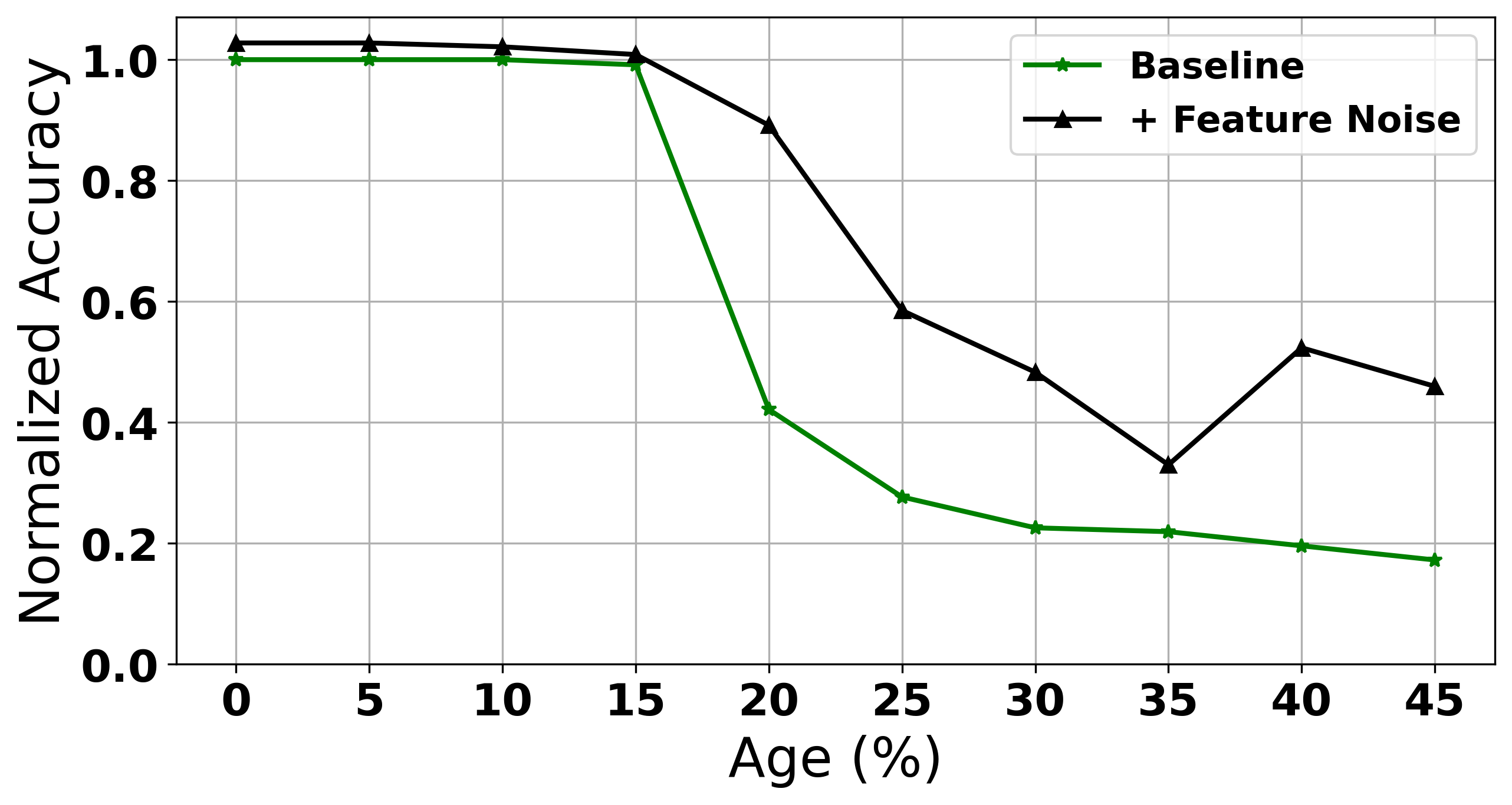}}
	\caption{Normalized accuracies for 0.8V and 0.6V when feature noise\index{feature noise} is applied during training.
  The tests are performed at the nominal clock periods (0.185 ns and 0.305 ns for 0.8V and 0.6V, respectively).
  }
	\label{fig:acc_featurenoise}
\end{figure}

The first approach for mimicking the noisy environment during training is to add noise to feature vectors. For this purpose, Gaussian noise is added to the DNN's feature vectors during training\index{feature noise}.
\begin{equation}
    f_t \leftarrow f_t + \mathcal{N}(\mu, \sigma_t^2)
\end{equation}
where $f_t$ is the feature vector, $\mu$ is the mean vector of the error between degraded feature vector and original vector, and $\sigma_t$ is the standard deviation vector of the error. To evaluate this methodology, the network in Sec.~\ref{sec:effects_age_DNN} was trained using feature noise\index{feature noise} to find the desired set of parameters for the DNN. The feature noise is estimated for each aging percentage. For this purpose, the network hardware was simulated with different percentages of aging at different voltage levels, i.e., 0.6V and 0.8V. An aggregate average model of feature noise was then used to train the network. The trained network was then evaluated for different percentages of aging on the hardware. The corresponding results are shown in Figure~\ref{fig:acc_featurenoise}. It could be seen that for all the percentages of aging at both 0.6V and 0.8V, the robustified network performs better. More specifically, for 0.8V at 45\% aging, a boost of around 52\% could be seen. In Figure~\ref{fig:acc_featurenoise} and all subsequent accuracy plots, ``Baseline'' refers to the performance of the original network (that was not trained using the aging-aware mitigation mechanisms) when tested at the aging level and voltage indicated in the plots.

\subsection{Gradient Noise (GN) During Training}
\label{subsec:AGN}
\begin{figure}[h]
	\centering
	\subfloat[0.8V]{\includegraphics[width=0.5\linewidth]{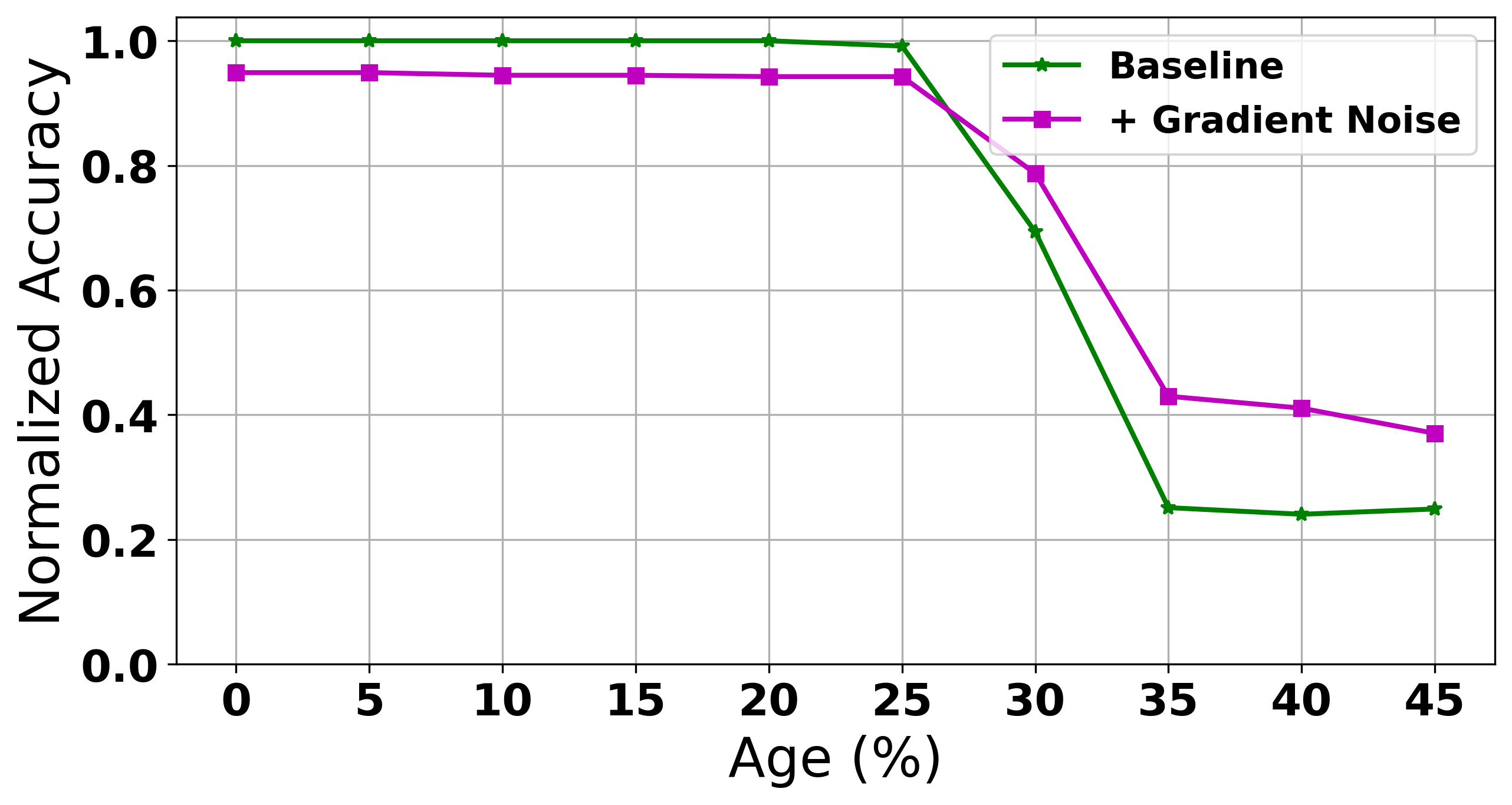}}
 \hfill
	\subfloat[0.6V]{\includegraphics[width=0.5\linewidth]{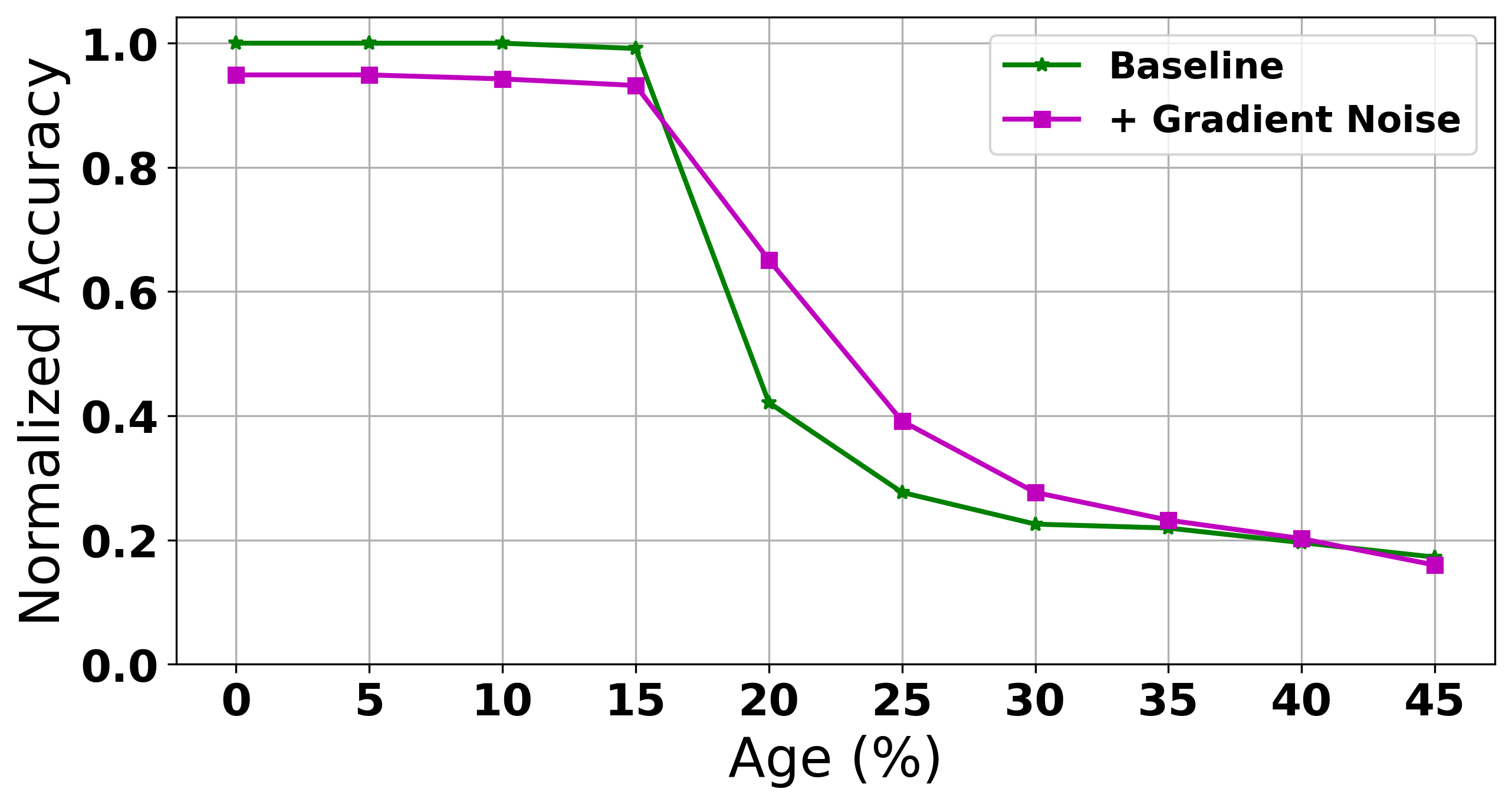}}
	\caption{Normalized accuracies for 0.8V and 0.6V when gradient noise\index{gradient noise} is applied during training.
  The tests are performed at the nominal clock periods (0.185 ns and 0.305 ns for 0.8V and 0.6V, respectively).
  }
	\label{fig:acc_gradnoise}
\end{figure}

Another approach for imitating the noisy training environment is to add the noise to the gradient vectors. For this purpose, Gaussian noise is added to the DNN's gradient during training\index{gradient noise}. Using an approach similar to \cite{neelakantan2015adding}:\begin{equation}
    g_t \leftarrow g_t + \mathcal{N}(0, \sigma_t^2)
\end{equation}
To evaluate this methodology, the network in Sec.\ref{sec:effects_age_DNN} is trained using GN to find the desired set of parameters for the DNN. The network hardware was simulated with different percentages of aging at different voltage levels, i.e., 0.6V and 0.8V. The network with gradient noise\index{gradient noise} is trained once and then it is evaluated for different percentages of aging on the hardware. The corresponding results are shown in Figure~\ref{fig:acc_gradnoise}. It could be seen that for 0.8V after 30\% aging, the robustified network performs reasonably well under the effect of aging. The reduced accuracy of the robustified network for 0\% to 30\% aging is the cost of the compromise between network robustification against aging and the nominal accuracy of the network. Also for 0.6V, the robustified network can be seen to improve the accuracy up to 25\% for 20\% aging.

\subsection{Combination of Feature and Gradient Noise\index{feature noise}\index{gradient noise}}
\label{subsec:COMBO}
\begin{figure*}
	\centering
	\subfloat[0.8V]{\includegraphics[width=0.5\linewidth]{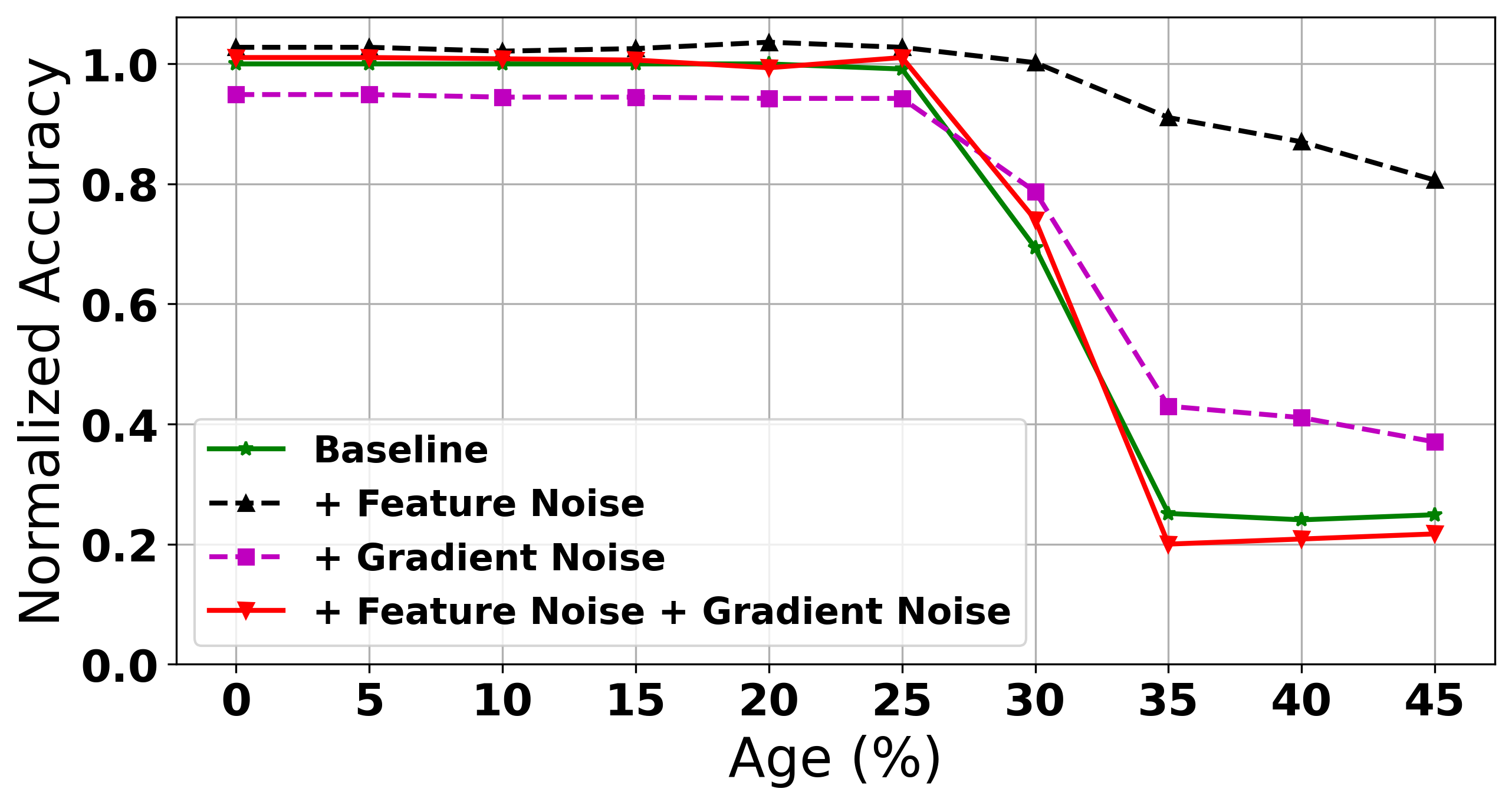}}
 \hfill
	\subfloat[0.6V]{\includegraphics[width=0.5\linewidth]{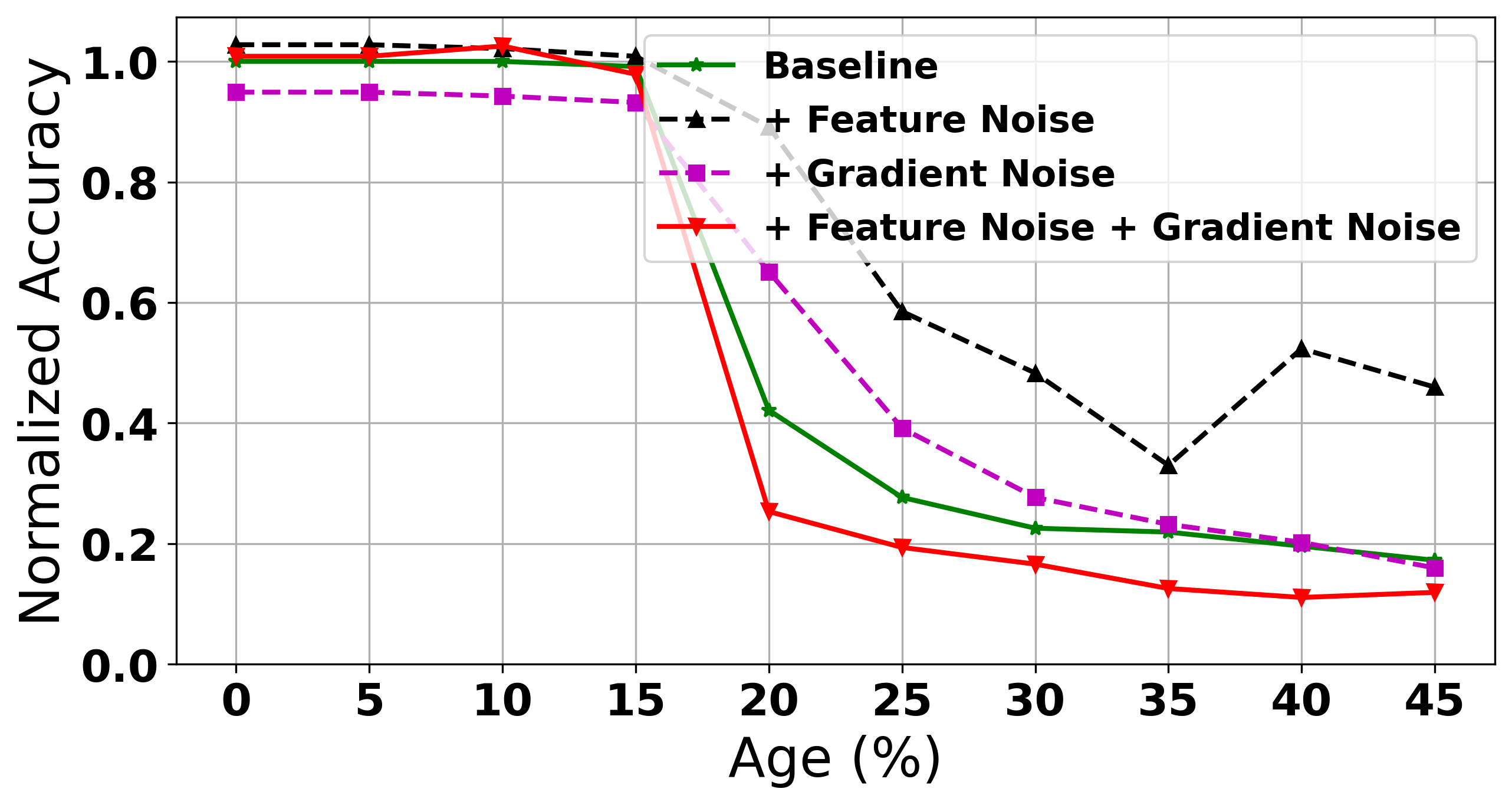}}
	\caption{Normalized accuracies for 0.8V and 0.6V when feature noise\index{feature noise} along with gradient noise\index{gradient noise} is applied.
  The tests are performed at the nominal clock periods (0.185 ns and 0.305 ns for 0.8V and 0.6V, respectively).
  }
	\label{fig:acc_combo}
\end{figure*}

The observations in \ref{subsec:feature_noise} and \ref{subsec:AGN} shows that noisy environment during training robustifies the network against aging. In this section, the effect of combining these  methods is discussed. The method was evaluated at 0.8V and 0.6V, with all experiments using the Adam optimizer. Figure~\ref{fig:acc_combo} displays the results, with the red line representing the robustified network's accuracy and the green line the baseline network. The method is not successful in mitigates aging effects at both 0.6V and 0.8V for varying aging percentages.

\subsection{Discussion}
\begin{figure*}
	\centering
	\subfloat[0.8V at 0.187 ns]{\includegraphics[width=0.5\linewidth]{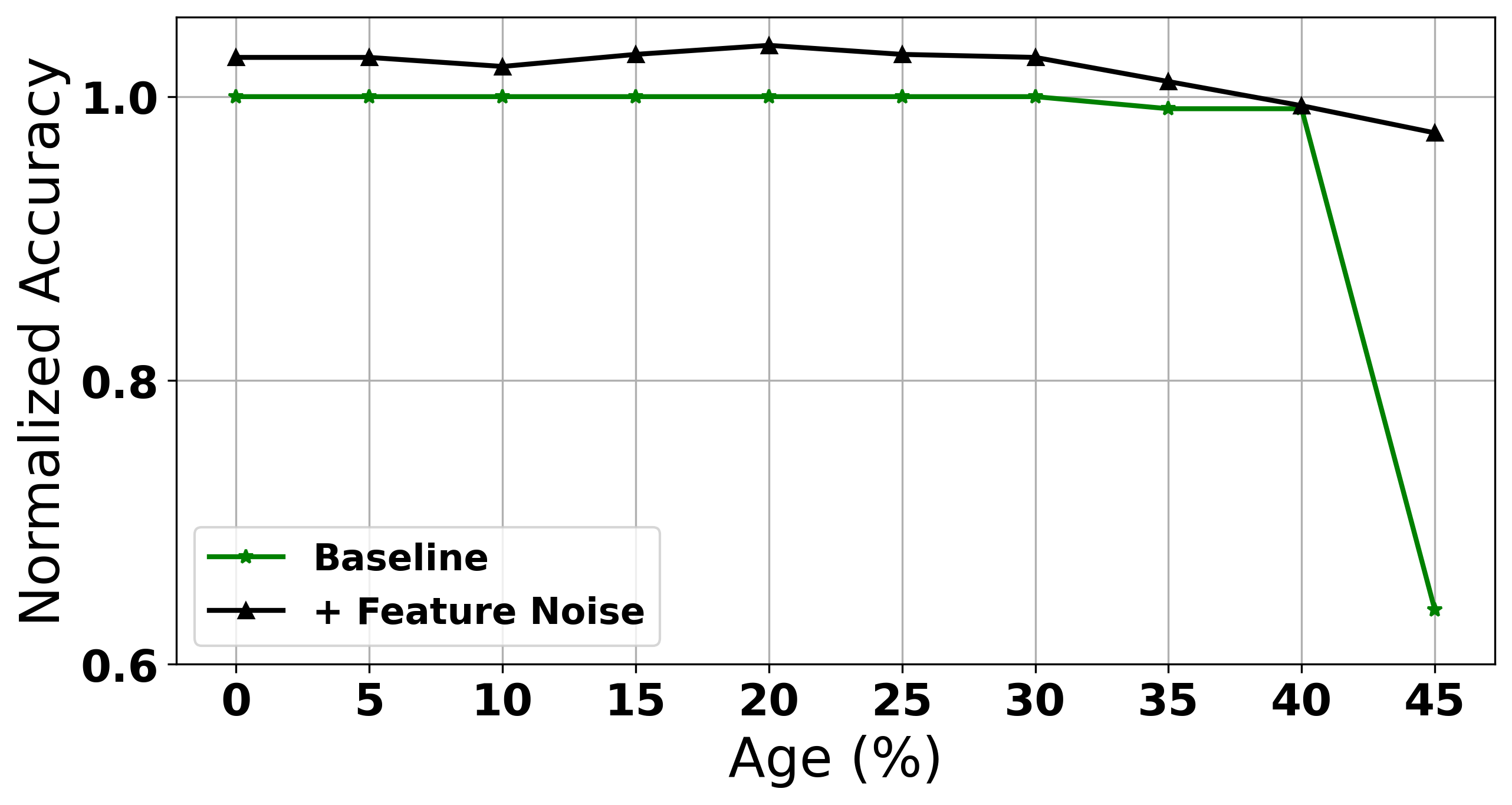}}
 \hfill
	\subfloat[0.6V at 0.313 ns]{\includegraphics[width=0.5\linewidth]{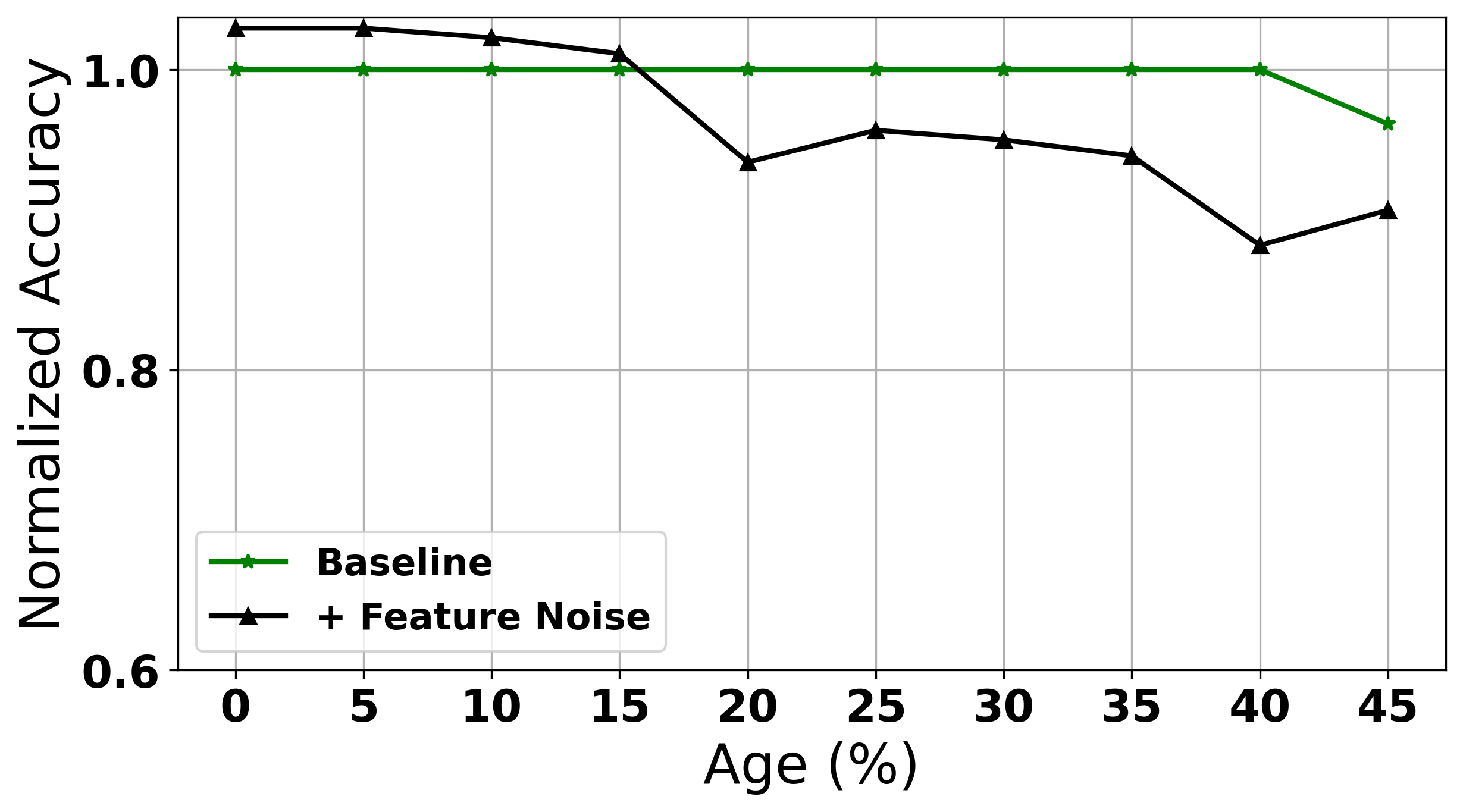}}
	\caption{Normalized accuracies for 0.8V and 0.6V when feature noise during training is applied. The clock periods used are 0.187 ns for 0.8V and 0.313 ns for 0.6V, corresponding to slightly lower clock frequencies than nominal (the nominal clock periods are 0.185 ns and 0.305 ns for 0.8V and 0.6V, respectively).
  }
	\label{fig:acc_featurenoise_clk}
\end{figure*}

\begin{figure}[h]
  \centering
  \subfloat[0.8V]{\includegraphics[width=0.5\linewidth]{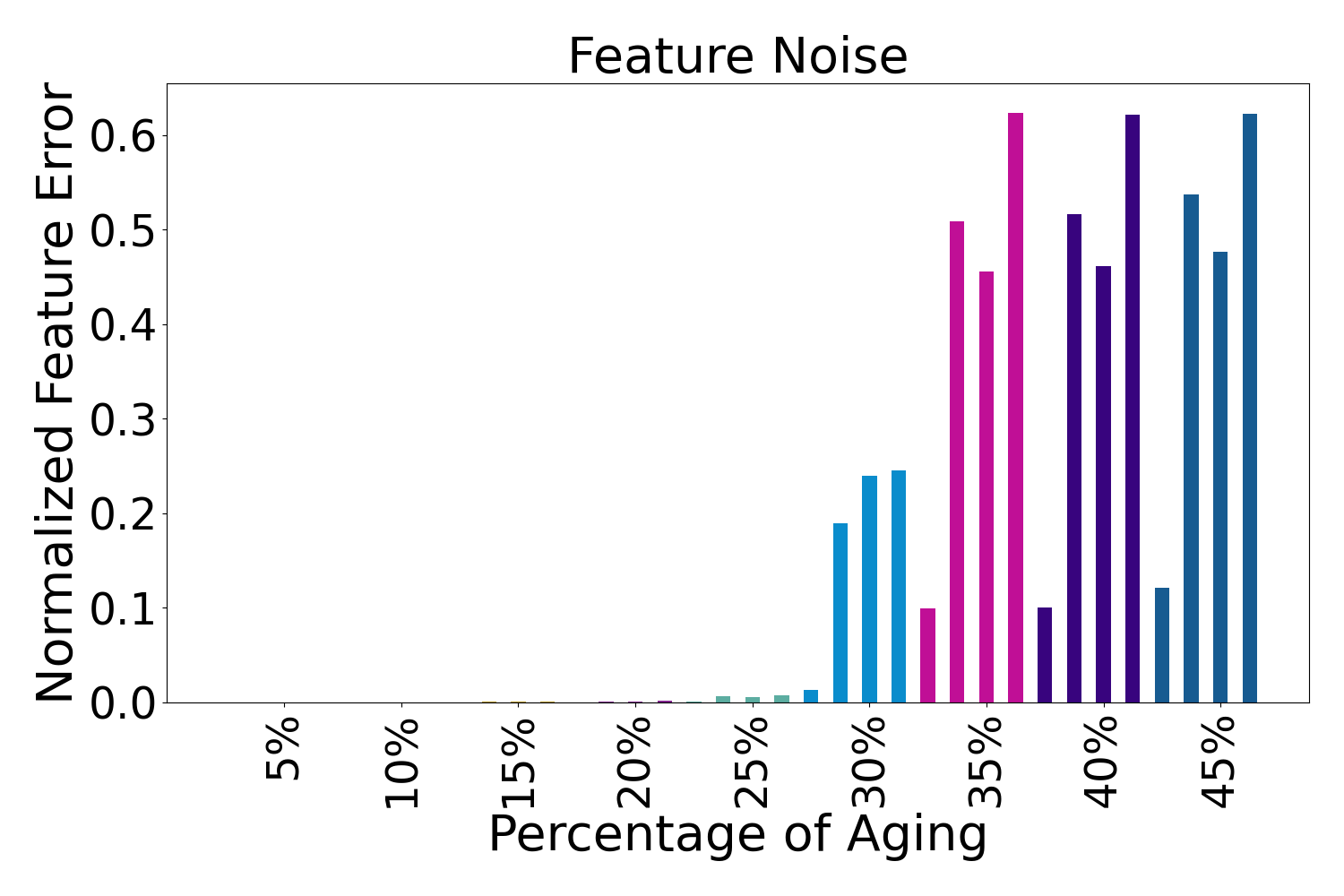}}
  \hfill
  \subfloat[0.6V]{\includegraphics[width=0.5\linewidth]{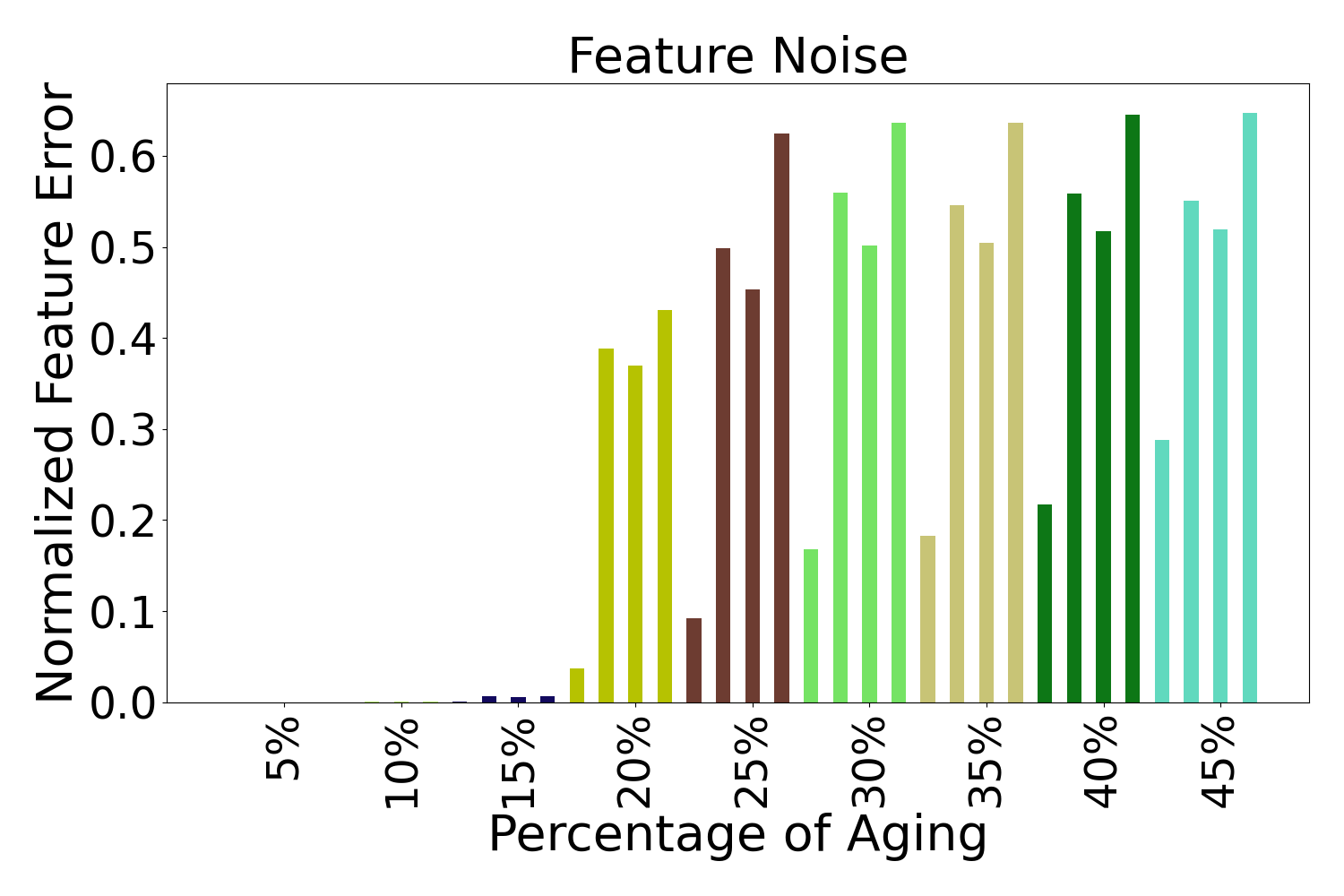}}
  \caption{
  Propagation of feature errors across multiple layers at different aging levels in a robustified network using the feature noise during training. Each bar illustrates the average \% of a feature vector's components deviating from 0\% aging. The four bars in each color group signify the errors for four distinct feature vectors, with each color set corresponding to a particular aging percentage.
  The tests are performed at the nominal clock periods (0.185 ns and 0.305 ns for 0.8V and 0.6V, respectively).
  }
  \label{fig:feature_errors_combo}
\end{figure}

The mitigation strategies introduced in the previous sections have different affect on the inference accuracy of DNNs. It was observed that Feature noise during training is the most effective method by increasing the inference accuracy of the network. Feature noise during training\index{feature noise} helps in narrowing the guardband\index{timing guardband} that can be employed on the chip. Figure \ref{fig:acc_featurenoise_clk} shows that by increasing the clock period to 0.187 ns (nominal clock period - 0.185 ns) at 0.8V, the inference accuracy can be recovered when compared to the baseline case for almost up to 50\% of aging. Similarly, when the voltage is 0.6V, by increasing the clock period to 0.313 ns, much of the inference accuracy can be recovered with the presented mitigation strategy. Also, observe that as noted in Sec.~\ref{sec:mitigation_clockfrequency}, the aging effect on the original network is also greatly reduced with clock frequency reduction\index{clock frequency scaling}. Figure \ref{fig:feature_errors_combo} illustrates errors at each layer across different aging states for 0.8V and 0.6V on a network trained using the Feature Noise during training. Errors propagate between layers, indicating the network's depth impact on mismatches. When compared to the feature mismatches in the original network (Figure \ref{fig:feature_errors}), the relative number of mismatches in the last layer is lower in Figure \ref{fig:feature_errors_combo}. This  indicates that the DNN is now more robust to the aging effects\index{aging-aware resilience}, increasing the inference accuracy.

\afterpage{\clearpage}

\section{Application of Short-Term Aging to Hardware Trojan Detection}
\label{sec:trojan_detection}
\begin{figure*}[t!]
  \centering
  \includegraphics[width=1.0\textwidth]{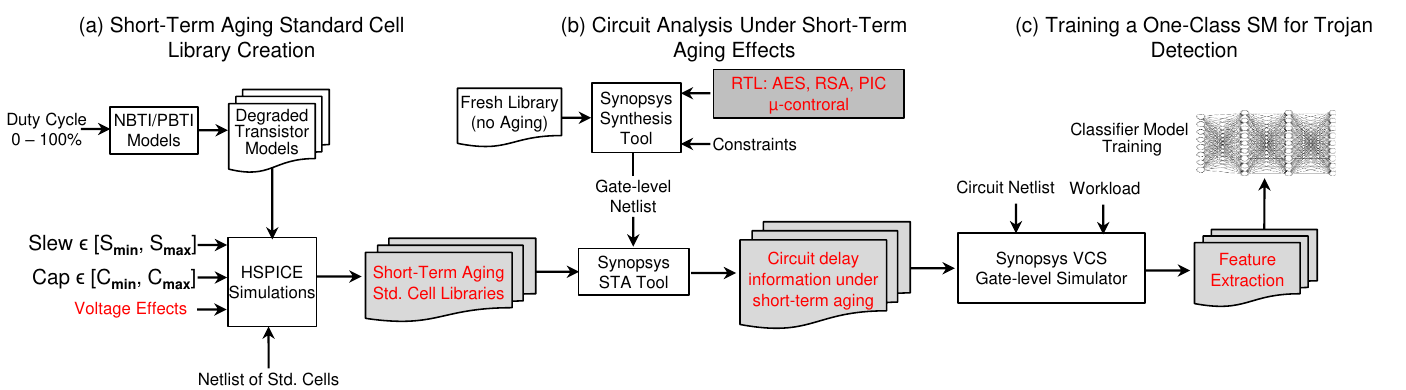}
  \caption{Overall pipeline~\cite{surabhi2022trojan} in the short-term aging-based Trojan detection methodology.}
  \label{fig:flowchart}
  \end{figure*}

While this article focused primarily on the effects of aging on DNNs and mitigations for these effects, the application of the short-term aging phenomenon for hardware Trojan detection in ICs\index{Hardware Trojan detection}\index{Trojan detection} is briefly discussed in this section. This technique for hardware Trojan detection has been studied in the context of multiple technology nodes in the literature including 45 nm CMOS~\cite{surabhi2020hardware,surabhi2020exposing} and 14 nm FinFET~\cite{surabhi2022trojan}.
Hardware Trojans are malicious modifications of circuits, inserted by adversaries at various stages of the manufacturing process to, for example, leak data, cause malfunctions, or denial of service in embedded systems using the ICs. Detecting
Hardware Trojans in ICs is challenging since Trojans are dormant most of the time, small relative to the rest of the IC, have unknown triggers corresponding to rare events that are hard to guess/reverse-engineer, typically not detectable using functional testing, and have small/negligible impact on side channels under normal operation of the IC.
Short-term aging, which is a transient effect observed when transistor aging is combined with voltage scaling, enables magnifying the observable effect of a Trojan without triggering it. These short-term degradations  vary the delays dynamically along various paths in the IC. These delay changes inside the IC manifest at the output during this transient phase as a fingerprint of errors in the output bits. The patterns of these bit errors enables discrimination between clean and Trojaned circuits.
The detection methodology is as follows. The RTL design of the clean
IC is used with a standard cell library (no
aging) during logic synthesis. The clean IC is synthesized to generate a
gate-level netlist using this library with operating conditions of nominal
voltage (e.g., 1.1 V) and temperature (e.g., 25 degree Celsius). A Trojan
inserted into the gate-level netlist of the clean IC yields a modified IC referred to as a Trojaned IC. The delay information at various operating voltage levels along with the aging stresses are obtained using Static Timing Analysis (STA) tool and Standard Delay Format (SDF) files.
Gate-level simulations are performed with the timing/delay information annotated
from the SDF files. A set of inputs are applied and the corresponding
input-output pairs are collected at every voltage level and aging stress. A rich
set of features is extracted from these input-output pairs. The features
of the clean IC are used to train a one class Support Vector Machine (SVM)\index{Support Vector Machine (SVM)}. This trained
classifier distinguishes the clean IC from a Trojaned IC. Figure~\ref{fig:flowchart} gives an overview of the approach.

\subsection{Feature Extraction and One Class SVM Classifier}
\label{sec:features}
The Trojan detection technique exercises the IC with several inputs, monitors
the IC outputs under test conditions (e.g., different short-term aging levels,
different voltages, different clock frequencies), and compares the observed
patterns of output bit errors relative to expected baseline outputs (under zero
aging and nominal clock).

The detection of Trojaned ICs depends on the bit error patterns at the output of
the IC when short-term aging is applied. The bit errors are extracted by
comparing the outputs of the IC against a known-good simulation of a Trojan-free
IC at normal operating conditions (i.e., no aging and no voltage scaling). A one class SVM classifier is used to learn the bit error patterns.

In order to train the SVM classifier on properties of the simulated ``golden
IC'', consider a set of $n_x$ inputs. Let $y_{ijk} = f_{ijk}(x)$ be the output
of an IC where $f_{ijk}$ maps an $m$-bit input $x\in X$ to $n$-bit output
$y_{ijk}\in Y$ at voltage $V_i$, aging state $A_j$, and opcode $O_k$. For circuits
without opcodes (e.g., cryptoaccelerators), the subscript $k$ is omitted.
Opcodes are highly relevant to microcontrollers such as the PIC and PULPino. Let
the range of supply voltages be $V_1, V_2,....., V_{n_V}$, the range of aging
states be $A_1, A_2,....., A_{n_A}$ and opcodes be $O_1, O_2,....., O_{n_O}$.
Let the expected output of a ``golden IC'' on input $x$ be $y_{O_k} =
g_{O_k}(x)$ (opcode dependent).
The function $g_{O_k}$ is  deterministic.
Manufacturing introduces process variations in
ICs. For the classifier to be robust to such variations, the machine learning model does not
simply learn the input-output mappings. Instead, it learns a model of the bit error patterns
at the outputs. For that purpose, different sets of inputs are used during
training and testing. In general, the methodology can be applied with a flexible
set of excitations (test conditions) that include different
short-term aging levels, voltages, and opcodes as discussed above. In addition,
different clock frequencies (i.e., overclocking) can be applied as part of the
test conditions. When overclocking is included, an additional subscript $c$
would be included (i.e., $f_{ijkc}$), indicating the $c^{th}$ clock frequency
applied. For simplicity of notation, the subscript $c$ is not explicitly included in the
discussion below.

In order to capture the bit error patterns from the outputs of an IC,
a rich set of features\index{feature extraction} is extracted from the outputs. For a given input $x$, the difference
between $y_{O_k} = g_{O_k}(x)$ and $y_{ijk} = f_{ijk}(x)$ at operating voltage $V_i$
and aging state $A_j$ is estimated using four features: 1) Number of
0$\rightarrow$1 bit flips\index{bit flips}, 2) Number of 1$\rightarrow$0  bit flips, 3) Weighted
combination of 0$\rightarrow$1 bit flips and 4) Weighted combination of
1$\rightarrow$0 bit flips considering their bit locations. The following ``bit
indicator functions'' $I_1(b)$, $I_0(b)$ are defined as subsets of $\mathcal{N}=\{0,...,n-1 \}$ to determine if a bit is ``1'' or ``0'' in a binary number $b$ of length $n$:
\begin{align}
I_1(b) = \{r\in {\cal N} | b\&(1<<r) \neq 0\} \\
I_0(b) = \{r\in {\cal N} | b\&(1<<r) = 0\}
\end{align}
where $\&$ is bit-wise AND and $<<$ is left shift. $I_1(b)$ and $I_0(b)$ are subsets of bit locations in $\mathcal{N}$ corresponding to 1 and 0 bits in $b$.

For a given input $x$, the feature vectors are generated for various operating conditions of supply voltages, aging states and opcodes, thus forming a 4-dimensional feature tensor of size $n_O \times n_V \times n_A \times n_H$. However, for accelerators, first dimension is always ``1'' as the opcodes are not involved. Hence, the size of a feature tensor for each input would be of size $1 \times n_V \times n_A \times n_H$ for such ICs.
Given an observed output $y$ and expected output $y_{O_k}$, the feature vector is defined as
\begin{align}
  h_1(y,y_{O_k}) &= \sum_{r\in(I_0(y_{O_k})\cap I_1(y))}1
  \\
  h_2(y,y_{O_k}) &= \sum_{r\in(I_1(y_{O_k})\cap I_0(y))}1
  \\
  h_3(y,y_{O_k}) &= \sum_{r\in(I_0(y_{O_k})\cap I_1(y))}(1<<r)
  \\
  h_4(y,y_{O_k}) &= \sum_{r\in(I_1(y_{O_k})\cap I_0(y))}(1<<r)
  \\
  H(y,y_{O_k}) &= [h_1(y,y_{O_k}),h_2(y,y_{O_k}),
  \nonumber\\
  &\quad h_3(y,y_{O_k}),h_4(y,y_{O_k})]^T \in {\cal R}^{n_H};n_H=4.
\end{align}
Using the feature vectors $H(.,.)$, a 4-dimensional feature tensor is computed
as described above. The $(i,j,k,l)^{th}$ element of the feature tensor is the
$l^{th}$ element of $H(y_{ijk},y_{O_k})$. With inclusion of overclocking, the feature tensor would, in
general, be 5-dimensional, i.e., of dimension $n_O \times n_V \times n_A \times
n_C \times n_H$ where $n_C$ is the number of clock frequencies used.

A series of inputs from the set $X$ is used to train/test the model. The
elements in the set $X$ are partially dependent on the IC as ATPG (Automatic Test Pattern Generation)
patterns are included in the set.
During training, the input set $X$ is divided into training set $X'$ and testing
set $X''$. Feature tensors for each input $x'_i \in X$ are generated, combined,
and used in training. The classifier has an autoencoder and a one-class SVM. The
feature tensors of all the training examples are input to the autoencoder to
extract a low-dimension feature vector from the middle layer of the autoencoder.
The four-layer autoencoder has two layers each of encoder and decoder with ReLU activations. The bottleneck layer of autoencoder is extracted and passed through a one-class SVM to train the model.
The SVM is trained as a one-class classifier for outlier detection using the set
of feature tensors for the training data. The trained classifier decides if a feature tensor computed from outputs measured from an IC under test is an outlier.
When testing an IC, inputs from the test set $X''$ are applied to the IC, and outputs
are collected. From the observed outputs, bit errors and feature vectors are
computed as discussed above. The generated feature tensors are passed through the trained autoencoder to obtain a
low-dimensional feature vector, which is then passed through the trained
one-class SVM for inlier/outlier estimation (i.e., to distinguish between a clean and Trojaned IC). The evaluation set $X''$ could be
disjoint or overlap with the training set $X'$. In any case, the anomaly
detector is input-independent and depends on function patterns of bit error
deviations under different aging states, clock periods, etc., rather than
comparing individual output values. Since the detector works on a feature tensor
derived from outputs calculated for a single given input, using a single input
is sufficient to evaluate inlier/outlier. However, multiple inputs (called a
bin) are used to increase precision, and a majority voting of inlier/outlier
determinations generates an aggregate inlier/outlier estimation. Furthermore,
majority voting over multiple bins further increases the accuracy.

\begin{figure}[!]
  \centering
  \includegraphics[width=\columnwidth]{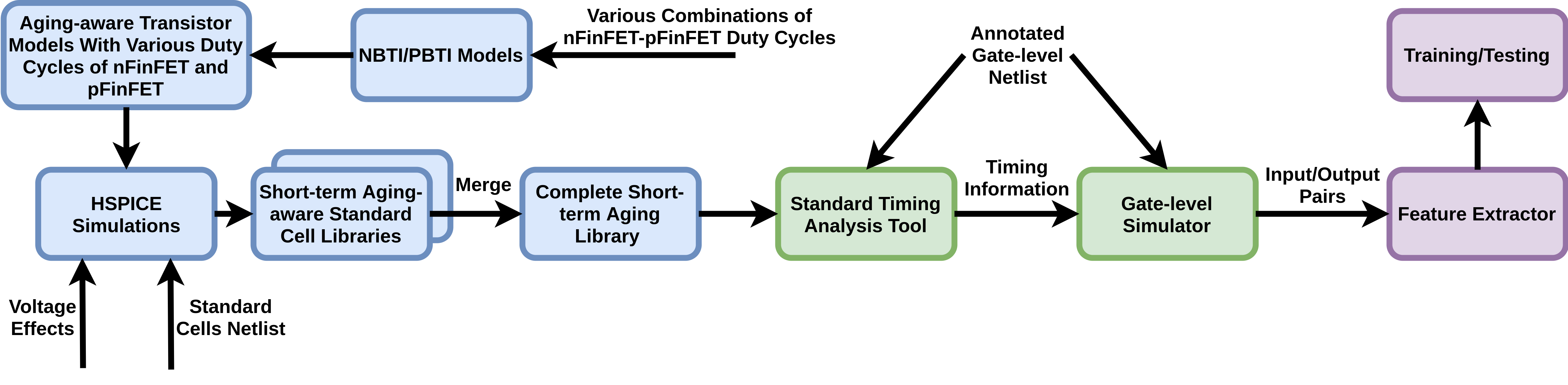}
  \caption{Overview~\cite{surabhi2022trojan} of short-term aging based Trojan detection approach: First, dynamic short-term aging-aware standard cell libraries are created by considering transistor aging and voltage effects for various duty cycles for FinFET transistors. Then, FinFET libraries are plugged into static timing analysis tools to extract delays of the annotated netlist. Finally, gate-level simulations are run to extract bit errors and other features to train a one-class SVM for detecting Trojans.}
  \label{fig:overall_methodology}
\end{figure}

The overall short-term aging-based Trojan detection methodology is illustrated in Figure~\ref{fig:overall_methodology}. To perform  gate-level simulations and collect input-output pairs for training the one-class classifier for Trojan detection, custom short-term aging-aware standard cell libraries with FinFET technology are created. These libraries are dynamic in nature; they contain models of transistor short-term aging of every cell at multiple levels of duty cycles. Combinations of pFinFET and nFinFET duty cycles of all cells are considered while creating short-term aging-aware standard cell libraries. This way, standard cells with a particular combination of pFinFET-nFinFET duty cycle according to probabilities of signals of the IC are identified. To extract signal probabilities of signals  in an IC, the circuit is exercised with a large number of inputs. The signal probabilities are mapped to duty cycles by observing the input signals to pFinFET and nFinFET of each gate. The netlist used for gate level simulations is annotated with duty cycles for all gates producing a duty cycle-annotated gate-level netlist. The annotated netlists and dynamic short-term aging-aware libraries are used to generate SDF files at various voltage levels. The IC is overclocked while collecting input-output pairs to improve Trojan detection accuracy\footnote{While short-term aging by itself is typically sufficient to detect Trojans in 45 nm technology~\cite{surabhi2020hardware,surabhi2020exposing}, it has been noted that in the case of 14nm FinFET, adding overclocking as an excitation can increase the hardware Trojan detection accuracy~\cite{surabhi2022trojan}. This is, in part, due to the lower nominal voltage of 14 nm FinFET technology (0.8 V) compared to older planar technologies such as 45 nm (1.1 V), thus reducing the range of voltages for scaling. In this case, short-term aging alone might not produce sufficient bit error patterns at the output especially when the Trojan is far away from the critical path. Combining short-term aging with overclocking increases Trojan observability. However, overclocking by itself does not produce enough bit error patterns at the output to detect Trojans. Instead, combining overclocking with short-term aging provides the most effective approach for hardware Trojan detection with high accuracy and low false positive rates.}. At each voltage level and clock period, a set of inputs is applied, and corresponding outputs are gathered. The input-output pairs yield features based on observed bit flips between observed outputs and nominal expected outputs based on the circuit's functionality. For microcontrollers, opcodes are part of inputs to the IC. A one-class SVM is trained using the features extracted from the simulation model of the clean IC to probabilistically discriminate clean and Trojaned IC.
To improve Trojan detection accuracy, a majority voting of inlier/outlier determinations across multiple inputs (``bin'') is used to generate an aggregate inlier/outlier estimate. Majority voting over multiple bins increases the accuracy.
When testing the hardware Trojan detection method, chip-to-chip and on-chip variations that arise due to global/local process variations and run-time variations induced by temperature changes on the die are also considered in \cite{surabhi2020hardware,surabhi2020exposing,surabhi2022trojan}.

\subsection{Performance of Hardware Trojan Detection}
Hardware Trojan detection performance in 14nm FinFET ICs~\cite{surabhi2022trojan} is summarized in this section. Analogous results for 45nm CMOS ICs can be found in \cite{surabhi2020hardware,surabhi2020exposing}.
The experimental results in this section use Synopsys Design Compiler and a 14 nm FinFET technology library (no aging) for synthesis. The libraries used are generated for operating voltages in 0.50 V -- 0.80 V in steps of 0.10 V and a temperature of 27 degree Celsius. Synthesis is performed once for each experiment and the library with 0.80 V and zero aging is used. To obtain optimized netlists, a timing constraint of zero clock period is used during synthesis. Synopsys PrimeTime is used to generate the SDF files corresponding to all libraries. ATPG patterns along with randomly generated patterns are used while constructing datasets. Synopsys Tetramax is used for generating ATPG patterns. The gate-level simulations for collecting the input-output pairs are done using Synopsys Verilog Compile Simulator (VCS) on the SDF annotated netlist. All simulations are run at the nominal clock periods of clean circuits as reported by Synopsys PrimeTime.
Several benchmark test cases from Trust-Hub~\cite{trusthub} (\url{https://trust-hub.org/#/benchmarks/chip-level-trojan}) are considered involving a variety of Trojans embedded into crypto-accelerators (RSA, AES) and a PIC microcontroller.
 In all experiments, Trojans are dormant wherever there is a trigger while collecting the data. In every experiment, Trojans are added at the netlist level. This is because RTL Trojans are easier to catch since the Trojan netlist alters dramatically from the clean version. While generating annotated gate-level netlists, the duty cycle details for Trojan-free and Trojaned netlists are computed separately to model the physical effects based on the transistor-level utilizations in the separate circuits. In all experiments, 4096 randomly generated vectors are utilized along with ATPG vectors as the input set. Synopsys Tetramax reported 100\% coverage for both the circuits. Gate-level simulations are used to construct datasets of input-output pairs at each operating voltage and clock period. The number of input-output pairs depends on the number of ATPG vectors of the circuit. Half of the collected data (i.e., \{4096 random vectors + number of ATPG vectors\}/2) is used to train the circuit. To ensure that the classifier does not learn input-output mappings instead of bit error patterns, the training and testing datasets are non-overlapping. The training is performed offline and it requires about 5 minutes on a 2019 laptop with Intel core i7 (10th generation) processor. For testing and inference phases, the computations are simple and  take a few seconds. The hardware Trojan detection performance across the several considered test cases are summarized in Table~\ref{metrics_classifier_finfet}.

\begin{figure*}[!htb]
 \centering
 \begin{tabular}{cc}
 \subfloat[Clean IC]{\includegraphics[width=0.5\textwidth]{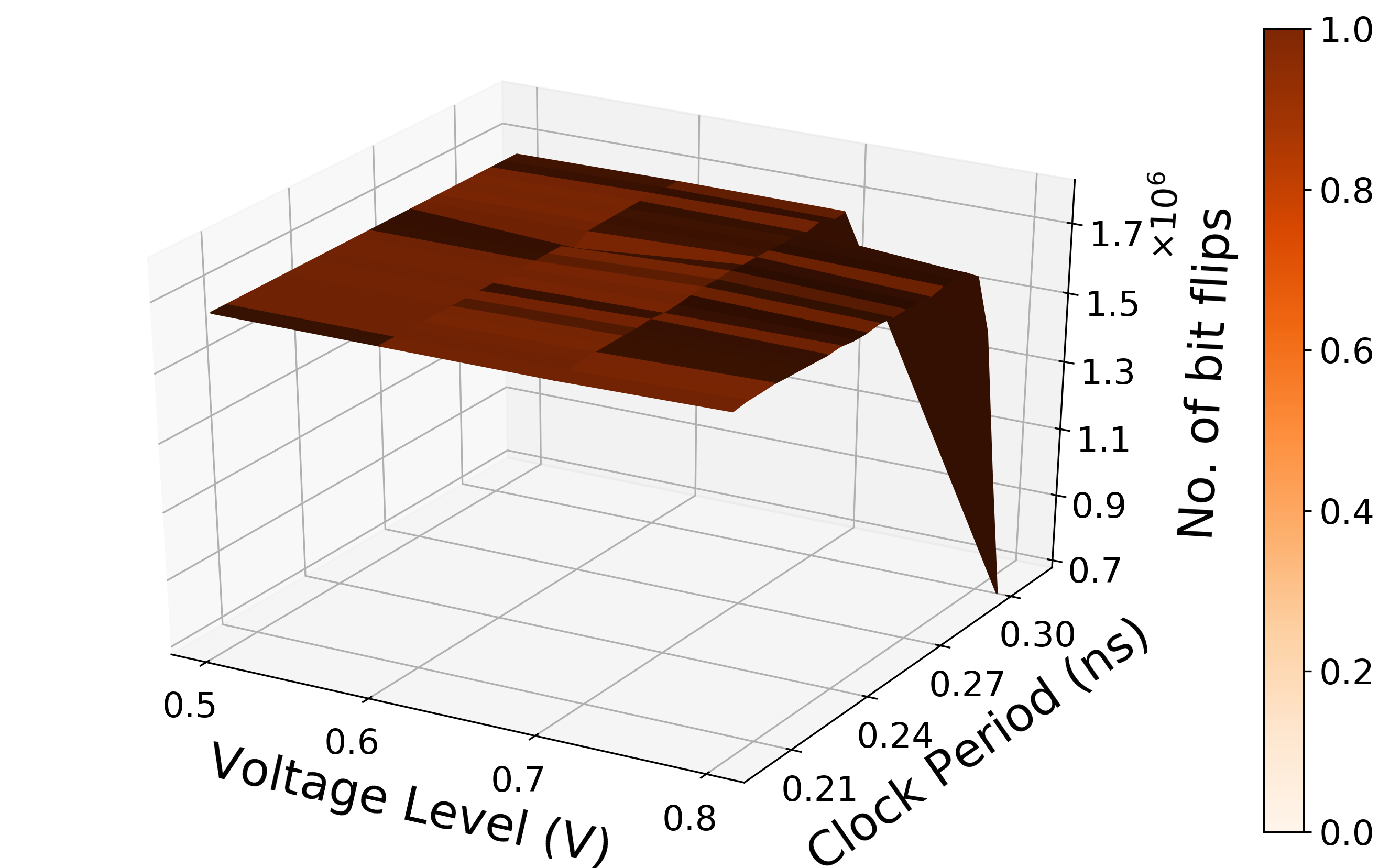}\label{fig:0to1_aes100}}
 \hspace{-0.05in}
   \subfloat[IC with Trojan]{\includegraphics[width=0.5\textwidth]{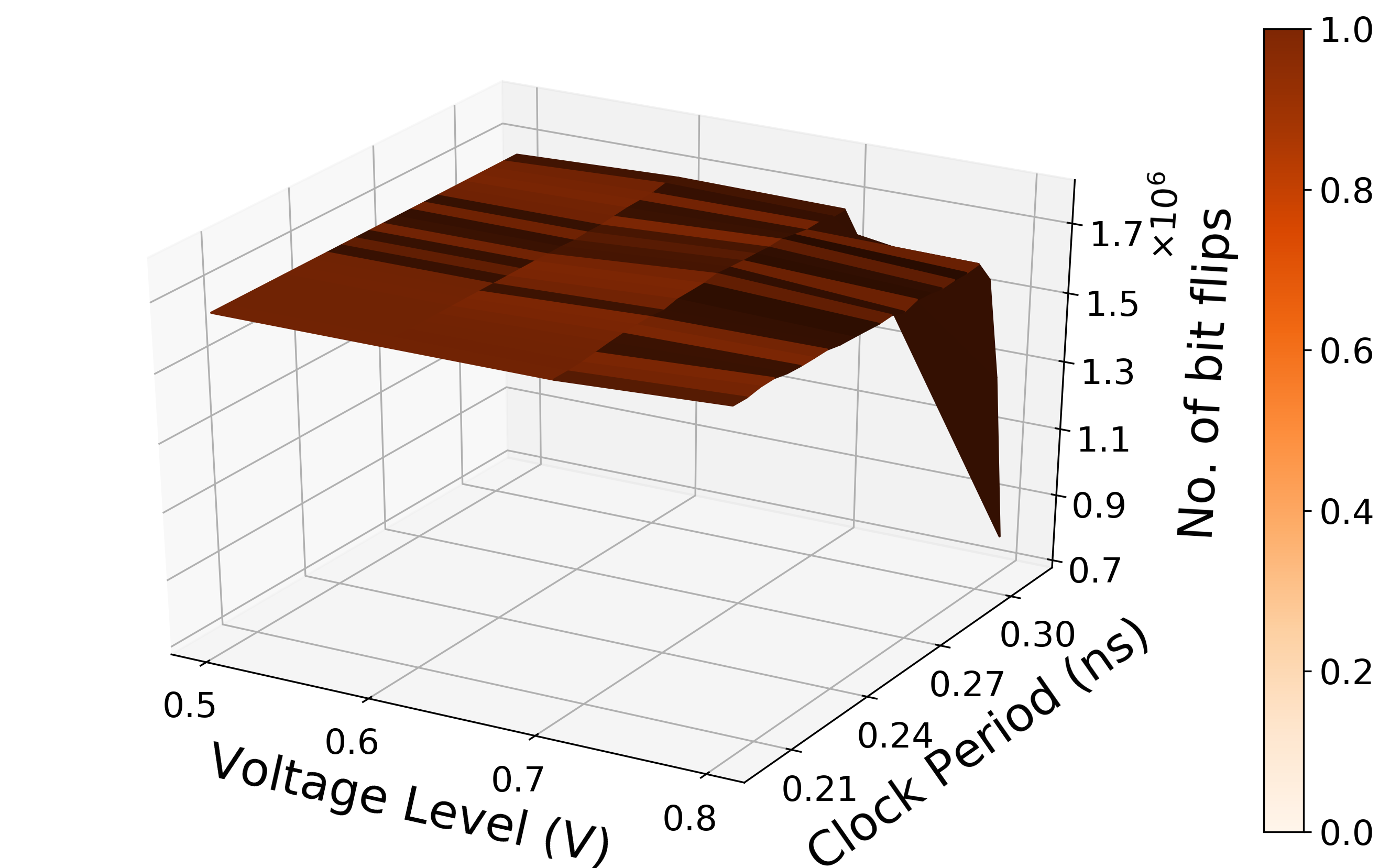}\label{fig:0to1_tr_aes100}}
 \end{tabular}
 \begin{tabular}{cc}
 \subfloat[Clean IC]{\includegraphics[width=0.5\textwidth]{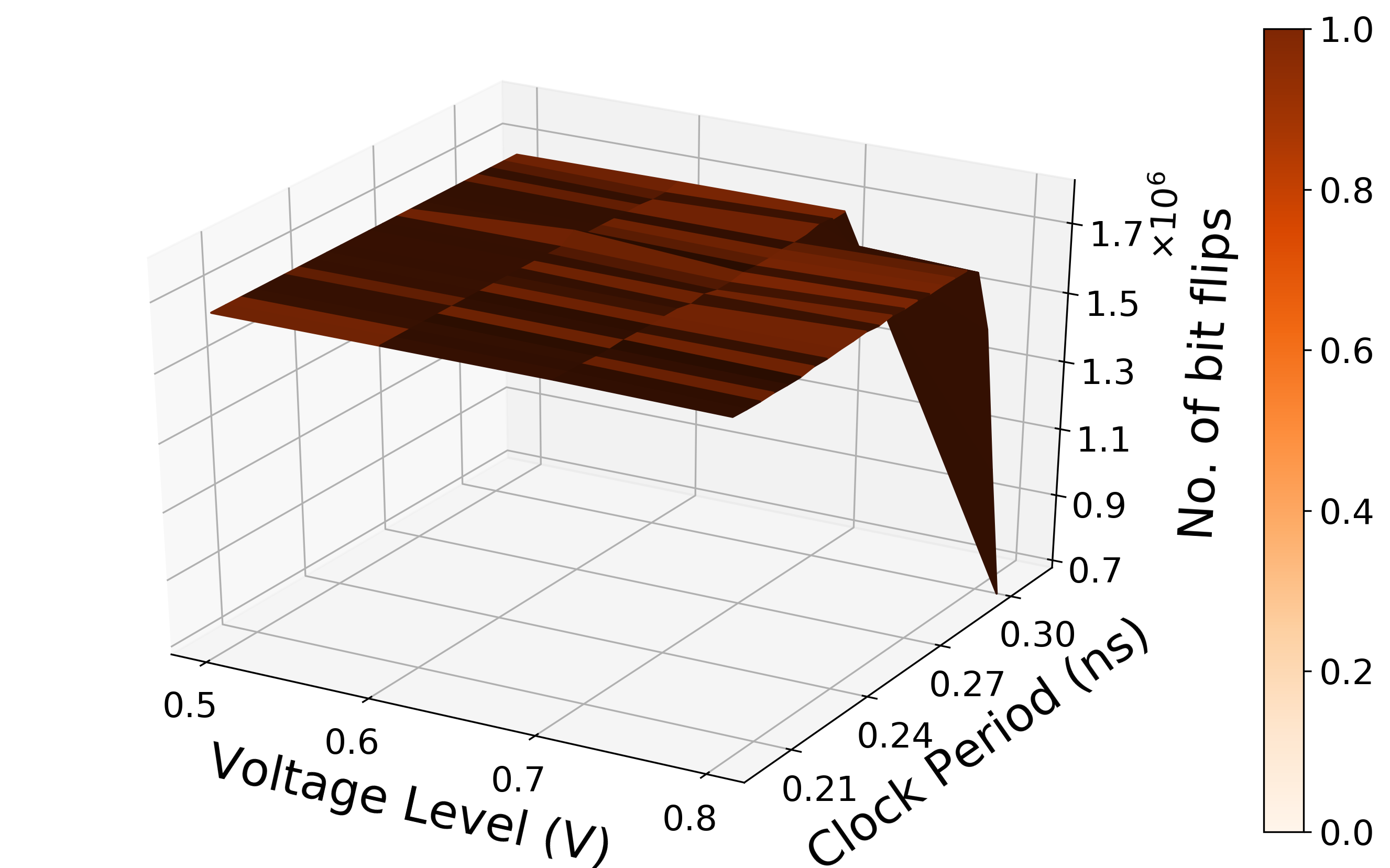}\label{fig:1to0_aes100}}
 \hspace{-0.05in}
   \subfloat[IC with Trojan]{\includegraphics[width=0.5\textwidth]{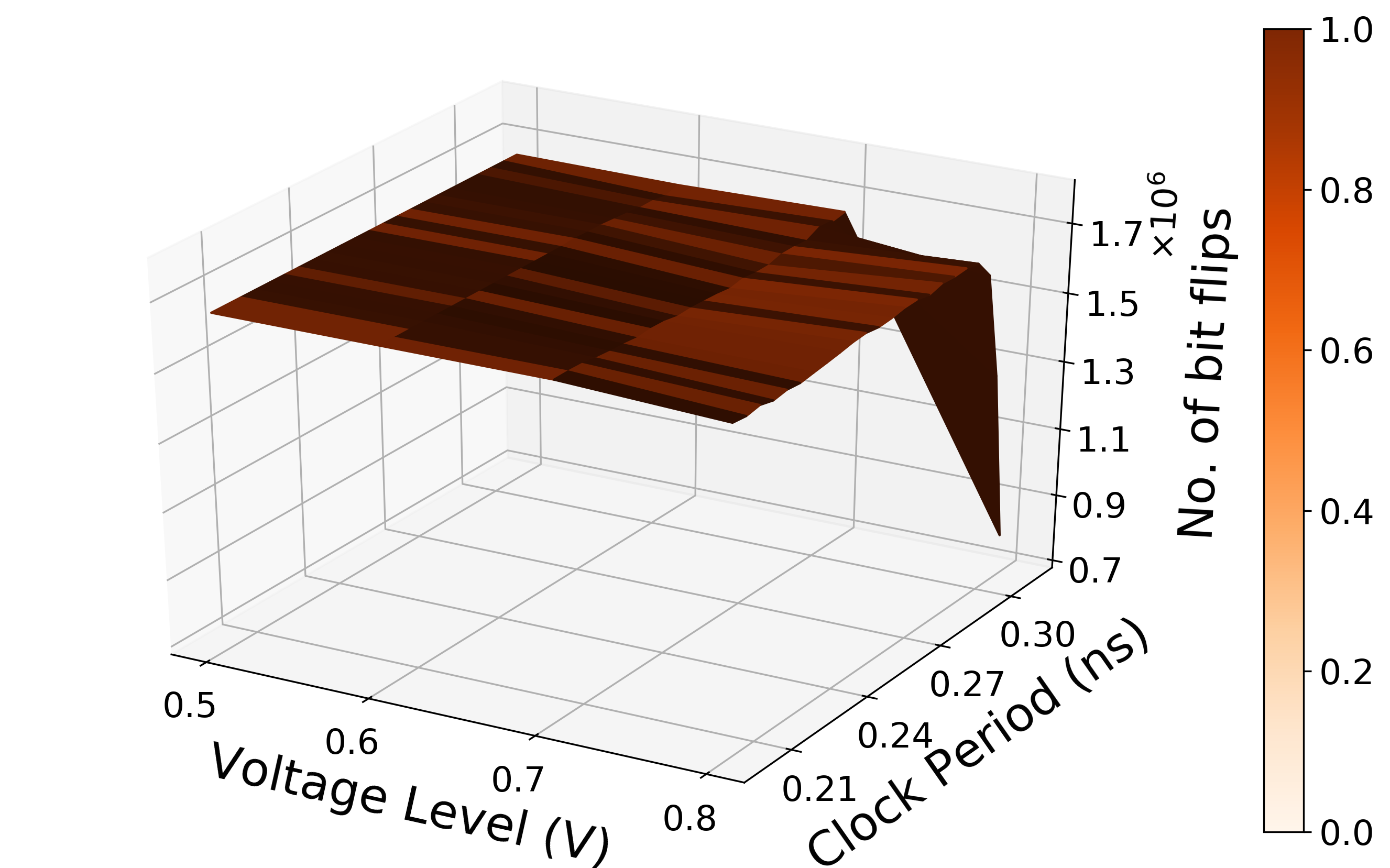}\label{fig:1to0_tr_aes100}}
 \end{tabular}
 \caption{(a, b) 0 $\rightarrow$ 1  (c, d)  1 $\rightarrow$ 0 bit flips over various clock periods and voltage levels for AES-T100. The color map indicates the number of inputs that have bit flips.}
 \label{fig:features1_aes100}
  \end{figure*}

\begin{figure*}[!htb]
 \centering
 \begin{tabular}{cc}
 \subfloat[Clean IC]{\includegraphics[width=0.5\textwidth]{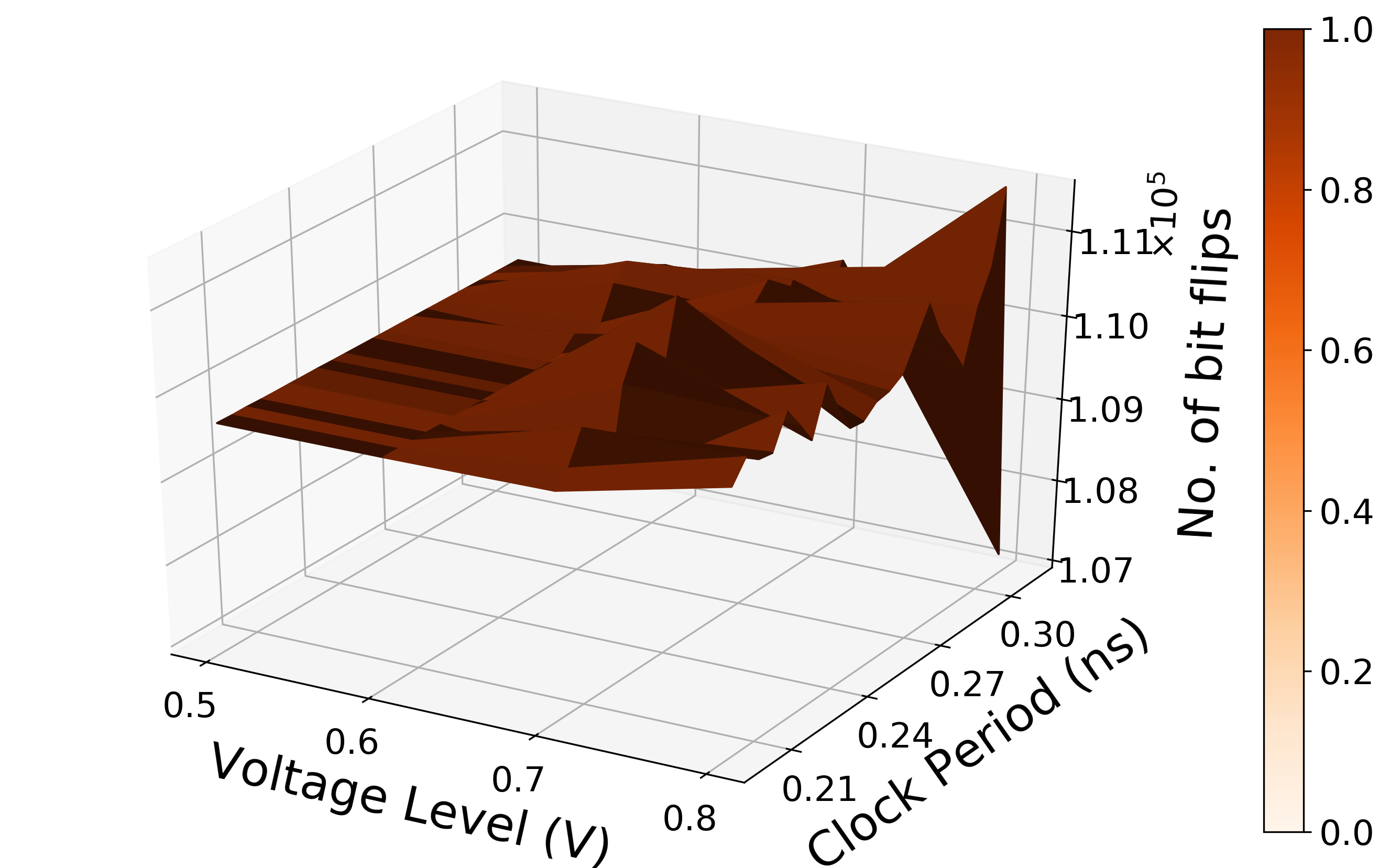}\label{fig:0to1_loc_aes100}}
 \hspace{-0.05in}
   \subfloat[IC with Trojan]{\includegraphics[width=0.5\textwidth]{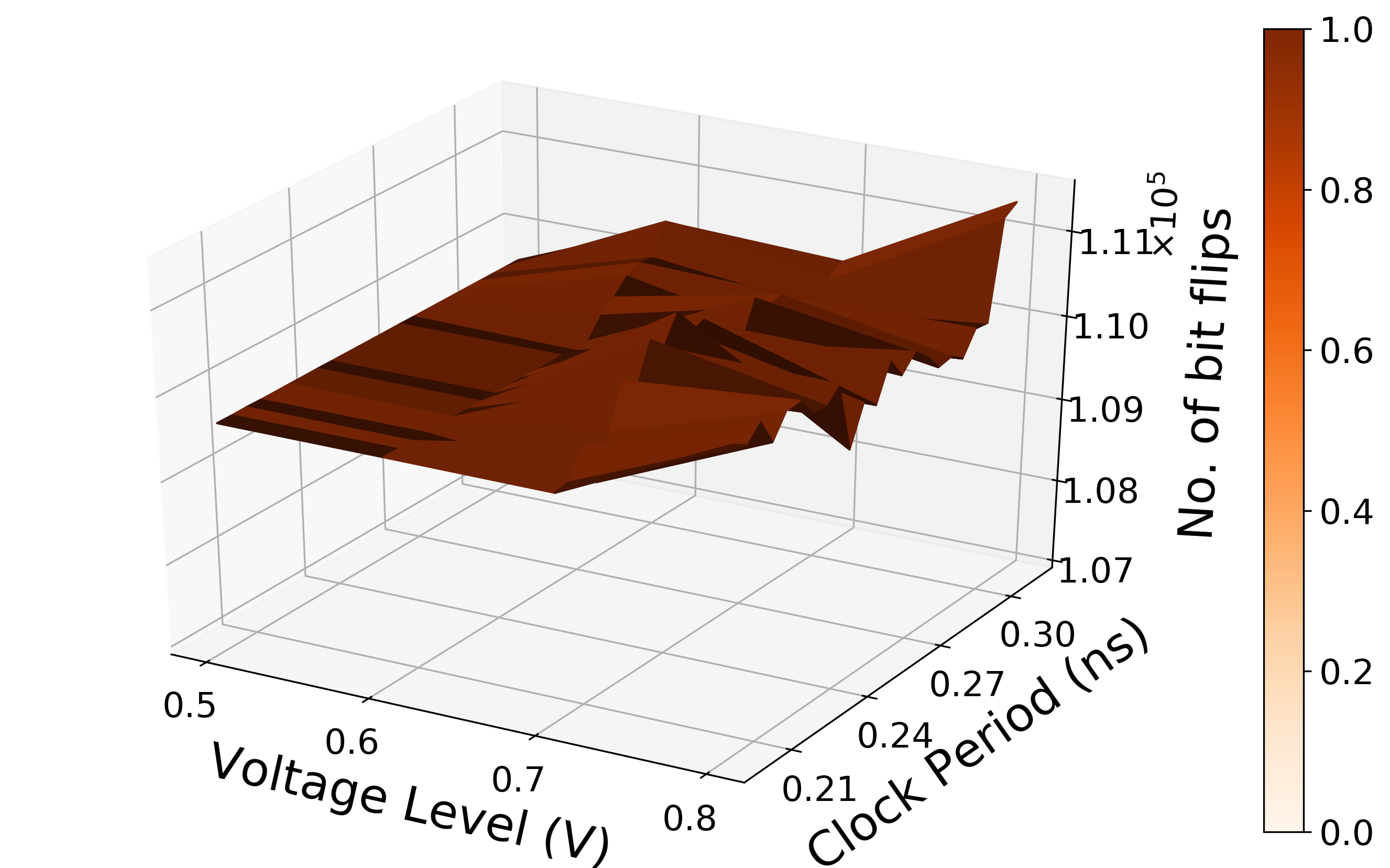}\label{fig:0to1_loc_tr_aes100}}
 \end{tabular}
 \begin{tabular}{cc}
 \subfloat[Clean IC]{\includegraphics[width=0.5\textwidth]{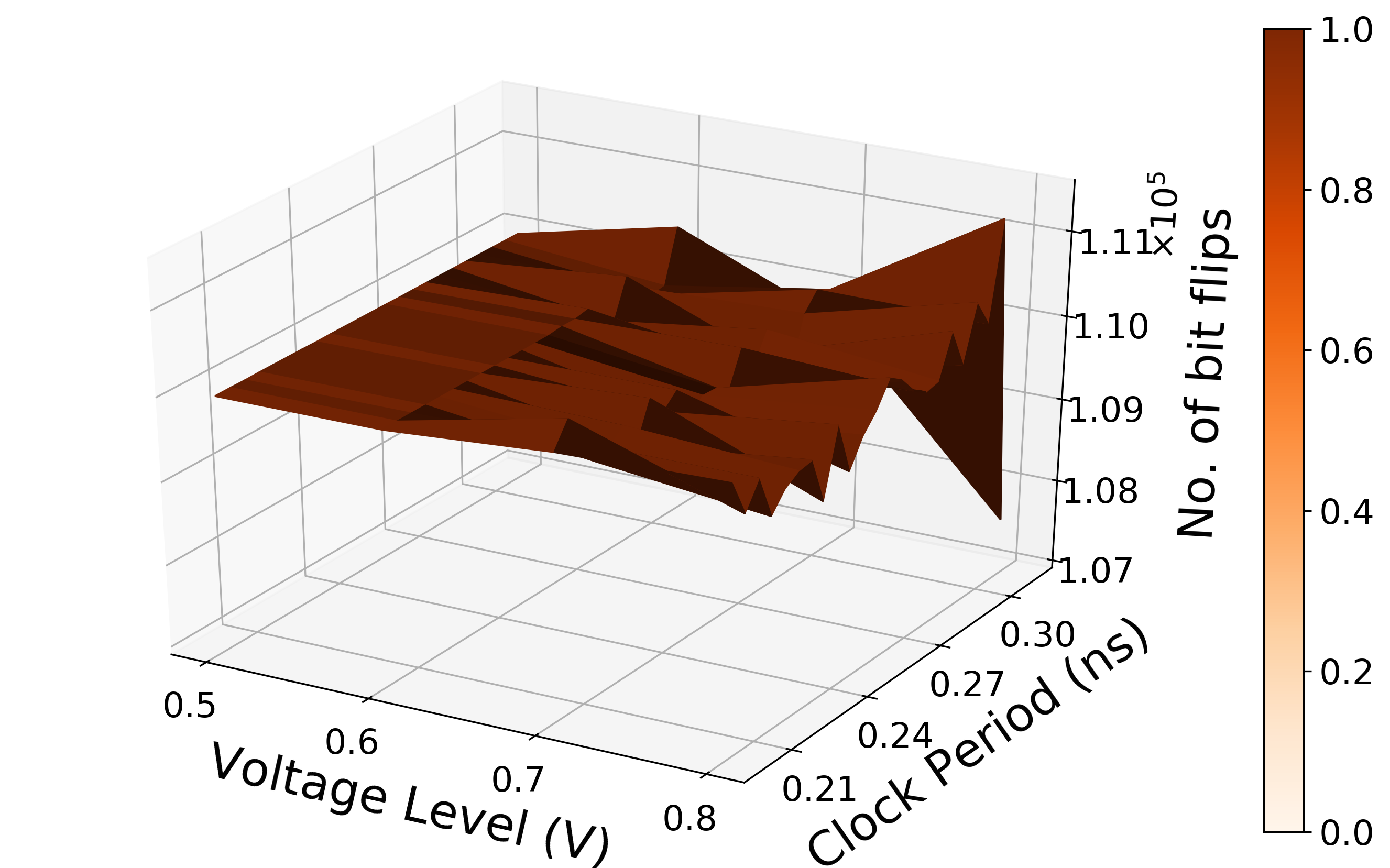}\label{fig:1to0_loc_aes100}}
 \hspace{-0.05in}
 \subfloat[IC with Trojan]{\includegraphics[width=0.5\textwidth]{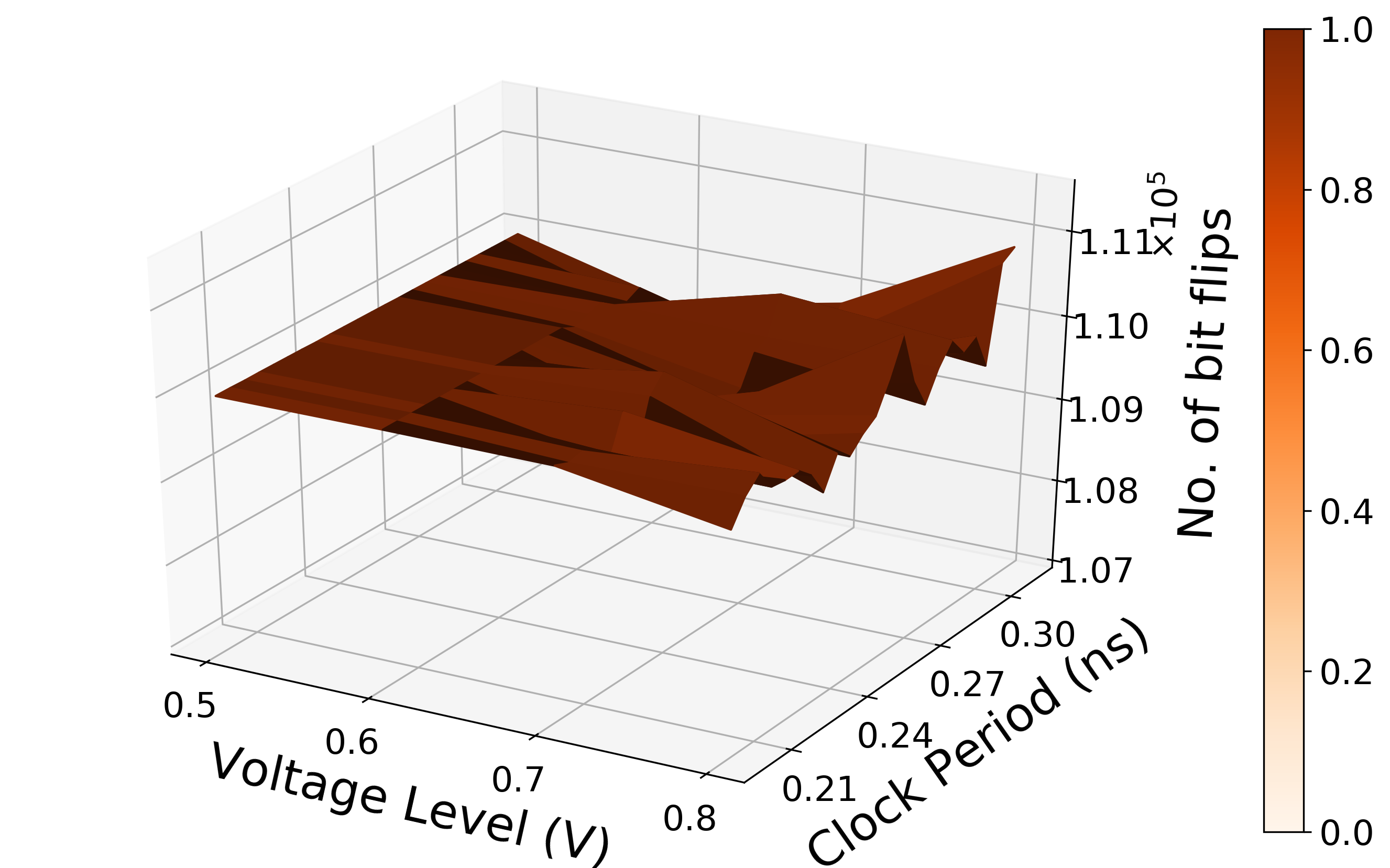}\label{fig:1to0_loc_tr_aes100}}
 \end{tabular}
 \caption{Weighted locations of 0 $\rightarrow$ 1 (a, b) 1 $\rightarrow$ 0 (c, d) bit flips over various clock periods and voltage levels for AES-T100. The color map indicates the location of bit flips.}
 \label{fig:features2_aes100}
  \end{figure*}

 To visually illustrate the types of features computed in Section~\ref{sec:features} for the purpose of hardware Trojan detection, an example for the AES-T100 test case is shown in Figures~\ref{fig:features1_aes100} and \ref{fig:features2_aes100}. Figures \ref{fig:0to1_aes100} and \ref{fig:0to1_tr_aes100} show the $0 \rightarrow 1$ bit flips across various voltage levels and clock periods. The color map shows the number of test inputs in which the bit flips occur. Darker color corresponds to more test inputs for which bit flips occur. Analogously, Figures \ref{fig:1to0_aes100} and \ref{fig:1to0_tr_aes100} visualize $1 \rightarrow 0$ bit flips across various voltage levels and clock periods. Although the change is not clearly visible from the figures, the classifier is able to learn the underlying patterns of bit flips. Figures \ref{fig:0to1_loc_aes100}, \ref{fig:0to1_loc_tr_aes100} show the weighted locations of $0 \rightarrow 1$ bit flips over different voltages and clock periods. Here, the color map indicates the location of the bit flips. Darker color corresponds to the most significant bits. Similarly, Figures \ref{fig:1to0_loc_aes100}, \ref{fig:1to0_loc_tr_aes100} show the weighted locations for $1 \rightarrow 0$ bit flips.

 \begin{table}[!h]
 \centering
 \caption{Performance of the one-class SVM classifier for Trojan detection for ICs with 14nm FinFET
   technology. Each bin has 5 inputs.}
 \label{metrics_classifier_finfet}
 \centering
 \begin{tabular}{|c|c|c|c|c|c|}
 \hline
 & Bins & Accuracy & Precision & Recall & F1-Score \\
 \hline
 \multirow{2}{*}{BasicRSA-T100} & 1 & 1.0000 & 0.9972 & 1.0000 & 0.9986 \\
 \cline{2-6}
  & 3 & 1.0000 & 0.9988 & 1.0000 & 0.9994 \\
  \hline
 \multirow{2}{*}{BasicRSA-T200} & 1 & 1.0000 & 0.9934 & 1.0000 & 0.9967 \\
 \cline{2-6}
  & 3 & 1.0000 & 0.9972 & 1.0000 & 0.9986 \\
  \hline
 \multirow{2}{*}{BasicRSA-T300} & 1 & 1.0000 & 0.9901 & 1.0000 & 0.9951 \\
 \cline{2-6}
  & 3 & 1.0000 & 0.9981 & 1.0000 & 0.9991 \\
  \hline
 \multirow{2}{*}{BasicRSA-T400} & 1 & 1.0000 & 0.9938 & 1.0000 & 0.9969 \\
 \cline{2-6}
  & 3 & 1.0000 & 0.9957 & 1.0000 & 0.9978 \\
 \hline
 \multirow{2}{*}{AES-T100} & 1 & 0.6796 & 0.6885 & 0.6796 & 0.6841 \\
 \cline{2-6}
  & 128 & 0.9988 & 1.0000 & 0.9987 & 0.9993 \\
 \hline
 \multirow{2}{*}{AES-T200} & 1 & 0.6546 & 0.6281 & 0.6547 & 0.6411 \\
 \cline{2-6}
  & 128 & 0.9921 & 0.9558 & 0.9975 & 0.9762 \\
 \hline
 \multirow{2}{*}{AES-T300} & 1 & 0.6898 & 0.6209 & 0.6809 & 0.6495 \\
 \cline{2-6}
  & 128 & 1.0000 & 0.9289 & 1.0000 & 0.9632 \\
 \hline
 \multirow{2}{*}{AES-T400} & 1 & 0.6616 & 0.6194 & 0.6616 & 0.6398 \\
 \cline{2-6}
  & 128 & 0.9956 & 0.9478 & 0.9966 & 0.9716 \\
 \hline
 \multirow{2}{*}{AES-T500} & 1 & 0.6736 & 0.6358 & 0.6736 & 0.6542 \\
 \cline{2-6}
  & 128 & 0.9975 & 0.9577 & 1.0000 & 0.9784 \\
 \hline
 \multirow{2}{*}{AES-T600} & 1 & 0.6524 & 0.6185 & 0.6524 & 0.6349 \\
 \cline{2-6}
  & 128 & 0.9759 & 0.9692 & 0.9863 & 0.9776 \\
 \hline
 \multirow{2}{*}{AES-T700} & 1 & 0.6583 & 0.6381 & 0.6584 & 0.6481 \\
 \cline{2-6}
  & 128 & 0.9568 & 0.9815 & 0.9667 & 0.9741 \\
 \hline
 \multirow{2}{*}{AES-T800} & 1 & 0.6528 & 0.6101 & 0.6528 & 0.6307 \\
 \cline{2-6}
  & 128 & 0.9961 & 0.9217 & 0.9985 & 0.9586 \\
 \hline
 \multirow{2}{*}{AES-T900} & 1 & 0.6588 & 0.6244 & 0.6588 & 0.6401 \\
 \cline{2-6}
  & 128 & 1.0000 & 0.9696 & 1.0000 & 0.9845 \\
 \hline
 \multirow{2}{*}{AES-T1000} & 1 & 0.6771 & 0.6598 & 0.6768 & 0.6682 \\
 \cline{2-6}
  & 128 & 0.9968 & 1.0000 & 0.9969 & 0.9984 \\
   \hline
   \multirow{2}{*}{PIC16F84-T100} & 1 & 0.9907 & 0.9858 & 0.9907 & 0.9882 \\
   \cline{2-6}
   & 8 & 1.0000 & 0.9894 & 1.0000 & 0.9947 \\
   \hline
 \end{tabular}
 \end{table}

\section{Conclusion}
\label{sec:conclusion}
 The effects of long-term aging\index{aging!long-term} and short-term aging\index{aging!short-term} in circuits (due to transistor aging) were discussed and shown to cause noticeable inference accuracy degradation on DNNs implemented in hardware. It is noted that device aging, which is an unavoidable issue with the dense integration of nanoscale transistors, can result in a 50\% decline in classification accuracy for a representative DNN after 20\% of aging once the DNN is deployed. Therefore, it is essential that hardware-implemented DNNs take into consideration the effects of aging.  Mitigations were presented to counter accuracy degradations in DNNs implemented in hardware caused by short- and long-term aging effects. To improve the overall inference accuracy of the DNN, these mitigations employ a mix of clock frequency scaling, feature noise application\index{feature noise}, gradient noise\index{gradient noise} and combination of feature and gradient noise developed based on modeling of the aging-related degradations. It was shown that feature noise addition during training results in higher DNN accuracy and ability to reliably operate at higher clock frequencies over the life of the hardware-implemented DNNs. Also, in the context of hardware Trojan detection, it was shown that short-term aging can be applied as an excitation mechanism to collect input-output bit error patterns that enable detection of modifications of integrated circuits indicating presence of hardware Trojans.

\bibliographystyle{plain}
\bibliography{references}

@misc{neelakantan2015adding,
  title={Adding gradient noise improves learning for very deep networks},
  author={Neelakantan, Arvind and Vilnis, Luke and Le, Quoc V and Sutskever, Ilya and Kaiser, Lukasz and Kurach, Karol and Martens, James},
  year={2015},
  eprint={1511.06807}
}

@ARTICLE{sze2017efficient,
  author={Sze, Vivienne and Chen, Yu-Hsin and Yang, Tien-Ju and Emer, Joel S.},
  journal={Proceedings of the IEEE},
  title={Efficient Processing of Deep Neural Networks: A Tutorial and Survey},
  year={2017},
  volume={105},
  number={12},
  pages={2295-2329}
}

@ARTICLE{chen2017eyeriss,
  author={Chen, Yu-Hsin and Krishna, Tushar and Emer, Joel S. and Sze, Vivienne},
  journal={IEEE Journal of Solid-State Circuits},
  title={Eyeriss: An Energy-Efficient Reconfigurable Accelerator for Deep Convolutional Neural Networks},
  year={2017},
  volume={52},
  number={1},
  pages={127-138}
}

@INPROCEEDINGS{moons2016tops,
  author={Moons, Bert and Verhelst, Marian},
  booktitle={Proceedings of IEEE Symposium on VLSI Circuits},
  title={A 0.3–2.6 {TOPS/W} precision-scalable processor for real-time large-scale {ConvNets}},
  year={2016},
  volume={},
  number={},
  pages={1-2},
  address={Honolulu, HI},
  publisher={IEEE}
}

@inproceedings{han2016eie,
author = {Han, Song and Liu, Xingyu and Mao, Huizi and Pu, Jing and Pedram, Ardavan and Horowitz, Mark A. and Dally, William J.},
title = {{EIE}: Efficient Inference Engine on Compressed Deep Neural Network},
year = {2016},
isbn = {9781467389471},
booktitle = {Proceedings of IEEE International Symposium on Computer Architecture},
pages = {243–254},
numpages = {12},
address = {Seoul, Republic of Korea},
publisher={IEEE}
}

@INPROCEEDINGS{whatmough2017soc,
  author={Whatmough, Paul N. and Lee, Sae Kyu and Lee, Hyunkwang and Rama, Saketh and Brooks, David and Wei, Gu-Yeon},
  booktitle={Proceedings of IEEE International Solid-State Circuits Conference},
  title={14.3 A 28nm {SoC} with a {1.2GHz 568nJ}/prediction sparse deep-neural-network engine with $>$0.1 timing error rate tolerance for {IoT} applications},
  year={2017},
  volume={},
  number={},
  pages={242-243},
  address={San Francisco, CA},
  publisher={IEEE}
}

@misc{han2015deep,
  title={Deep compression: Compressing deep neural networks with pruning, trained quantization and huffman coding},
  author={Han, Song and Mao, Huizi and Dally, William J},
  year={2015},
  eprint={1510.00149}
}

@misc{gysel2016hardware,
  title={Hardware-oriented approximation of convolutional neural networks},
  author={Gysel, Philipp and Motamedi, Mohammad and Ghiasi, Soheil},
  year={2016},
  eprint={1604.03168}
}

@ARTICLE{moons2017energy,
  author={Moons, Bert and Verhelst, Marian},
  journal={IEEE Journal of Solid-State Circuits},
  title={An Energy-Efficient Precision-Scalable ConvNet Processor in 40-nm {CMOS}},
  year={2017},
  volume={52},
  number={4},
  pages={903-914}
}

@ARTICLE{wang2007compact,
  author={Wang, Wenping and Reddy, Vijay and Krishnan, Anand T. and Vattikonda, Rakesh and Krishnan, Srikanth and Cao, Yu},
  journal={IEEE Transactions on Device and Materials Reliability},
  title={Compact Modeling and Simulation of Circuit Reliability for 65-nm {CMOS} Technology},
  year={2007},
  volume={7},
  number={4},
  pages={509-517}
}

@INPROCEEDINGS{liu2019analysis,
  author={Liu, Wenye and Chang, Chip-Hong},
  booktitle={Proceedings of IEEE International Symposium on Circuits and Systems},
  title={Analysis of Circuit Aging on Accuracy Degradation of Deep Neural Network Accelerator},
  year={2019},
  volume={},
  number={},
  pages={1-5},
  address={Sapporo, Japan},
  publisher={IEEE}
}

@INPROCEEDINGS{salamin2021reliability,
  author={Salamin, Sami and Zervakis, Georgios and Spantidi, Ourania and Anagnostopoulos, Iraklis and Henkel, Jörg and Amrouch, Hussam},
  booktitle={Proceedings of IEEE Design, Automation \& Test in Europe Conference \& Exhibition},
  title={Reliability-Aware Quantization for Anti-Aging {NPUs}},
  year={2021},
  volume={},
  number={},
  pages={1460-1465},
  address={Grenoble, France},
  publisher={IEEE}
}

@inproceedings{sootkaneung2017thermal,
  title={Thermal effect on performance, power, and {BTI} aging in {FinFET}-based designs},
  author={Sootkaneung, Warin and Howimanporn, Suppachai and Chookaew, Sasithorn},
  booktitle={Proceedings of IEEE Euromicro Conference on Digital System Design},
  pages={345--351},
  year={2017},
  address={Vienna, Austria},
  publisher={IEEE}
}

@ARTICLE{mahapatra2013comparative,
  author={Mahapatra, S. and Goel, N. and Desai, S. and Gupta, S. and Jose, B. and Mukhopadhyay, S. and Joshi, K. and Jain, A. and Islam, A. E. and Alam, M. A.},
  journal={IEEE Transactions on Electron Devices},
  title={A Comparative Study of Different Physics-Based {NBTI} Models},
  year={2013},
  volume={60},
  number={3},
  pages={901-916}
}

@INPROCEEDINGS{jiao2017assessment,
  author={Jiao, Xun and Luo, Mulong and Lin, Jeng-Hau and Gupta, Rajesh K.},
  booktitle={Proceedings of IEEE/ACM International Conference on Computer-Aided Design},
  title={An assessment of vulnerability of hardware neural networks to dynamic voltage and temperature variations},
  year={2017},
  volume={},
  number={},
  pages={945-950},
  address={Irvine, CA},
  publisher={IEEE}
}

@INPROCEEDINGS{arechiga2018effect,
  author={Arechiga, Austin P. and Michaels, Alan J.},
  booktitle={Proceesings of IEEE Annual Computing and Communication Workshop and Conference},
  title={The effect of weight errors on neural networks},
  year={2018},
  address={Las Vegas, NV},
  publisher={IEEE},
  volume={},
  number={},
  pages={190-196}
}

@INPROCEEDINGS{arechiga2018robustness,
  author={Arechiga, Austin P. and Michaels, Alan J.},
  booktitle={Proceedings of IEEE High Performance extreme Computing Conference},
  title={The Robustness of Modern Deep Learning Architectures against Single Event Upset Errors},
  year={2018},
  address={Waltham, MA},
  publisher={IEEE},
  volume={},
  number={},
  pages={1-6}
}

@inproceedings{hong2019terminal,
  title={Terminal Brain Damage: Exposing the Graceless Degradation in Deep Neural Networks Under Hardware Fault Attacks},
  author={Hong, Sanghyun and Frigo, Pietro and Kaya, Yigitcan and Giuffrida, Cristiano and Dumitras, Tudor},
  booktitle={Proceedings of USENIX Security Symposium},
  pages={497--514},
  year={2019},
  address={Santa Clara, CA},
  publisher={USENIX Association}
}

@INPROCEEDINGS{neggaz2018reliability,
  author={Neggaz, Mohamed A. and Alouani, Ihsen and Lorenzo, Pablo R. and Niar, Smail},
  booktitle={Proceedings of IEEE International Conference on Computer Design},
  title={A Reliability Study on {CNNs} for Critical Embedded Systems},
  year={2018},
  volume={},
  number={},
  pages={476-479},
  address={Orlando, FL},
  publisher={IEEE}
}

@INPROCEEDINGS{vansanten2016agingaware,
  author={van Santen, Victor M. and Amrouch, Hussam and Parihar, Narendra and Mahapatra, Souvik and Henkel, Jörg},
  booktitle={Proceedings of IEEE Design, Automation \& Test in Europe Conference \& Exhibition},
  title={Aging-aware voltage scaling},
  year={2016},
  volume={},
  number={},
  pages={576-581},
  address={Dresden, Germany},
  publisher={IEEE}
}

@INPROCEEDINGS{amrouch2016reliability,
  author={Amrouch, Hussam and Khaleghi, Behnam and Gerstlauer, Andreas and Henkel, Jörg},
  booktitle={Proceedings of ACM/EDAC/IEEE Design Automation Conference},
  title={Reliability-aware design to suppress aging},
  year={2016},
  month={June},
  address={Austin,TX},
  publisher={IEEE},
  volume={},
  number={},
  pages={1-6}
}

@book{chauhan2015finfet,
  title={{FinFET} modeling for {IC} simulation and design: using the {BSIM-CMG} standard},
  author={Chauhan, Yogesh Singh and Lu, Darsen Duane and Venugopalan, Sriramkumar and Khandelwal, Sourabh and Duarte, Juan Pablo and Paydavosi, Navid and Niknejad, Ai and Hu, Chenming},
  year={2015},
  publisher={Academic Press},
  address={Cambridge, MA}
}

@article{mishra2018simulation,
  title={A simulation study of {NBTI} impact on 14-nm node {FinFET} technology for logic applications: Device degradation to circuit-level interaction},
  author={Mishra, Subrat and Amrouch, Hussam and Joe, Jerin and Dabhi, Chetan K and Thakor, Karansingh and Chauhan, Yogesh S and Henkel, Joerg and Mahapatra, Souvik},
  journal={IEEE Transactions on Electron Devices},
  volume={66},
  number={1},
  pages={271--278},
  year={2018},
  publisher={IEEE}
}

@article{parihar2017bti,
  title={{BTI} analysis tool—Modeling of {NBTI DC, AC} stress and recovery time kinetics, nitrogen impact, and {EOL} estimation},
  author={Parihar, Narendra and Goel, Nilesh and Mukhopadhyay, Subhadeep and Mahapatra, Souvik},
  journal={IEEE Transactions on Electron Devices},
  volume={65},
  number={2},
  pages={392--403},
  year={2017},
  publisher={IEEE}
}

@misc{finfet_pdk,
author  = {Si2},
title ={15nm Open-cell Library and 45nm {FREEPDK}},
note     = {\url{https://si2.org/open-cell-library}},
year    = {2021},
organization={Si2}
}

@INPROCEEDINGS{reagen2018ares,
  author={Reagen, Brandon and Gupta, Udit and Pentecost, Lillian and Whatmough, Paul and Lee, Sae Kyu and Mulholland, Niamh and Brooks, David and Wei, Gu-Yeon},
  booktitle={Proceedings of ACM/ESDA/IEEE Design Automation Conference},
  title={Ares: A framework for quantifying the resilience of deep neural networks},
  year={2018},
  volume={},
  number={},
  pages={1-6},
  address={San Francisco, CA},
  publisher={IEEE}
}

@inproceedings{li2017understanding,
  title={Understanding error propagation in deep learning neural network {(DNN)} accelerators and applications},
  author={Li, Guanpeng and Hari, Siva Kumar Sastry and Sullivan, Michael and Tsai, Timothy and Pattabiraman, Karthik and Emer, Joel and Keckler, Stephen W},
  booktitle={Proceedings of IEEE International Conference for High Performance Computing, Networking, Storage and Analysis},
  pages={1--12},
  year={2017},
  address={Denver, CO},
  publisher={IEEE}
}

@INPROCEEDINGS{vipin2019zynet,
  author={Vipin, Kizheppatt},
  booktitle={Proceedings of International Conference on Field-Programmable Technology},
  title={{ZyNet}: Automating Deep Neural Network Implementation on Low-Cost Reconfigurable Edge Computing Platforms},
  year={2019},
  volume={},
  number={},
  pages={323-326},
  address={Tianjin, China},
  publisher={IEEE}
}

@inproceedings{salami2018resilience,
  title={On the resilience of {RTL NN} accelerators: Fault characterization and mitigation},
  author={Salami, Behzad and Unsal, Osman S and Kestelman, Adrian Cristal},
  booktitle={Proceedings of IEEE International Symposium on Computer Architecture and High Performance Computing},
  pages={322--329},
  year={2018},
  address={Lyon, France},
  publisher={IEEE}
}

@INPROCEEDINGS{sabbagh2019evaluating,
  author={Sabbagh, Majid and Gongye, Cheng and Fei, Yunsi and Wang, Yanzhi},
  booktitle={Proceedings of IEEE International Conference on Embedded Software and Systems},
  title={Evaluating Fault Resiliency of Compressed Deep Neural Networks},
  year={2019},
  volume={},
  number={},
  pages={1-7},
  address={Las Vegas, NV},
  publisher={IEEE}
}

@book{mahapatra2016fundamentals,
  title={Fundamentals of bias temperature instability in mos transistors},
  author={Mahapatra, Souvik},
  year={2016},
  publisher={Springer},
  address={New Delhi, India}
}

@INPROCEEDINGS{schorn2018accurate,
  author={Schorn, Christoph and Guntoro, Andre and Ascheid, Gerd},
  booktitle={Proceedings of IEEE Design, Automation \& Test in Europe Conference \& Exhibition},
  title={Accurate neuron resilience prediction for a flexible reliability management in neural network accelerators},
  year={2018},
  volume={},
  number={},
  pages={979-984},
  address={Dresden, Germany},
  publisher={IEEE}
}

@INPROCEEDINGS{choi2019sensitivity,
  author={Choi, Wonseok and Shin, Dongyeob and Park, Jongsun and Ghosh, Swaroop},
  booktitle={Proceedings of ACM/IEEE Design Automation Conference},
  title={Sensitivity based Error Resilient Techniques for Energy Efficient Deep Neural Network Accelerators},
  year={2019},
  address={Las Vegas, NV},
  publisher={IEEE},
  volume={},
  number={},
  pages={1-6}
}

@INPROCEEDINGS{zhao2017aep,
  author={Zhao, Lei and Zhang, Youtao and Yang, Jun},
  booktitle={Proceedings of IEEE/ACM International Conference on Computer-Aided Design},
  title={{AEP}: An error-bearing neural network accelerator for energy efficiency and model protection},
  year={2017},
  volume={},
  number={},
  pages={1047-1053},
  address={Irvine, CA},
  publisher={IEEE}
}

@INPROCEEDINGS{jia2018calibrating,
  author={Jia, Kaige and Liu, Zheyu and Wei, Qi and Qiao, Fei and Liu, Xinjun and Yang, Yi and Fan, Hua and Yang, Huazhong},
  booktitle={Proceedings of ACM/ESDA/IEEE Design Automation Conference},
  title={Calibrating Process Variation at System Level with In-Situ Low-Precision Transfer Learning for Analog Neural Network Processors},
  year={2018},
  volume={},
  number={},
  pages={1-6},
  address={San Francisco, CA},
  publisher={IEEE}
}

@inproceedings{hacene2019training,
  title={Training modern deep neural networks for memory-fault robustness},
  author={Hacene, Ghouthi Boukli and Leduc-Primeau, Fran{\c{c}}ois and Soussia, Amal Ben and Gripon, Vincent and Gagnon, Fran{\c{c}}ois},
  booktitle={Proceedings of IEEE International Symposium on Circuits and Systems},
  pages={1--5},
  year={2019},
  address={Sapporo, Japan},
  publisher={IEEE}
}

@inproceedings{kim2018matic,
  title={{MATIC}: Learning around errors for efficient low-voltage neural network accelerators},
  author={Kim, Sung and Howe, Patrick and Moreau, Thierry and Alaghi, Armin and Ceze, Luis and Sathe, Visvesh},
  booktitle={Proceedings of IEEE Design, Automation \& Test in Europe Conference \& Exhibition},
  pages={1--6},
  year={2018},
  month={},
  address={Dresden, Germany},
  publisher={IEEE}
}

@INPROCEEDINGS{temam2012defect,
  author={Temam, Olivier},
  booktitle={Proceedings of IEEE Annual International Symposium on Computer Architecture},
  title={A defect-tolerant accelerator for emerging high-performance applications},
  year={2012},
  volume={},
  number={},
  pages={356-367},
  address={Portland, OR},
  publisher={IEEE}
}

@INPROCEEDINGS{zhang2018analyzing,
  author={Zhang, Jeff Jun and Gu, Tianyu and Basu, Kanad and Garg, Siddharth},
  booktitle={Proceedings of IEEE VLSI Test Symposium},
  title={Analyzing and mitigating the impact of permanent faults on a systolic array based neural network accelerator},
  year={2018},
  volume={},
  number={},
  pages={1-6},
  address={San Francisco, CA},
  publisher={IEEE}
}

@INPROCEEDINGS{nguyen2019stdrc,
  author={Nguyen, Duy-Thanh and Ho, Nhut-Minh and Chang, Ik-Joon},
  booktitle={Proceedings of ACM/IEEE Design Automation Conference},
  title={{St-DRC}: Stretchable {DRAM} Refresh Controller with No Parity-overhead Error Correction Scheme for Energy-efficient DNNs},
  year={2019},
  volume={},
  number={},
  pages={1-6},
  address={Las Vegas, NV},
  publisher={IEEE}
}

@INPROCEEDINGS{pandey2019greentpu,
  author={Pandey, Pramesh and Basu, Prabal and Chakraborty, Koushik and Roy, Sanghamitra},
  booktitle={Proceedings of ACM/IEEE Design Automation Conference},
  title={{GreenTPU}: Improving Timing Error Resilience of a Near-Threshold Tensor Processing Unit},
  year={2019},
  volume={},
  number={},
  pages={1-6},
  address={Las Vegas, NV},
  publisher={IEEE}
}

@INPROCEEDINGS{chandramoorthy2019resilient,
  author={Chandramoorthy, Nandhini and Swaminathan, Karthik and Cochet, Martin and Paidimarri, Arun and Eldridge, Schuyler and Joshi, Rajiv V. and Ziegler, Matthew M. and Buyuktosunoglu, Alper and Bose, Pradip},
  booktitle={Proceedings of IEEE International Symposium on High Performance Computer Architecture},
  title={Resilient Low Voltage Accelerators for High Energy Efficiency},
  year={2019},
  address={Washington, DC},
  publisher={IEEE},
  volume={},
  number={},
  pages={147-158}
}

@manual{primetime,
title = {PrimeTime User Guide},
date = {2020},
language = {English},
version = {Version R-2020.09-SP5},
organization = {Synopsys Inc.},
note = {Version R-2020.09-SP5},
year = {2020},
}

@article{surabhi2020hardware,
  title={Hardware trojan detection using controlled circuit aging},
  author={Surabhi, Virinchi Roy and Krishnamurthy, Prashanth and Amrouch, Hussam and Basu, Kanad and Henkel, J{\"o}rg and Karri, Ramesh and Khorrami, Farshad},
  journal={IEEE Access},
  volume={8},
  pages={77415--77434},
  year={2020},
  publisher={IEEE}
}

@article{surabhi2020exposing,
  title={Exposing hardware Trojans in embedded platforms via short-term aging},
  author={Surabhi, Virinchi Roy and Krishnamurthy, Prashanth and Amrouch, Hussam and Henkel, J{\"o}rg and Karri, Ramesh and Khorrami, Farshad},
  journal={IEEE Transactions on Computer-Aided Design of Integrated Circuits and Systems},
  volume={39},
  number={11},
  pages={3519--3530},
  year={2020},
  publisher={IEEE}
}

@ARTICLE{surabhi2022trojan,
  author={Surabhi, Virinchi Roy and Krishnamurthy, Prashanth and Amrouch, Hussam and Henkel, Jörg and Karri, Ramesh and Khorrami, Farshad},
  journal={IEEE Transactions on Computers},
  title={Trojan Detection in Embedded Systems With {FinFET} Technology},
  year={2022},
  volume={71},
  number={11},
  pages={3061-3071}
}

@ARTICLE{surabhi2023golden,
  author={Surabhi, Virinchi Roy and Krishnamurthy, Prashanth and Amrouch, Hussam and Henkel, Jörg and Karri, Ramesh and Khorrami, Farshad},
  journal={IEEE Transactions on Computer-Aided Design of Integrated Circuits and Systems},
  title={Golden-Free Robust Age Estimation to Triage Recycled {ICs}},
  year={2023},
  volume={},
  number={},
  pages={1-1}
}

@INPROCEEDINGS{trusthub,
author={H. {Salmani} and M. {Tehranipoor} and R. {Karri}},
booktitle={Proceedings of the IEEE International Conference on Computer Design (ICCD)},
title={On design vulnerability analysis and trust benchmarks development},
year={2013},
volume={},
number={},
pages={471-474},
doi={10.1109/ICCD.2013.6657085},
ISSN={1063-6404},
month={Oct},}

\end{document}